\documentclass[notitlepage,aps, showpacs]{revtex4-1}

\usepackage{graphicx,epsfig}              
\usepackage[fleqn]{amsmath}   
\usepackage{mathtools}           
\usepackage{amsfonts}              
\usepackage{amsthm}  
\usepackage{lineno}
\usepackage{color}

\numberwithin{equation}{section}

\newcommand{\beqs}{\begin{eqnarray}}
\newcommand{\eeqs}{\end{eqnarray}}

\theoremstyle{definition}

\theoremstyle{remark}

\DeclareMathOperator\arctanh{arctanh}

\DeclareMathOperator\si{Si}

\DeclareMathOperator\ci{Ci}

\begin{document}

\title{One-Dimensional semirelativistic Hamiltonian with multiple Dirac delta potentials}

\author{Fatih Erman}
\affiliation{Department of Mathematics, \.Izmir Institute of Technology, Urla
35430, \.Izmir, Turkey}

\email{fatih.erman@gmail.com}

\author{Manuel Gadella}
\affiliation{
Departamento de F\'isica Te\'orica, At\'omica y \'Optica and IMUVA. Universidad de Valladolid, Campus Miguel Delibes, Paseo Bel\'en 7, 47011, Valladolid, Spain}

\email{manuelgadella1@gmail.com}

\author{Haydar Uncu}
\affiliation{Department of Physics, Adnan Menderes University, 09100, Ayd{\i}n, Turkey}

\email{huncu@adu.edu.tr}

\begin{abstract}
In this paper, we consider the one-dimensional semirelativistic Schr\"{o}dinger equation for a particle interacting with $N$ Dirac delta potentials. Using the heat kernel techniques, we establish a resolvent formula in terms of an $N \times N$ matrix, called the principal matrix. This matrix essentially includes all the information about the spectrum of the problem. We study the bound state spectrum by working out the eigenvalues of the principal matrix. With the help of the Feynman-Hellmann theorem, we analyze how the bound state energies change with respect to the parameters in the model. We also prove that there are at most $N$ bound states and explicitly derive the bound state wave function. The bound state problem for the two-center case is particularly investigated. We show that the ground state energy is bounded below, and there exists a self-adjoint Hamiltonian associated with the resolvent formula. Moreover, we prove that the ground state is nondegenerate. The scattering problem  for $N$ centers is analyzed by exactly solving the semirelativistic Lippmann-Schwinger equation. The reflection and the transmission coefficients are numerically and asymptotically computed for the two-center case.  
We observe the so-called threshold anomaly for two symmetrically located centers. The semirelativistic version of the Kronig-Penney model is shortly discussed, and the band gap structure of the spectrum is illustrated. The bound state and scattering problems in the massless case are also discussed. Furthermore, the reflection and the transmission coefficients for the two delta potentials in this particular case are analytically found. Finally, we solve the renormalization group equations and compute the beta function nonperturbatively. 
\end{abstract}


\maketitle

%%%%%%%%%%%%%%%%%%%%%%%%%%%%%%%%%%%%%%%%%%%%%%%%%%%%%%%%%%%%%%%%%%%%%

\section{Introduction}
\label{Introduction}

In nonrelativistic quantum mechanics, Dirac delta potentials are one class of  exactly solvable models, and they are useful to describe very short interactions between a single particle and a fixed heavy source. For this reason, they are also called contact or point interactions if Dirac delta function is pointlike. It is a good approximation to use them when the wavelength of the particle is much larger than the range of the potential. Besides their simplicity,  
they have a vast amount of applications for modeling real physical systems (see the recent review \cite{Belloni Robinett} and the books \cite{Demkov, albeverio} and references therein). A well-known model utilizing Dirac delta potentials in nonrelativistic quantum mechanics
is the so-called Kronig-Penney model \cite{KPmodel}, and it is actually a reference model in describing the band gap structure of metals in solid state physics \cite{Kittel}.

Moreover, pointlike Dirac delta potentials in two and three dimensions are known as simple pedagogical toy models in understanding several nontrivial concepts, originally introduced in quantum field theory, namely dimensional transmutation, regularization, renormalization, asymptotic freedom, etc. \cite{Thorn, BF, Hagen, PC, GT, MG, Jackiw, MT, Huang}. 
It is also a nontrivial subject from a purely mathematical point of view. 
One approach to define them properly is based on the theory of self-adjoint extensions of symmetric operators. This allows us to define rigorously the formal Hamiltonian for Dirac delta potentials as a self-adjoint extension of the local free kinetic energy operator \cite{albeverio, Albeverio 2000}.

As is well known, the relativistic extensions of the Schr\"{o}dinger equation, namely the Klein-Gordon and the Dirac equations, require the introduction of antiparticles, so they are inconsistent with the single particle theory. However, they describe the dynamics of quantum fields whose excitations are bosons or fermions.  
In other words, the Klein-Gordon and the Dirac equations indeed belong to the domain of quantum field theory. In contrast to the Klein-Gordon and the Dirac equations, the eigenvalue equation for the semirelativistic kinetic energy operator $\sqrt{P^2+m^2}$ does not require antiparticles since it has only positive energy solutions. Historically, it appeared as an approximation to the Bethe-Salpeter formalism \cite{SB51, S52} in describing the bound states in the context of relativistic quantum field theory. For this reason, this Hamiltonian $\sqrt{P^2 +m^2}$ is known as the free spinless Salpeter Hamiltonian. 
Moreover, widely used and rather successful models in phenomenological meson physics have been constructed by considering spinless Salpeter Hamiltonians with several potentials \cite{Kowalski, BM, BS}. 
It is important to emphasize that only the relativistic dispersion relation is imposed here, whereas 
relativistic invariance is not fully required (e.g., all the momentum integral measures are just $d p$). Therefore, the Salpeter Hamiltonian is a good approximation to relativistic systems in the domain, where the particle creations and annihilations are not allowed. On the other hand, the use of potentials for the interaction of two or more particles violates the principle of relativity even at the classical level. This is due to the fact that the message in the change of the position of the particle has to be received instantaneously by the other particle \cite{CJS, Leutwyler}. Nevertheless, it has been proposed in \cite{Al-Hashimi} that Dirac delta potentials could be an exception, and the following one-dimensional Salpeter Hamiltonian is considered:
\beqs
H= \sqrt{P^2+m^2} - \lambda \; \delta(x) \;. \label{Salpeter with delta}
\eeqs
Here $\lambda$ is the coupling constant or the strength of the interaction, and
the nonlocal kinetic energy operator (free part of the above Hamiltonian) is defined in momentum space as multiplication by $\sqrt{p^2+m^2}$ \cite{LiebLoss}. 
Similar to its nonrelativistic version in higher dimensions, this model has been used in order to illustrate  some quantum field theoretical  concepts in a simpler relativistic quantum mechanics context \cite{Al-Hashimi}.
This model was actually first discussed from the mathematical point of view as a self-adjoint extension of pseudodifferential operators in \cite{AlbeverioKurasov}. Moreover, an extension of the method developed in \cite{Al-Hashimi} to the derivative of the Dirac delta potentials has been  studied in \cite{Al-Hashimi2}.

In this paper, we study the generalization of the work \cite{Al-Hashimi} to finitely many Dirac delta potentials. Our formal one-dimensional spinless Salpeter Hamiltonian with $N$ Dirac delta potentials located at some fixed points $a_i$ is 
\beqs
H=\sqrt{P^2 +m^2} - \sum_{i=1}^{N} \lambda_i \; \delta(x-a_i) \;, \label{Hamiltonianformal}
\eeqs
where $\lambda_i$'s are the coupling constants (the strengths of the interaction), which are assumed to be positive throughout the paper.  We also assume that $a_i \neq a_j$ for $i \neq j$. This potential can be generated by $N$ heavy particles located at some certain fixed points. Then, a single particle interacts with these heavy particles through the Dirac delta potentials at those points. 

This is a very toy model of a relativistic particle trapped in one dimension, and it interacts with some impurities (in the massless case, it could be the photons trapped in one dimension that interact with the impurities).
Similar to the one-center (one delta potential) case, this problem must also require renormalization. 
Our approach here is to find the formal resolvent $(H-E)^{-1}$ or Green's function expression (see \cite{Park} for the nonrelativistic case) of the above Hamiltonian (\ref{Hamiltonianformal}) by renormalizing the coupling constant through the heat kernel techniques with emphasis on some general results on the spectrum of the problem.  
Green's function approach is rather useful since it includes all the  information about the spectrum of the Hamiltonian.   
The method we use here has been constructed in the nonrelativistic version of the model on two- and three-dimensional manifolds \cite{point Dirac on manifolds1, point Dirac on manifolds2} and in the nonrelativistic many-body version of it in \cite{turgutmanybody}.  A one-dimensional nonrelativistic many-body version of the model (\ref{Hamiltonianformal}), where the particles are interacting through the two-body Dirac delta potentials, is known as the  Lieb-Liniger model \cite{LiebLiniger} and has been studied in great detail in the literature \cite{McGuire, Yang1, Yang2, Calogero}.

It is well known that the heat kernel is a very useful tool in studying one-loop divergences, anomalies, asymptotic expansions of the effective action, and the Casimir effect in quantum field theory  \cite{Vas} and also in quantum gravity \cite{Avramidi}. Here, we claim that it can be used as a regularization of the above formal Hamiltonian (\ref{Hamiltonianformal}). This is essentially due to the fact that the heat kernel $K_t(x,y)$ converges to the Dirac delta function in the distributional sense so that the Hamiltonian can be regularized by replacing it with the heat kernel. One advantage of using the heat kernel is it may allow possible extensions to consider more general elliptic pseudodifferential free Hamiltonians (it may even include some regular potentials) since the only requirement to remove the divergent part is to have the information of short time asymptotic expansion of the heat kernel \cite{Gilkey}.
By renormalizing the coupling constant after the heat kernel regularization through the resolvent formalism, we obtain an explicit expression for the resolvent - a kind of Krein's formula \cite{Albeverio 2000}. It is given in terms of  an $N \times N$ holomorphic (analytic) matrix  [on the region $\Re{(E)}<m$]. This matrix is called the principal matrix (this terminology is originally introduced in \cite{Rajeev} for several toy field theoretical models), and it is essentially the only thing we need for discussing the bound states and scattering analysis of the problem. The results we obtain by the heat kernel regularization for the $N=1$ case are the same as the ones obtained by using the dimensional regularization in \cite{Al-Hashimi}.

After the renormalization procedure, we also address some formal issues that arise from physically important questions. For instance, one has to check whether the renormalization of the coupling constant is sufficient to remove all the divergences in the model so that we have physically meaningful results at the end. It is not obvious that the renormalization procedure guarantees that the ground state energy of our model is bounded from below. Here,
we show that this is indeed the case (this is necessary for every physical system \cite{LiebThirring}) and prove that there exists a unique self-adjoint operator associated with the resolvent formula we find. The issues about the self-adjointness can also be shown in the more abstract self-adjoint extension theory in mathematics literature (for one center, see \cite{AlbeverioKurasov, AlbeverioFassari}).

The discrete or the bound state spectrum of the one-dimensional spinless free Salpeter Hamiltonian perturbed by one and two Dirac delta potentials has been  rigorously discussed in \cite{AlbeverioFassari}. We obtain essentially the same results on the bound state spectrum for the two-center case. Additionally, we show some general results on the number of bound states for an arbitrary number of centers and study how the bound state energies change with respect to the parameters in the model by working out the principal matrix. We find an explicit expression for the bound state wave function for an arbitrary number of centers. Actually, no matter how many Dirac delta potentials there are in our system, the bound state wave function is calculated from the contour integration of the resolvent around its isolated simple poles. This wave function is shown to be pointwise bounded except at the location of the centers, as expected for any system in quantum mechanics \cite{simon}. Although it diverges at the points where Dirac delta potentials are located, it is still square integrable. However, the expectation value of the free Hamiltonian for the bound states is divergent. This is not surprising, and it basically tells us that the bound state wave function of the system does not belong to the domain of the free Hamiltonian. This gives us an intuitive idea why these interactions are defined  through the self-adjoint extension theory. 
Although we do not expect any degeneracy for bound states in one-dimensional quantum mechanics \cite{Shankar}, this may not be true for singular potentials \cite{Loudon} and for the semirelativistic Salpeter equation. Therefore, the non-degeneracy of the ground states in the context of Salpeter Hamiltonians is not obvious. In this paper, we show that the ground state of our model is nondegenerate as long as the distance between the centers is finite; then, the wave function for the ground state can be chosen to be positive.

We solve the semirelativistic Lippmann-Schwinger equation for the scattering problem of the $N$ Dirac delta potential. This is one of the main results of the paper. The reflection coefficient $R(k)$ and transmission coefficient $T(k)$ are explicitly calculated in closed analytical forms. We find the behavior of the reflection and transmission coefficients as functions of the energy of the incoming particle numerically. We also make an asymptotic approximation and obtain an analytical expression for the reflection and transmission coefficients when $k|a_i-a_j|$ is sufficiently large. 
In particular, for two centers located symmetrically around the origin, the results for the reflection and the transmission coefficients obtained from the asymptotic approximation is completely consistent with the one obtained numerically.
We see that the reflection and transmission coefficients behave like those in the nonrelativistic case. For example, the transmission coefficient $T(k)$ has some sharp peaks around certain values of $k$ where it becomes unity. These peaks  have been interpreted as resonances by some authors \cite{L, LS, AKSS} in the nonrelativistic case. However, they should not be confused with resonances as unstable states in quantum mechanics \cite{Arno}.

Furthermore, we realize one novel behavior of the reflection coefficient near very small values of $k/m$. It is surprising that  the reflection coefficient suddenly vanishes as the kinetic energy of the incoming particles goes to zero for a certain choice of the parameters. This phenomenon is actually known as the threshold anomaly in one-dimensional nonrelativistic quantum mechanics \cite{Senn}. The underlying reason for such an anomaly  is essentially the appearance of a bound state very close to the threshold energy (starting point of the continuum spectrum). The reflection coefficient generally goes to unity as we decrease the energy of the incoming particles. However, if physically meaningful continuum wave functions can be constructed for all values of $x$, and the potential $V(x)$ is symmetric [$V(-x)=V(x)$] and vanishes outside a finite region, and if it supports a bound state at threshold, then the reflection coefficient goes to zero at the threshold.  This is stated as a theorem in \cite{Senn} and is valid only for the above class of potentials in the nonrelativistic quantum mechanics.
Here, we show that the threshold anomaly also appears even in the semirelativistic case, where renormalization is required.
We find numerically and approximately (through asymptotic expansion) those critical values of the parameters that lead to the threshold anomaly. We show that these critical values of the parameters are those values for which the second bound state appears at threshold energy $E=m$.
In the massless case $m=0$, we can analytically obtain the reflection and transmission coefficients and show that the threshold anomaly occurs precisely at those values of the parameters for which the new bound state at threshold ($E=0$ in the massless case) appears. However, this anomaly in the massless case is slightly different from the nonrelativistic and semirelativistic massive case. The reflection coefficient always approaches zero as the energy of the incoming particles goes to zero no matter which values of the parameters are chosen. In any case, an anomalous behavior is observed as a sudden change in the reflection coefficient.  Moreover, we consider the semirelativistic version of the Kronig-Penney model, and the band gaps  in the spectrum are illustrated by examining the transmission coefficient.   
We also study the nonrelativistic limits of the bound state and scattering solutions, and these limits are consistent with the nonrelativistic results.

The model under this study is shown to be asymptotically free; that is, the scattered particle becomes free as its energies become higher and higher. In the massless case,  
the Hamiltonian initially does not contain any intrinsic energy scale due to the dimensionless coupling constants in natural units. However, a new set of parameters, namely the bound state energies to  each center, is introduced after the renormalization procedure. The appearance of the dimensional parameters is called dimensional transmutation and fixes the energy scale of the system. This can be interpreted as the simplest example of anomaly or quantum mechanical symmetry breaking, as in the nonrelativistic version of the problem  \cite{Jackiw}.  
We also derive the renormalization group equations and find the fixed points of the single beta function for the full system. All these issues have already been addressed in the context of a single Dirac delta potential, and studying such nontrivial concepts in quantum field theory in a single particle relativistic theory has been one of the main motivation in \cite{Al-Hashimi} from pedagogical reasons. We expect that extending the single center problem to many centers  may help to understand some nontrivial concepts in quantum field theory and bridge the huge gap between the quantum field theory and relativistic quantum mechanics.

The paper is organized as follows. In Sec. \ref{Renormalization of Relativistic Finitely Many Dirac delta Potentials through Heat Kernel}, we derive a resolvent formula for our model in terms of an $N\times N$ matrix (called principal matrix) by using the heat kernel as a regularization. In Sec. \ref{On Bound States}, we discuss the bound state spectrum and prove that the eigenvalues of the principal matrix are decreasing functions of energy, and we show that we have at most $N$ bound states. We also study how the eigenvalues of the principal matrix 
change with respect to the bound state energy of the $i$ th delta center (coming from the renormalization condition) and with respect to the distance between the centers. In Sec. \ref{Lower Bound on the Ground State Energy}, the ground state energy is shown to be bounded from below using the Ger\v{s}gorin theorem. Then, we briefly give a proof that there exists a unique self-adjoint operator associated with the resolvent formula obtained in the renormalization procedure (technical details are given in Appendix B).  In Sec. \ref{Bound State Wave Function For $N$ Centers}, we find the bound state wave function by computing the contour integral of the resolvent around one of its isolated simple poles  and discuss its nonrelativistic limit. In Sec. \ref{Pointwise Bound on the Bound State Wave Function and Expectation Value of the Free Hamiltonian}, we show that the bound state wave function is exponentially pointwise bounded and diverges at the location of the centers. Then, we point out that the wave function is square integrable, but the expectation value of the free Hamiltonian in the bound state is divergent. In Sec. \ref{Non-degeneracy of the Ground State}, we prove that the ground state is nondegenerate unless the centers are infinitely far away from each other, and the wave function for the ground state can be chosen strictly positive. In Sec. \ref{Scattering Problem for $N$ Centers}, we exactly solve the semirelativistic Lippmann-Schwinger equation for the problem and find an explicit expression for the reflection and transmission coefficients. Furthermore, we discuss the threshold anomaly and show that it occurs near the border of the continuum energy spectrum.  In Sec. \ref{Massless Case}, we analytically study the bound state and scattering problem in the massless case. Finally, we derive the renormalization group equations and compute the $\beta$ function for the model in Sec. \ref{RGEquationssection} and shortly introduce a possible extension of the model in Sec. \ref{A Possible Extension of the Model}. Section \ref{Conclusions} contains our conclusions, and Appendix A includes the proof of the analyticity of the principal matrix.

%%%%%%%%%%%%%%%%%%%%%%%%%%%%%%%%%%%%%%%%%%%%%%%%%%%%%%%%%%%%%%%%%%%%%

\section{Renormalization of Relativistic Finitely Many Dirac delta Potentials through Heat Kernel}
\label{Renormalization of Relativistic Finitely Many Dirac delta Potentials through Heat Kernel}
We consider the time-independent Schr\"{o}dinger equation (also called Salpeter equation) for the Hamiltonian (\ref{Hamiltonianformal})  
\beqs \label{1}
 \langle x | H | \psi \rangle & = & 
\langle x | H_0 |\psi \rangle  -\sum_{i=1}^N \lambda_i \delta(x-a_i) \psi(x) 
= \langle x | \left( H_0 - \sum_{i=1}^N \lambda_i | a_i \rangle
\langle a_i |  \right)| \psi \rangle = E \; \psi(x) \,,
\end{eqnarray}
where $H_0=\sqrt{P^2+m^2}$ and the kets $|a_i\rangle$ are the eigenkets of the position operator with eigenvalue $a_i$. The second equality in Eq. (\ref{1}) is just the consequence of the property of Dirac delta function, $\delta(x-a_i) \psi(x) = \delta(x-a_i) \psi(a_i)$.  
We will use the units such that $\hbar=c=1$ throughout the paper.  We first find the regularized resolvent for the regularized version of the above Hamiltonian. We propose that the regularized Hamiltonian is
\beqs
H_{\epsilon} = H_0 - \sum_{i=1}^{N} \lambda_i(\epsilon) |a_i^{\epsilon} \rangle \langle a_i^{\epsilon}| \;,
\eeqs
where we have introduced short ``time" cutoff $\epsilon$ through the heat kernel $K_{\epsilon/2} (x,a_i)=\langle x| a_i^{\epsilon} \rangle$ and made the coupling constants explicitly dependent $\epsilon$. The heat kernel is defined as the fundamental solution
to the following heat equation \cite{LiebLoss}:
\beqs   H_0 \; K_t(x,y) = - {\partial K_t(x,y)
\over
\partial t} \;. \label{relativistic heat equation}\eeqs
The expression $|a_i^{\epsilon} \rangle \langle a_i^{\epsilon}|$ written in Dirac's bra-ket notation is just the projection operator onto the space spanned by $|a_{i}^{\epsilon} \rangle$ in $L^2(\mathbb{R})$.   
The reason why the heat kernel works for the regularization of the problem is based on the fact that it converges to the Dirac delta function in the distributional sense as the cutoff is removed, i.e., $\langle x| a_i^{\epsilon} \rangle \rightarrow \langle x | a_i \rangle = \delta(x-a_i)$ as $\epsilon \rightarrow 0^+$. In other words, we recover the original Hamiltonian when the cutoff goes to zero.

To find the regularized resolvent $R_\epsilon (E) =(H_\epsilon -E)^{-1}$, we will solve the following inhomogenous equation: 
\beqs \label{reseq} \left( H_0  - \sum_{j=1}^N \lambda_i(\epsilon) | a_j^{\epsilon} \rangle
\langle a_j^\epsilon | -E \right) | \psi \rangle = | \rho  \rangle \;, \eeqs
assuming complex number $E \not \in \mathrm{Spec}(H_{0})$. Let $|f_i^\epsilon \rangle= \sqrt{\lambda_i(\epsilon)} |a_i^\epsilon \rangle$ or $\langle x |f_i^\epsilon \rangle = \sqrt{\lambda_i(\epsilon)}  K_{\epsilon/2} (x,a_i)$. Then, after acting
with the operator $(H_0 - E)^{-1}$ on both sides from left, we obtain 
\beqs \label{psi}
|\psi \rangle  = \sum_{j=1}^{N} (H_0-E)^{-1} | f_j^\epsilon \rangle \langle f_j^\epsilon | \psi \rangle + (H_0-E)^{-1} | \rho \rangle \;.
\eeqs
If we project this onto
$\langle f_i^\epsilon |$, we get
\beqs \label{Tequation} \sum_{j = 1}^N T_{ij}(\epsilon, E) \langle
f_j^{\epsilon} | \psi \rangle =  \langle f_i^\epsilon |\left( H_0 - E
\right)^{-1} | \rho \rangle \;, \eeqs
where 
\beqs \label{Tmatrix} T_{ij} (\epsilon, E) =
\begin{cases}
\begin{split}
1 - \langle f_i^\epsilon | \left( H_0 - E \right)^{-1}|f_i^\epsilon \rangle
\end{split}
& \textrm{if $i = j$} \\
\begin{split}
- \; \langle f_i^\epsilon |\left( H_0 - E \right)^{-1} |
f_j^\epsilon \rangle
\end{split}
& \textrm{if $i \neq j$}.
\end{cases}
\eeqs
By solving $\langle
f_j^{\epsilon} | \psi \rangle$ from the above matrix equation (\ref{Tequation}) and substituting it into Eq. (\ref{psi}), we obtain the regularized resolvent
\beqs
R_\epsilon(E) = (H_0-E)^{-1} + \left( H_0 - E \right)^{-1} \left( \sum_{i,j=1}^N | f_i^\epsilon \rangle
\left[T^{-1}(\epsilon, E) \right]_{ij} \langle f_j^\epsilon | \right) \left( H_0 - E
\right)^{-1}\;.
\eeqs
We now go back to the original variables and define a new matrix (called regularized principal matrix)
\beqs \label{Phimatrixepsilon} \Phi_{ij} (\epsilon, E) =
\begin{cases}
\begin{split}
{1 \over \lambda_i(\epsilon)}  - \langle a_i^\epsilon | \left( H_0 - E \right)^{-1}|a_i^\epsilon \rangle
\end{split}
& \textrm{if $i = j$} \\
\begin{split}
- \; \langle a_i^\epsilon |\left( H_0 - E \right)^{-1} |
a_j^\epsilon \rangle
\end{split}
& \textrm{if $i \neq j$} \;,
\end{cases}
\eeqs
so that we get
\beqs
R_\epsilon(E) = (H_0-E)^{-1} + \left( H_0 - E \right)^{-1} \left( \sum_{i,j=1}^N | a_i^\epsilon \rangle
\left[\Phi^{-1}(\epsilon, E) \right]_{ij} \langle a_j^\epsilon | \right) \left( H_0 - E
\right)^{-1}\;.
\eeqs
We can express the resolvent of the free Hamiltonian in terms of the heat kernel associated with $H_0$ in the following way. The integral representation of the resolvent of  $H_0$ is given by \cite{Pazy_Semigroups}
\beqs
(H_0-E)^{-1}=\int_{0}^{\infty} dt \; e^{-t(H_0-E)} \;. \label{intrep}
\eeqs
For $H_0=\sqrt{P^2 +m^2}$, we have $||e^{-t \sqrt{P^2+m^2}}|| \leq e^{-m t}$ for all $t \geq 0$. Then,  the integral (\ref{intrep}) exists if $\Re{(E)}<m$. Equivalently, the above integral can be expressed as $R_0(x,y|E)=\langle x | (H_0-E)^{-1} | y \rangle = \int_{0}^{\infty} dt \; K_t(x,y) \; e^{t E}$ by sandwiching it with $\langle x|$ and $|y \rangle$. The expression of Green's function as an integral of the heat kernel was first used in quantum field theory by Fock \cite{Fock} and Schwinger \cite{Schwinger1}. Hence, it follows that  
\beqs
\langle a_i^\epsilon |\left( H_0 - E \right)^{-1} |
a_j^\epsilon \rangle &=& \int_{-\infty}^{\infty} \int_{-\infty}^{\infty} \; d x \; dy \; K_{\epsilon/2}(x, a_i)   \int_{0}^{\infty} dt \; K_t(x,y) \; e^{t E} \; K_{\epsilon/2}(y,a_j) \cr &=& \int_{0}^{\infty} dt \; K_{t+ \epsilon} (a_i,a_j) \; e^{t E} \;,
\end{eqnarray}
where we have used the semigroup property  of the heat kernel 
\beqs
\int_{-\infty}^{\infty} d z \; K_{t_1}(x,z) K_{t_2}(z,y) = K_{t_1 + t_2}(x,y) \;, \label{semigroupheat}
\eeqs 
for all $x,y$ and $t_1, t_2 \geq 0$. If we now take the limit $\epsilon \rightarrow 0^+$, before taking the integral above with respect to $x$ and $y$, and assume that the function $\int_{0}^{\infty} d t \; e^{t E} K_t(x,y)$ belongs to some class of test functions of each variable $x$ and $y$ for $\Re{(E)}<m$,  we obtain $\langle a_i^\epsilon |\left( H_0 - E \right)^{-1} |
a_j^\epsilon \rangle \rightarrow \int_{0}^{\infty} d t \; e^{t E} K_t(a_i,a_j)$ as $\epsilon \rightarrow 0^+$.

The integral $\int_{0}^{\infty} d t \; K_t (a_i,a_i) e^{t E}$ in the diagonal part of the matrix (\ref{Phimatrixepsilon}) is actually divergent, whereas the integrals in the off-diagonal terms are convergent. This can be shown as follows.

The explicit expression of the heat kernel associated with the operator $\sqrt{P^2 +m^2}$ is given in \cite{LiebLoss} by the following formula:
\beqs
K_t(x,y)={m t \over \pi \sqrt{(x-y)^2 + t^2}} \; K_1 (m \sqrt{(x-y)^2 + t^2}) \;, \label{heatkernel}
\eeqs
for any $x,y \in \mathbb{R}$ and $t>0$. Here, $K_1$ is the modified Bessel function of the first kind. This is easily derived by using the so-called subordination identity
\beqs
e^{-t A} = {t \over 2 \sqrt{\pi}} \int_{0}^{\infty} d u \; {e^{-t^2/4u -u A^2} \over u^{3/2}} \;,
\eeqs
for $A=\sqrt{P^2 +m^2}$. 

For large values of $t$, the diagonal part of the principal matrix is convergent for $\Re{(E)} < m$ due to the asymptotic behavior of the Bessel function $K_1(m t) \sim {m \over \pi} \sqrt{{\pi \over 2 m t}} e^{- t m}$ as $t \rightarrow \infty$ \cite{Lebedev}. Moreover,    
the exponential upper bound of the Bessel function 
\beqs
K_1(x) < e^{-x/2} \left({1 \over x} + {1 \over 2}\right) \;, \label{besselupperbound}
\eeqs
for all $x>0$, which was given in \cite{point Dirac on manifolds2} by using its integral representation,  guarantees that the integral $\int_{0}^{\infty} dt \; K_{t} (a_i,a_j) \; e^{t E}$ is finite.
However, the integral in the diagonal part of matrix (\ref{Phimatrixepsilon}) is divergent due to the asymptotic behavior 
\beqs
K_1(m t) \sim {1\over m t} \;, \label{asymtoticBessel} 
\eeqs
as $t \rightarrow 0$ \cite{Lebedev}.

Let us temporarily consider the one-center case ($N=1$) for simplicity. Suppose that the $i$th center is isolated from all other centers. Then the regularized principal matrix is just a single function for the $i$th center and reads  
\beqs
\Phi_{ii}(\epsilon, E) = {1 \over \lambda_i(\epsilon)} - \int_{0}^{\infty} d t \; K_{t+\epsilon} (a_i,a_i) \; e^{t E} \;,
\eeqs
for any $i=1, \ldots, N$.  If we choose the bare running coupling constants 
\beqs \label{couplingconstantrenormalization}
{1 \over \lambda_i(\epsilon)} = {1 \over \lambda_i^R (M_i)} + \int_{0}^{\infty} d t \; K_{t+\epsilon} (a_i,a_i) \; e^{t M_i} \;,
\eeqs
where $M_i$ is the renormalization scale and we take the limit as $\epsilon \rightarrow 0^+$, we obtain
a nontrivial finite expression for the resolvent for  a single delta potential,
\beqs \label{renormalized resolvent 1 center}
R(E) = (H_0-E)^{-1} + \left(H_0 - E \right)^{-1}  | a_i \rangle
\left[\Phi^{-1}_{ii}(E) \right]\langle a_i|  \left( H_0 - E
\right)^{-1}\;,
\eeqs
where  the function $\Phi_{ii}$ is
\beqs \label{Phi one center} \Phi_{ii} (E) = {1 \over \lambda_i^R(M_i)}
+ \int_{0}^{\infty} d t  \; K_{t} (a_i,a_i) \; (e^{t M_i} -e^{t E}) \;,
\eeqs
for all $i$ and $\Re{(E)}<m$. 
Since the poles of the resolvent are the bound state energies, and the above resolvent formula includes the reciprocal of the function $\Phi_{ii}(E)$, its zeros determine the bound state spectrum of the model. 

The above renormalization scale $M_i$ could possibly be eliminated in favor of a physical parameter by imposing the renormalization condition. For instance, the renormalization scale can be chosen to be equal to the bound state energy of the particle to the $i$th center, say  $E_B^i$ (it must be less than $m$ for bound states), so that
\beqs
\Phi_{ii}(E_B^i)=0 \;.
\eeqs

Therefore, for bound state problems, it is very convenient to choose the renormalization scale to be the bound state energy by setting $1/\lambda_i^R = 0$ so that we eliminate the unphysical scale $M_i$.

If we apply the same argument to the several center case, we end up with the following resolvent formula:
\beqs \label{renormalized resolvent}
R(E) = (H_0-E)^{-1} + \left(H_0 - E \right)^{-1} \left( \sum_{i,j=1}^N | a_i \rangle
\left[\Phi^{-1}(E) \right]_{ij} \langle a_j| \right) \left( H_0 - E
\right)^{-1}\;,
\eeqs
where 
\beqs \label{Phimatrixheatkernel} \Phi_{ij} (E) =
\begin{cases}
\begin{split}
 \int_{0}^{\infty} d t \; K_{t} (a_i,a_i) \; (e^{t E_{B}^{i}} - e^{t E})
\end{split}
& \textrm{if $i = j$} \\
\begin{split}
- \int_{0}^{\infty} d t  \; K_{t} (a_i,a_j) \; e^{t E}
\end{split}
& \textrm{if $i \neq j$} \;,
\end{cases}
\eeqs
defined on the complex $E$ plane, where $\Re{(E)}<m$. We shall call the matrix $\Phi_{ij}(E)$ the principal matrix. The above formula can be extended onto the largest possible subset of the complex plane by analytic continuation. Here it is important to note that the principal matrix satisfies $\Phi^{\dagger}(E)=\Phi(E^*)$. The resolvent formula (\ref{renormalized resolvent}) is a kind of Krein's formula \cite{Albeverio 2000} and is expressed in terms of the heat kernel. This implies that it is a rather general formula in the sense that the heat kernel for the Salpeter free Hamiltonian may in principle be replaced by a much more general heat kernel associated with a free pseudo-differential operator. In particular, the formula contains the massless case $m=0$. In this case,  the heat kernel associated with $H_0=|P|$ is given by \cite{LiebLoss}
\beqs
K_t(x,y)= {1 \over \pi} \left( {t \over t^2 +(x-y)^2} \right) \;. \label{heatkernelmassless}
\eeqs
The principal matrix (\ref{Phimatrixheatkernel}) can also be expressed in the momentum space by  using the completeness relation $\int_{-\infty}^{\infty} {d p \over 2\pi} \; |p\rangle \langle p| =1$, 
\beqs
K_t(a_i,a_j) &=& \langle a_i | e^{-t \sqrt{P^2 +m^2}} |a_j \rangle = \int_{-\infty}^{\infty} {d p \over 2 \pi} \; e^{i p (a_i-a_j)} \; e^{-t \sqrt{p^2 +m^2}} \;.
\eeqs
Substituting this into Eq. (\ref{Phimatrixheatkernel}) and changing the order of integrations, we obtain 
\beqs \label{Phimatrixmomentum} \Phi_{ij} (E) =
\begin{cases}
\begin{split}
 \int_{-\infty}^{\infty} {d p \over 2\pi} \; \left( {1 \over \sqrt{p^2+m^2} -E_{B}^{i}} - {1 \over \sqrt{p^2+m^2} -E} \right)
\end{split}
& \textrm{if $i = j$} \\ \\
\begin{split}
- \int_{-\infty}^{\infty} {d p \over 2 \pi}  \;  {e^{i p (a_i-a_j)} \over \sqrt{p^2+m^2} -E }
\end{split}
& \textrm{if $i \neq j$} \;,
\end{cases}
\eeqs
where $\Re{(E)}<m$. The integral in the diagonal terms can be directly evaluated 
\beqs \label{Phi1a} & & \Phi_{ii}(E)  =  {E_B^i \over \pi \sqrt{m^2 -(E_B^i)^2}} \left( {\pi \over 2} + \arctan {E_B^i \over \sqrt{m^2-(E_B^i)^2}}\right)  - {E \over \pi \sqrt{m^2 -E^2}} \left( {\pi \over 2} + \arctan {E \over \sqrt{m^2-E^2}}\right) \;.
\eeqs 
The off-diagonal elements are actually the free resolvent kernels, and these integrals have been expressed in the following form by using the residue theorem in \cite{Al-Hashimi, AlbeverioKurasov}:
\beqs
\Phi_{ij} (E) = 
\begin{cases}
\begin{split}
-  {1 \over \pi} \int_{m}^{\infty} d \mu \; e^{- \mu |a_i-a_j|} \; {\sqrt{\mu^2 -m^2} \over \mu^2 -m^2 + E^2}
\end{split}
& \textrm{if $\Re{(E)}<0$} \\ \\
\begin{split}
- i \; {e^{i \sqrt{E^2 -m^2} |a_i-a_j|} \over \sqrt{1-{m^2 \over E^2}}} - {1 \over \pi} \int_{m}^{\infty} d \mu \; e^{- \mu |a_i-a_j|} \; {\sqrt{\mu^2 -m^2} \over \mu^2 -m^2 + E^2}
\end{split}
& \textrm{if $\Re{(E)}>0$} \;,
\end{cases} \label{Phioffdiagonalexplicit}
\eeqs 
where $i \neq j$ and $\Im{(E)}>0$. Here the integral over the variable $\mu$ comes from the integration over the branch cut along $[i m, i \infty)$. Expressing the integral in the off-diagonal part of the principal matrix (\ref{Phimatrixmomentum}) by Eq. (\ref{Phioffdiagonalexplicit}) is very useful when we study the spectrum of the problem.

%%%%%%%%%%%%%%%%%%%%%%%%%%%%%%%%%%%%%%%%%%%%%%%%%%%%%%%%%%%%%%%%%%%%%

\section{On the Bound State Spectrum}
\label{On Bound States}

Since the bound state spectrum can be found from the poles of resolvent, the bound states energies should only come from the points of the real $E$ axis such that the principal matrix is not invertible; i.e., the bound state energies must be the solution of the characteristic equation for the principal matrix
\beqs \label{detbound}
\det \Phi(E) =0 \;.
\eeqs
This is essentially the result of the fact that free resolvent has no point or bound state spectrum, and it has only a continuous spectrum starting from $m$ on the real $E$ axis. Equation (\ref{detbound}) is rather difficult to solve in general since it is a transcendental equation.

Let us recall the following terminology introduced for the single center problem in \cite{Al-Hashimi}. We call the bound state   
\begin{itemize} 
\item[(a)] weak if $0 < E < m$;
\item[(b)] strong if $-m < E <0$;
\item[(c)] ultrastrong if $E < -m$.  
\end{itemize}
It must be emphasized here that the bound state energy is already fixed in the single center case by $E_B$ from the renormalization condition. 

To study the bound state spectrum, we may use an alternative but a much more useful approach in determining the general behavior of the bound states. We first notice that the solutions of Eq. (\ref{detbound}) are actually zeros of the eigenvalues of the principal matrix.  Let 
\beqs \Phi(E) A(E)=\omega(E) A(E) \;, \eeqs
be the eigenvalue equation for the principal matrix. For real values of $E$, the principal matrix is Hermitian due to the symmetry property of the heat kernel $K_t(a_i,a_j)=K_t(a_j,a_i)$ so all its eigenvalues are real valued and depend on the real variable $E$.  
We are now going to show that the eigenvalues of the principal matrix are decreasing functions of $E$. For simplicity, we will show this fact for the nondegenerate case without loss of generality (it can be generalized to the degenerate case as well).  
To prove this, we first need to show that the principal matrix is holomorphic (analytic) on the complex plane $\Re{(E)}<m$. Since it is a little technical issue, we give the proof of it in Appendix A by following a similar idea given in \cite{numberofboundstates}. This allows us to interchange the order of integration and the derivative so we can take derivatives under the integral signs.

Using  the Feynman-Hellmann theorem \cite{feynman, hellmann}, the derivative of the $k$th eigenvalue $\omega^k$ is given by 
\beqs
   {\partial \omega^k(E)\over \partial E} & = & \sum_{i,j=1}^{N}
(A^{k}_{i}(E))^* \; {\partial \Phi_{ij}(E) \over \partial E} \; A^{k}_{j}(E) \;. \label{feynmanhelmann} \eeqs
Inserting
\beqs {\partial \Phi_{ij}(E) \over \partial E} = - \int_{0}^{\infty} d t \; t \;  K_t(a_i,a_j) \; 
e^{t E} \;, \label{derivative of phi} \eeqs
into Eq. (\ref{feynmanhelmann}), we obtain
\beqs  {\partial \omega^k(E) \over \partial E} & = & - \sum_{i,j=1}^{N} (A^{k}_{i}(E))^* \int_{0}^{\infty} d t \; t \; e^{t E} K_t(a_i, a_j) \; A^{k}_{j}(E) \;. \eeqs
Then, using the semigroup property of the heat kernel (\ref{semigroupheat}) and changing the  integration variables $t=t_1+t_2$ and $u=t_1-t_2$, and integrating over the new variable $u$, we find
\beqs
 {\partial \omega^k(E) \over \partial E} & = &  - \int_{-\infty}^{\infty} d x \;
\sum_{i,j=1}^{N} (A^{k}_{i}(E))^*  \left( \int_{0}^{\infty} d t_1 \; e^{t_1 E} K_{t_1}(x, a_i) \right) \; \left( \int_{0}^{\infty} d t_2 \; e^{t_2 E} K_{t_2}(x, a_j) \right) \;  A^{k}_{j}(E) \cr & = & - \int_{-\infty}^{\infty} d x \;
\left| \sum_{i=1}^{N} A_{i}^{k}(E) \; \int_{0}^{\infty} d t \; K_{t}(x,a_i) \; e^{t  E} \right|^2 < 0 \;.
\label{derivative of lambda evaluated nu} \eeqs
The above fact implies that the eigenvalues of the principal matrix are decreasing functions of $E$.
As a consequence of this fact, there are at most $N$ bound states (including the weak, strong, and ultrastrong ones) since there are at most $N$ distinct eigenvalues that cross the $E$ axis $N$ times at most.

Moreover, the zeros of the eigenvalues shift to the right as we increase $E_B^i$. This is physically expected and can be proved by the following argument: First we can show by following the same arguments above that ${\partial \omega^k \over \partial E_B^i} >0$ for fixed values of $E$ and adjacent distances between the centers.  This tells us that for a given $E$, the $k$th eigenvalue $\omega^k$ is shifted upward as we increase $E_B^i$. Then, the zero of each $k$th eigenvalue $\omega^k$ is shifted toward the larger values of $E$. It is important to notice that no matter how small the values of $E_B^i$ are, the zeros of the eigenvalues cannot be arbitrarily small, i.e., the ground state energy must be bounded from below. We will prove this in the next section. 

It is also interesting to study the behavior of the eigenvalues as functions of the distance between the centers. From the explicit expression of the principal matrix (\ref{Phioffdiagonalexplicit}), all its off-diagonal elements are decreasing functions of $|a_i-a_j|$ in magnitude. This means that all the off-diagonal terms vanish as $|a_i-a_j| \rightarrow \infty$. Hence, the principal matrix eventually becomes a diagonal matrix so that its eigenvalues are its diagonal elements. In other words, $\omega^k \rightarrow \Phi_{kk}$. If they converge to the same diagonal term $\Phi_{kk}$ (this is the case only if all $E_B^i$s are the same), then we have degenerate bound states.

The two-center case ($N=2$):

Let us consider now the particular case where we have twin ($E_B^1 =E_B^2=E_B$) centers located symmetrically around the origin ($a_1=-a_2=-a$). Equation (\ref{detbound}) in this particular case simply turns out to be 
\beqs
\Phi_{ii}(E)= \pm \Phi_{ij}(E) \;, \hspace{1cm}  \mathrm{for} \; \mathrm{all} \; \; i,j=1,2 \;.
\label{2deltaboundstateeq}\eeqs
The bound state energies are the solutions to the above transcendental equation for the region $E<m$. It is easy to see from Eqs. (\ref{Phi1a}) and (\ref{Phioffdiagonalexplicit}) that 
the diagonal and the off-diagonal elements of the principal matrix are always decreasing functions of $E$ for all $E<m$. This means that a solution to the equation $\Phi_{ii}(E)= \Phi_{ij}(E)$ may or may not exist. However, there is always one and only one solution to the equation $\Phi_{ii}(E)= - \Phi_{ij}(E)$ since the right-hand side is a positive increasing function of $E$, whereas the left-hand side is a decreasing function of $E$. It is also important to emphasize that the diagonal part of the principal matrix is positive when $E<E_B$ and negative when $E>E_B$. Combining all these arguments implies that there is at least one solution to Eq. (\ref{2deltaboundstateeq}). Because of this fact, we can call the solution to the equation $\Phi_{ii}(E)= - \Phi_{ij}(E)$ the ground state, whereas the solution to the equation $\Phi_{ii}(E)=\Phi_{ij}(E)$ is the excited state.

\begin{figure}[h!]
\centering
\begin{minipage}{6cm}
\includegraphics[scale=0.5]{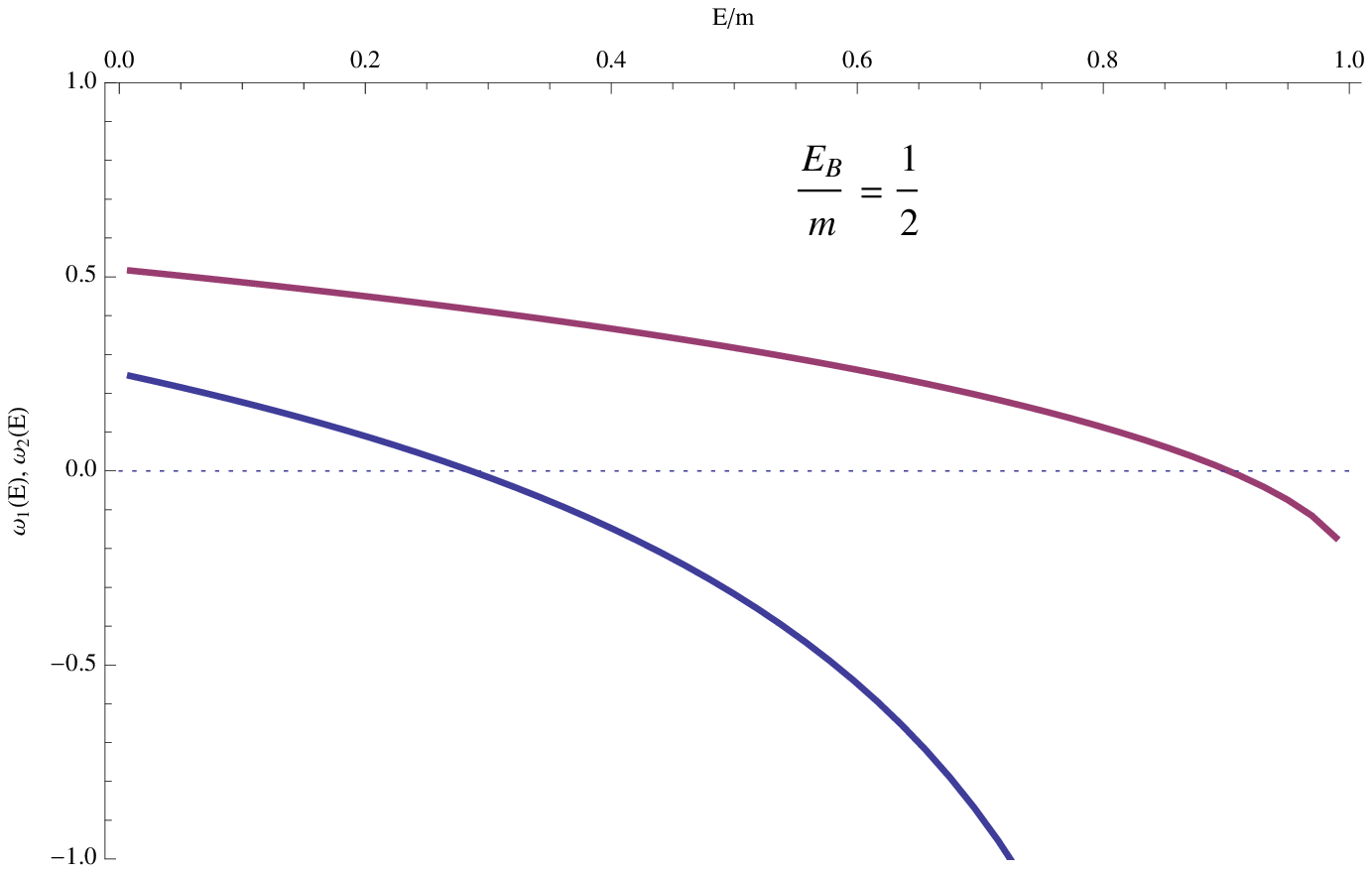}
\end{minipage}
\qquad \qquad
\begin{minipage}{6cm}
\includegraphics[scale=0.5]{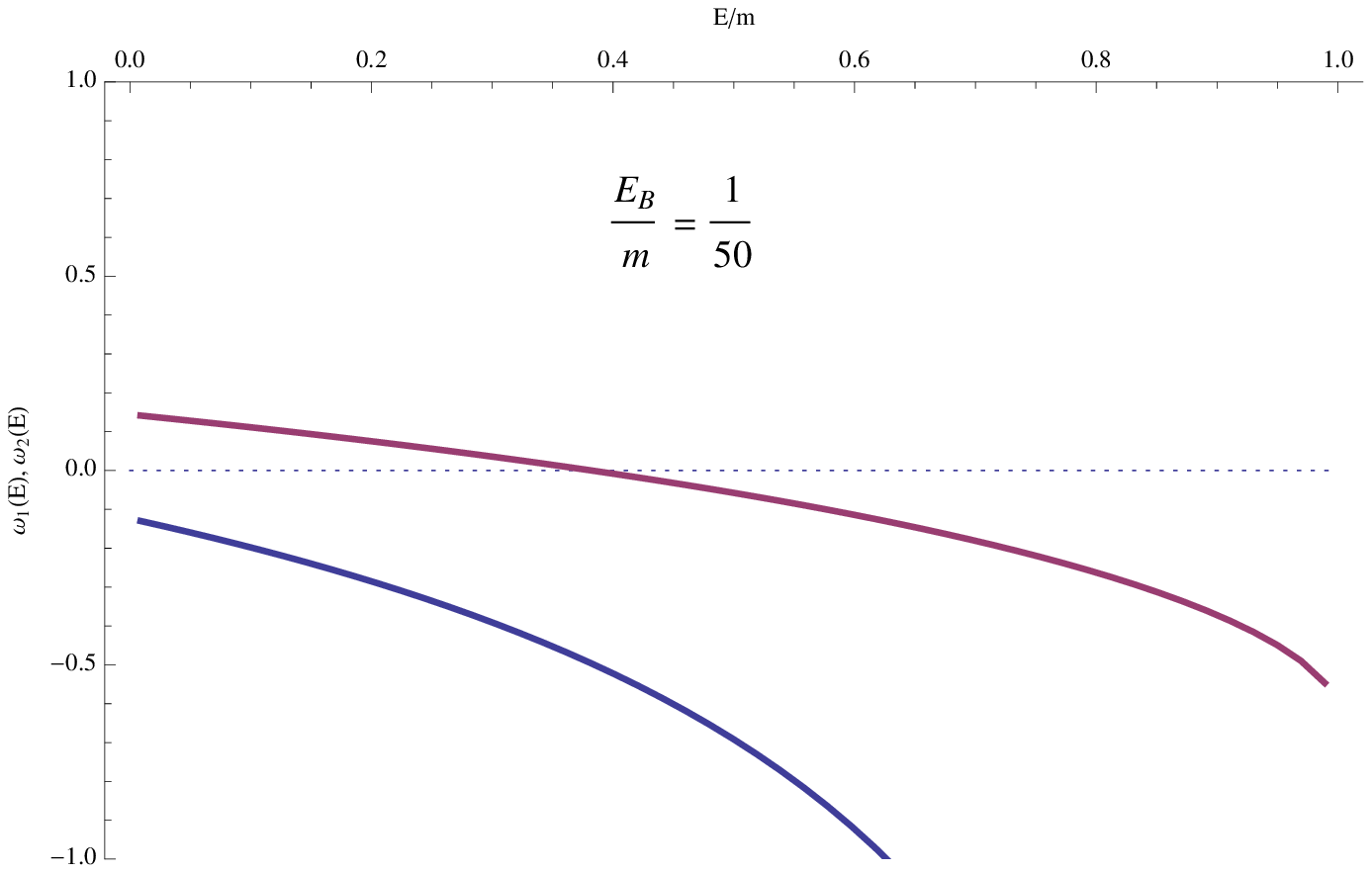}
\end{minipage}
\caption{Eigenvalues $\omega_1$ and $\omega_2$ as a function of $E/m$ for different values of $E_B/m$ (assuming that delta centers are twin, i.e., $E_B^1=E_B^2$ for simplicity) and $2 m a=1$. Here $a_1=-a$ and $a_2=a$.} \label{omega12versusE} 
\end{figure}

We can test all these arguments by finding the eigenvalues of the principal matrix numerically.  Evaluating the integral in the off-diagonal elements (\ref{Phioffdiagonalexplicit}) of the principal matrix numerically by \textit{Mathematica}, we can find its eigenvalues and plot them as a function of $E/m$ for the given values  of $E_B/m$ and $2ma$, as shown in Fig. \ref{omega12versusE}. This shows that the eigenvalues are decreasing functions of $E$ and the bound state energies are shifting to its larger values as $E_B/m$ increases, as expected.

\begin{figure}[htp]
\centering
\includegraphics[scale=0.5]{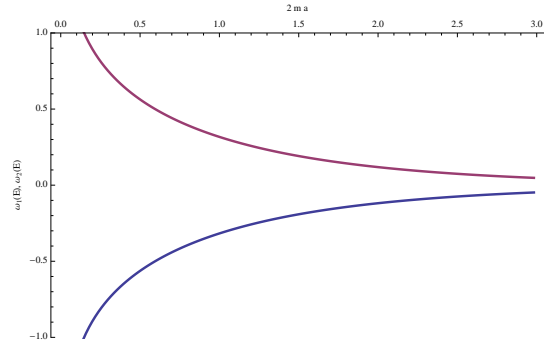}
\caption{Eigenvalues $\omega_1$ and $\omega_2$ as a function of $2 m a$ for the values $E_B/m=1/2$ and $E/m=1/2$.} \label{eigenvaluesversusa1}
\end{figure}

\begin{figure}[htp]
\includegraphics[scale=0.5]{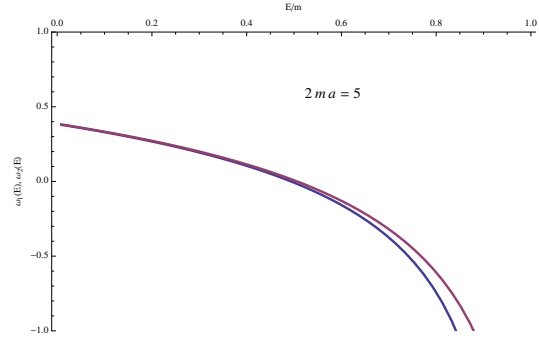}
\caption{Eigenvalues $\omega_1$ and $\omega_2$ as a function of $E/m$ for $2ma=5$ and $E_B/m=1/2$.} \label{omega12versusE2} 
\end{figure}

Moreover, as shown above for the general case, we confirm from Figs. \ref{eigenvaluesversusa1} and \ref{omega12versusE2} that $\omega^1 \rightarrow \omega^2$ as the distance between the centers goes to infinity. When the centers are infinitely far away from each other and $E_B^1=E_B^2$, then we have only one bound state so that the ground state becomes degenerate in this limiting case. 

This behavior has been already observed in \cite{AlbeverioFassari} (see Fig. 7 there) and illustrated by directly studying the flow of the bound state energies. Here we show this by working out the eigenvalues of the principal matrix. 

\begin{figure}[htp]
\centering
\begin{minipage}{5cm}
\includegraphics[scale=0.5]{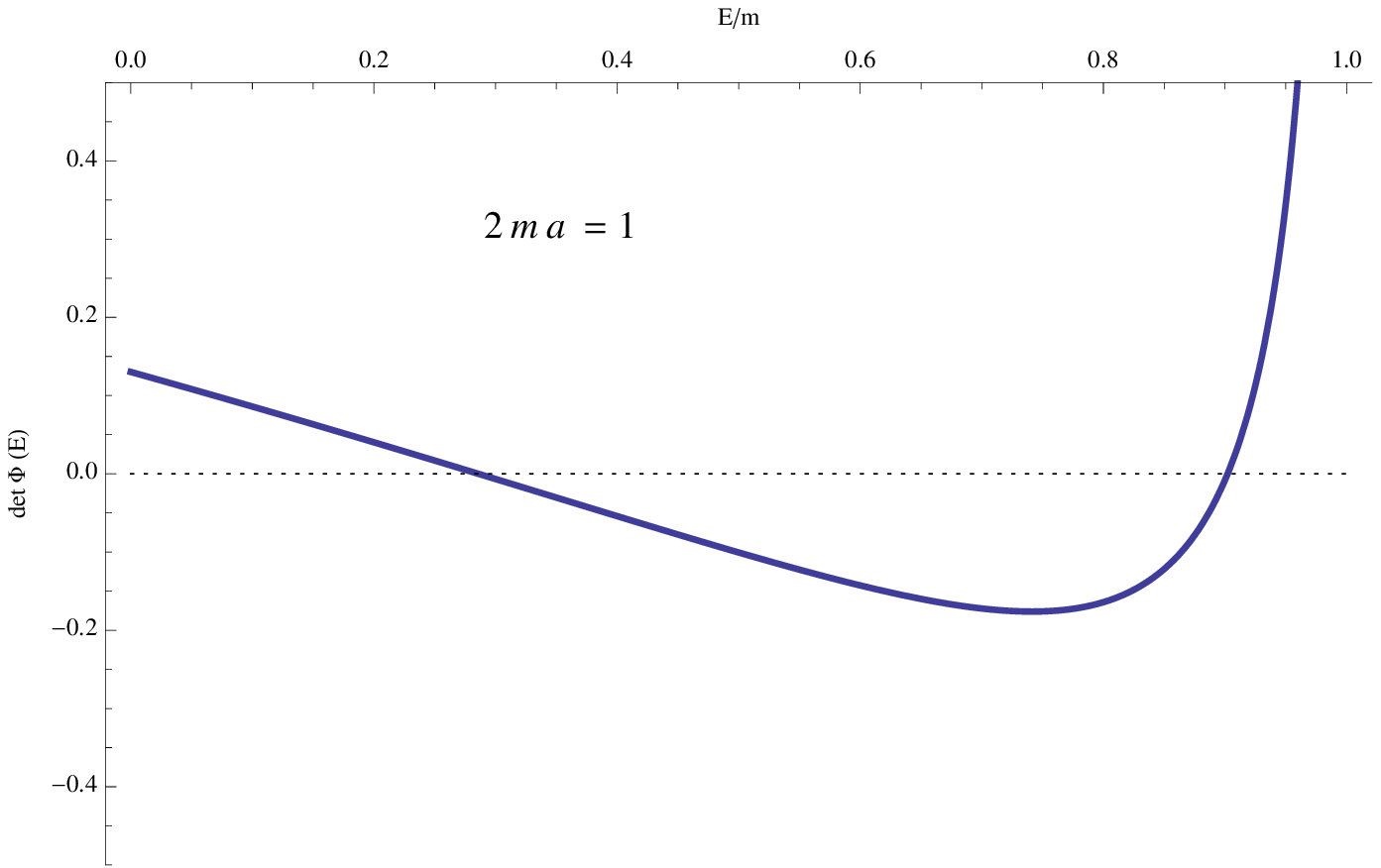}
\end{minipage}
\qquad \qquad \qquad \qquad  \qquad  
\begin{minipage}{5cm}
\includegraphics[scale=0.5]{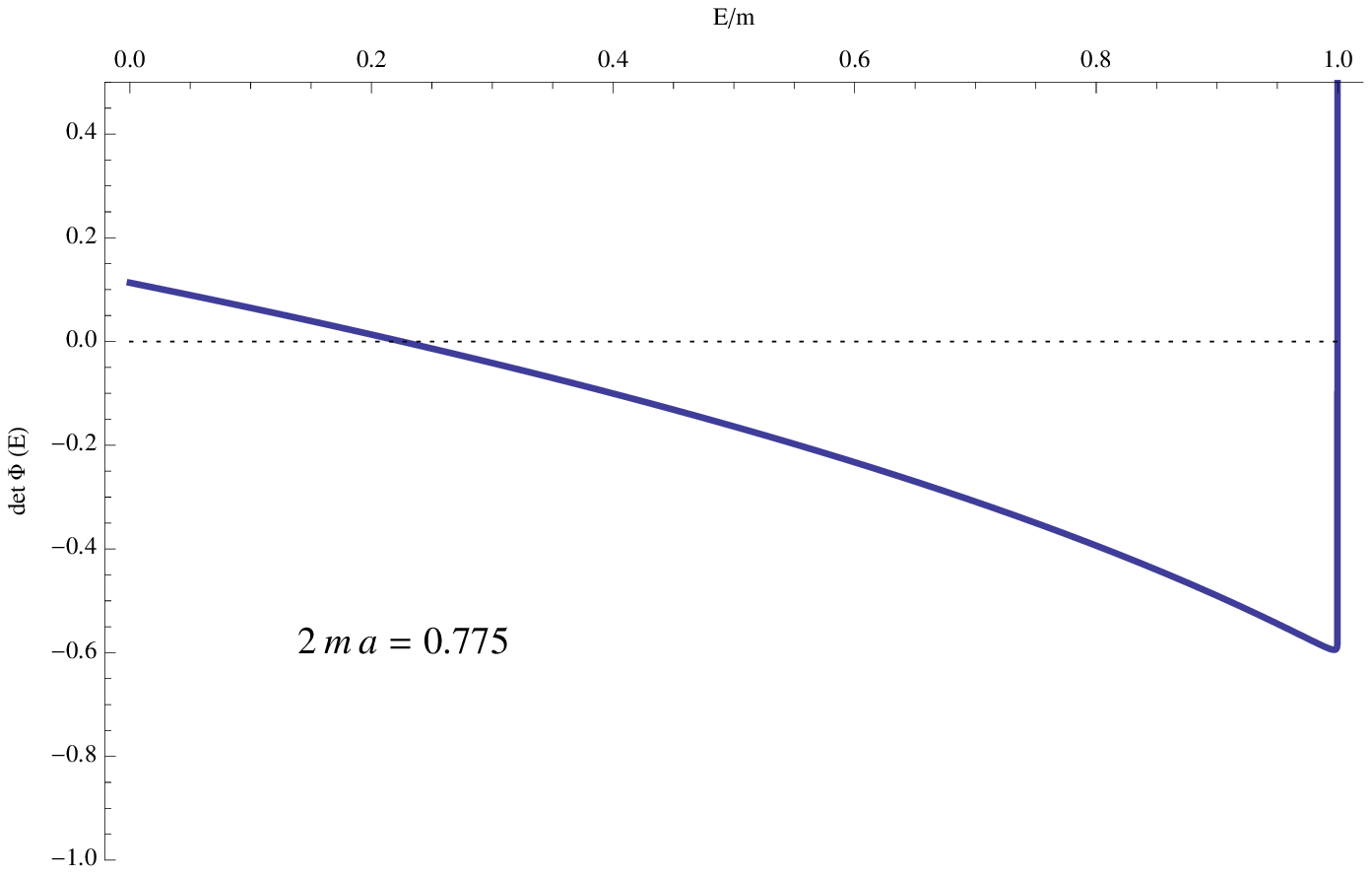}
\end{minipage} \\
\begin{minipage}{5cm}
\includegraphics[scale=0.5]{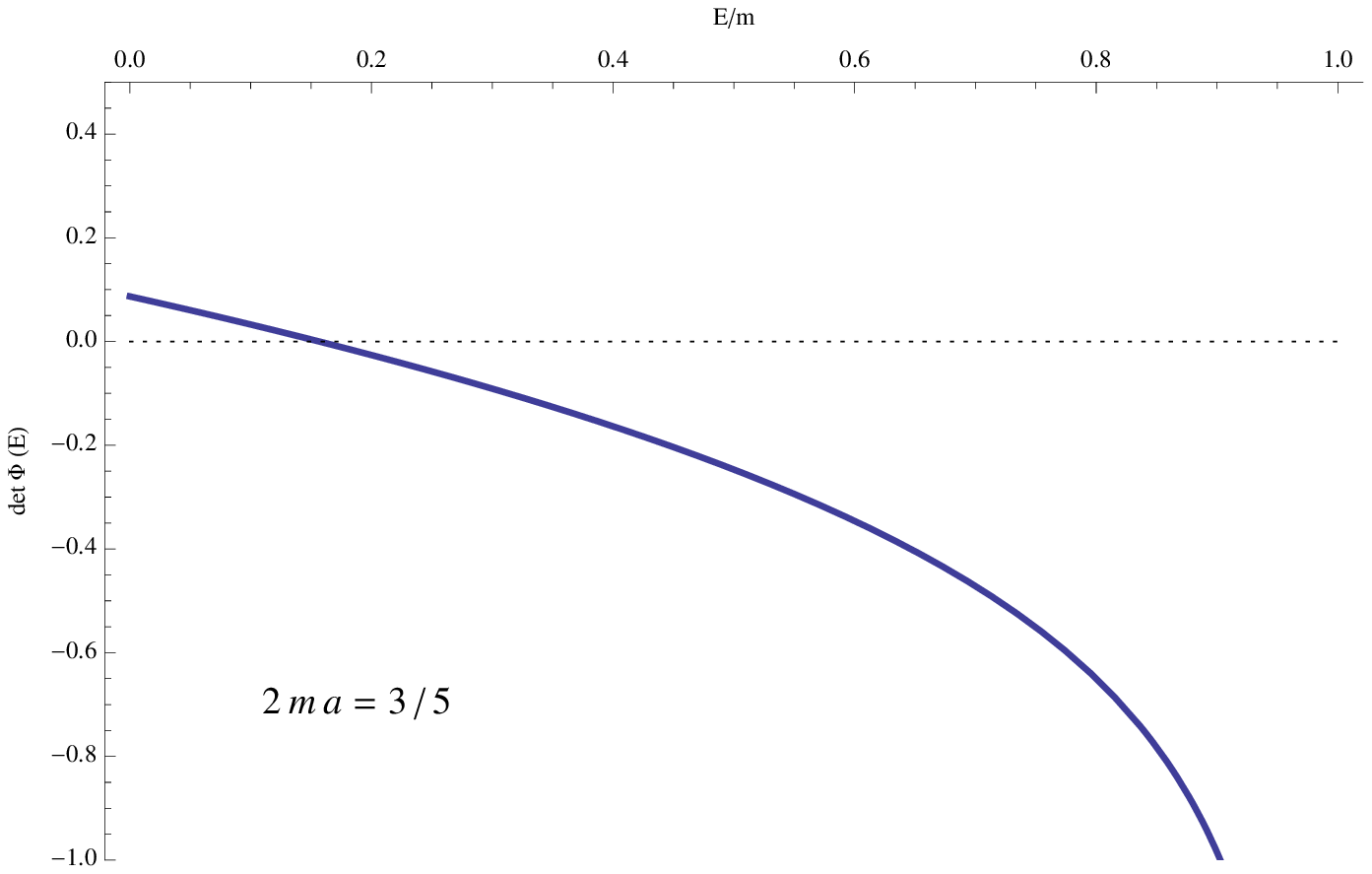}
\end{minipage}
\qquad \qquad \qquad\qquad \qquad 
\begin{minipage}{5cm}
\includegraphics[scale=0.5]{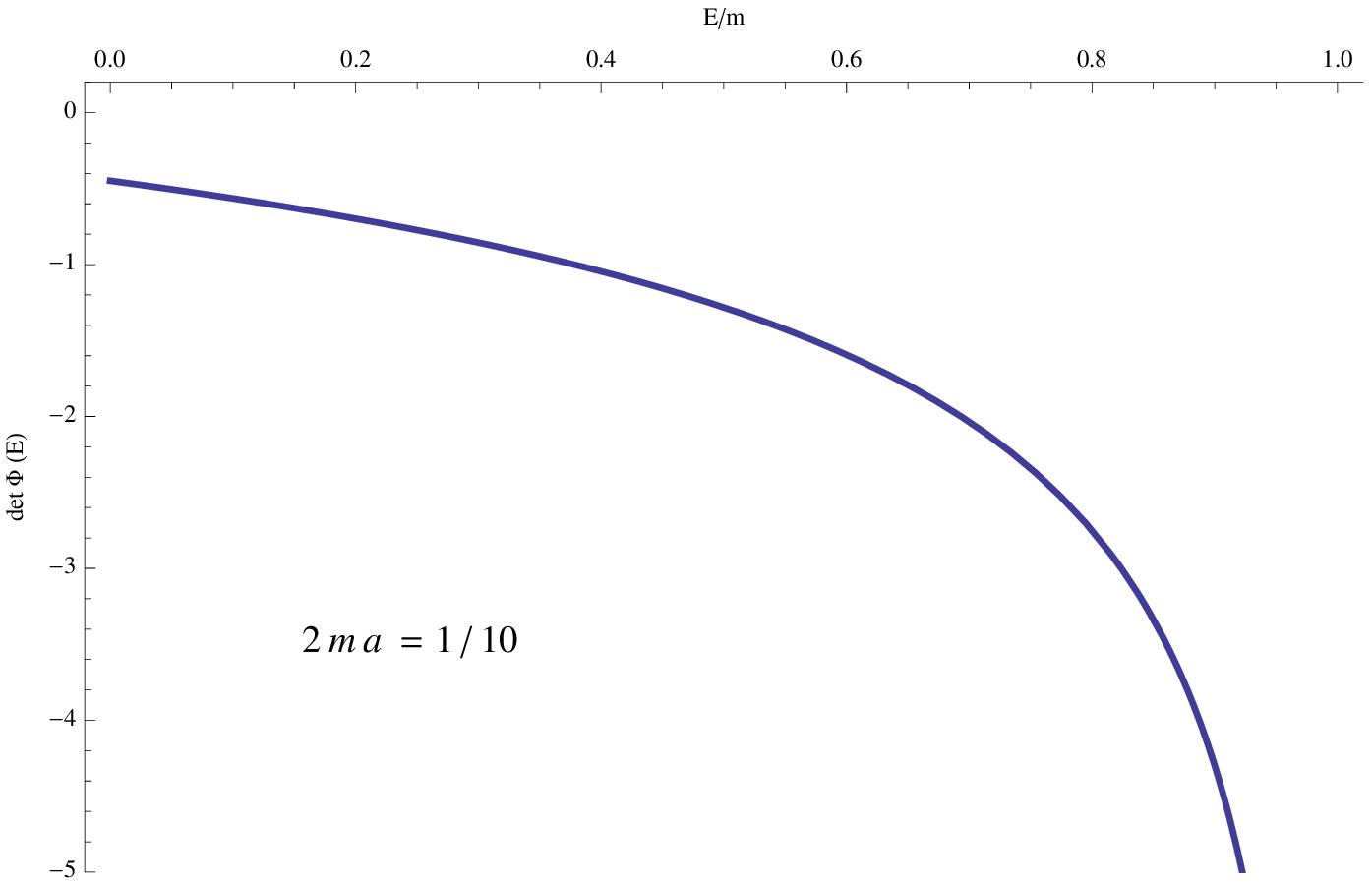}
\end{minipage}
\caption{The determinant of the principal matrix as a function of $E/m$ for different values of $2ma$. Here $E_B/m=1/2$.} \label{det2}
\end{figure}

Let us also analyze the zeros of  determinant of the principal matrix by plotting it for different values of the parameters. The graphs in Fig. \ref{det2} are very convenient to determine how many bound states there are for certain values of the parameters.
As can be seen in Fig. \ref{det2}, there are two (weak) bound states when $2ma=1$, only one (weak) bound state when $2ma=3/5$, and no (weak) bound state but possibly (strong or ultrastrong) a bound state exists when $2ma=1/10$. It is worth emphasizing that for a rather fine-tuned value of the parameter $2 m a$ at $0.775$, a new bound state very close to the threshold energy $E=m$ (at the border of the continuum states) appears. This point will be important when we study the scattering problem.

%%%%%%%%%%%%%%%%%%%%%%%%%%%%%%%%%%%%%%%%%%%%%%%%%%%%%%%%%%%%%%%%%%%%%

\section{A Lower Bound on the Ground State Energy}
\label{Lower Bound on the Ground State Energy}

After renormalization, we still need to prove that the ground state energy is bounded from below.   The essential idea of the proof is similar to the one given for the two- and three-dimensional nonrelativistic case in \cite{point Dirac on manifolds2}.  However, it is worthwhile going through the proof in our simple semirelativistic system where a single particle interacts with $N$ external Dirac delta potentials. A much more interesting case is, of course, associated with the model where the particles are interacting through two-body Dirac delta potentials and the stability of matter in this context is rather an important issue \cite{LiebThirring}. Once we understand the problem for a single particle, it may help to guide us to find a lower bound on the ground state energy of the semirelativistic many-body system.

Let us first recall the Ger\v{s}gorin theorem \cite{horn} in matrix analysis, which states
that all eigenvalues $\omega$ of an $N \times N$ matrix are
located in the union of $N$ disks
\beqs \label{gerstheorem} \bigcup_{i=1}^{N} \{|\omega-
\Phi_{ii}| \leq \sum_{i \neq j =1}^{N} |\Phi_{ij}|  \} \;. \eeqs
Let  $E_*$ be the lower bound of the ground state energy, and then for all $E<E_*$ none of the Ger\v{s}gorin disks contain the zero eigenvalue, i.e., 
\beqs \label{gersh}  |\Phi_{ii}(E)|
> \sum_{i \neq j}^{N}|\Phi_{ij}(E)| \;, \eeqs
for all $E<E_*$ and $i$. Our goal is to find this critical value $E_*$ by solving the above inequality. Unfortunately, this is not possible analytically. Nevertheless, we can still find a less sharper critical value by the following argument.

From the explicit expression of the principal matrix given in Eq. (\ref{Phimatrixheatkernel}), it is easy to see that 
\beqs {\partial |\Phi_{ii} (E)| \over \partial E} =
\begin{cases}
\begin{split}
 - \int_{0}^{\infty} d t \; K_{t} (a_i,a_i) \; t \; e^{t E} < 0 \;,
\end{split}
& \mathrm{when} \; E < E_B^i\\
\begin{split}
 \int_{0}^{\infty} d t  \; K_{t} (a_i,a_i) \; t \;  e^{t E} > 0 \;, 
\end{split}
& \mathrm{when} \; E > E_B^i \;,
\end{cases}
\eeqs
and $ {\partial |\Phi_{ij} (E)| \over \partial E} = \int_{0}^{\infty} d t \; K_{t} (a_i,a_i) \; t \; e^{t E} >0$ for all $E$. It follows from this fact that the critical value only exists when $E<E_B^i$. In this case,
$|\Phi_{ii}(E)|$ is a decreasing function of $E$ and $|\Phi_{ij}|$ is a increasing function of $E$. Note that we are looking for the values of $E$ for which the above inequality (\ref{gersh}) is satisfied. Hence, if we find a lower bound for $|\Phi_{ii}|$ and an upper bound for $|\Phi_{ij}|$, namely
\beqs |\Phi_{ii}(E)| & \geq & \min_{1 \leq i \leq n} |\Phi_{ii}(E)| \;, \cr \sum_{i
\neq j}^{N}|\Phi_{ij}(E)| & \leq & (N-1)\max_{1 \leq j \leq N}  |\Phi_{ij}(E)| \;,
\eeqs
the condition (\ref{gersh}) is implied by the stronger
requirement
\beqs \min_{1 \leq i \leq n} |\Phi_{ii}(E)|>  (N-1)\max_{1 \leq j \leq N}  |\Phi_{ij}(E)|\;.
\label{lowerbound condition delta} \eeqs
Once we obtain the value of $E$, which saturates this inequality, it is satisfied for
all $E$ below this critical value. Consequently, there cannot be any solution beyond this
critical value, and the ground state energy must be
larger than that critical value. Let $\mu = \min_{i} E_{B}^{i}$ and $d= \min_{j} |a_i-a_j|$ for all $i$. Then,
\beqs
\min_{i} |\Phi_{ii}(E)| & = &  \int_{0}^{\infty} d t \; K_{t} (a_i,a_i) \; (e^{t \mu} - e^{t E}) \;, \cr
\max_{j} |\Phi_{ij}| & = & \int_{0}^{\infty} d t \; { m \, t \over \pi \sqrt{d^2 +t^2}} \; K_1(m\sqrt{d^2 +t^2}) \; e^{t E} \;,
\eeqs
for $E<\mu$.

Now we follow the above line of arguments until we obtain an analytical solution. For that purpose, let us find a lower bound for $\min_{i} |\Phi_{ii}(E)|$ and an upper bound for $\max_{j} |\Phi_{ij}|$. Using the integral representation of the Bessel function $K_1$ \cite{Lebedev}
\beqs
K_1(x)=\int_{0}^{\infty} d t \; \cosh t \; e^{-x \cosh t}  \;,
\eeqs
and the bounds $\cosh t \geq {e^t /2}$, and $\cosh t \leq {1 + e^t \over 2}$, we have
\beqs
K_1(x) \geq e^{-x/2} \; \int_{0}^{\infty}  d t \; {e^t \over 2} \; e^{-x  {e^t \over 2} } \;.
\eeqs
By making the change of variables $u=e^t$, we obtain a lower bound for the Bessel function
\beqs
K_1(x) \geq {e^{-x/2} \over 2} \; \int_{1}^{\infty} d u \; e^{-x  {u \over 2} }  = {e^{-x} \over x} \;, \label{bessellowerbound}
\eeqs
for all $x>0$. Using the upper bound of the Bessel function (\ref{besselupperbound}), we have
\beqs
\min_{i} |\Phi_{ii}(E)| > {1 \over \pi} \log \left( {m - E \over m- \mu}\right) \;,
\eeqs
where $E<E_B <m$. Then, it is easy to see that
\beqs \label{offdiagonalupperbound}
\sum_{i
\neq j}^{N}|\Phi_{ij}(E)| & \leq &  (N-1) \int_{0}^{\infty} d t \; { m \, t \over \pi \sqrt{d^2 +t^2}} \; e^{t E} \; e^{-{m \over 2} \sqrt{d^2 +t^2}} \left({1 \over m \sqrt{d^2 +t^2}} +{1 \over 2}\right) \;.
\eeqs
Since $e^{-{m \over 2} \sqrt{d^2 +t^2}} \leq e^{-{m \over 2} \; t}$ and $\sqrt{d^2 +t^2} \geq t$ for all $t$, we get
\beqs \label{offdiagonalupperbound2}
\sum_{i
\neq j}^{N}|\Phi_{ij}(E)| & \leq &  (N-1) \left( {1 \over \pi d^2} \; \int_{0}^{\infty} d t \; t\; e^{-t \;({m \over 2}-E)} \; + {m \over 2 \pi \, d} \; \int_{0}^{\infty} d t \; t \; e^{-t \;({m \over 2}-E)} \right) \cr & < & (N-1) \left[ {1 \over (E-m)^2} \; \left( {1 \over \pi d^2} + {m \over 2 \pi d} \right) \right] \;.
\eeqs
This leads to the need for imposing the following inequality:
\beqs
 {1 \over \pi} \log \left( {m - E \over m- \mu}\right) < (N-1) \left[ {1 \over (E-m)^2} \; \left( {1 \over \pi d^2} + {m \over 2 \pi d} \right) \right] \;.
\eeqs
The value of $E$ that saturates this inequality can be found analytically now, so that we conclude for all $N\geq 1$ that
\beqs
E_{gr} \geq m- \left[ {2 \pi (N-1) C(m,d) \over W \left({2 \pi (N-1) C(m,d) \over (m-\mu)^2} \right) } \right]^{1/2} \;,
\eeqs
where $W$ is the Lambert $W$ function \cite{Lambert}, defined by the solution of $
y \; e^{y}=x$ and $C(m,d)= \left( {1 \over \pi d^2} + {m \over 2 \pi d} \right)$.

%%%%%%%%%%%%%%%%%%%%%%%%%%%%%%%%%%%%%%%%%%%%%%%%%%%%%%%%%%%%%%%%%%%%%

\section{The Hamiltonian After Renormalization}
\label{Existence of Hamiltonian}

Although we do not know what the form of the Hamiltonian after the renormalization procedure is, we can ask whether there is a self-adjoint operator associated with the resolvent formula. We show that there exists a unique self-adjoint Hamiltonian associated with the resolvent formula (\ref{renormalized resolvent}). This problem has been discussed from the self-adjoint extension point of view in \cite{AlbeverioKurasov} and could also be proved by other methods. Here, our approach is to renormalize the model by heat kernel techniques, formally obtain an explicit formula for the resolvent, and then show that this formula for the resolvent corresponds to a unique densely defined self-adjoint Hamiltonian without going into rather technical domain issues of unbounded operators. We think this proof can be useful if we extend this model into many-body or field theoretical models. The self-adjointness of the Hamiltonian after the renormalization procedure is very crucial from the physical point of view since only self-adjoint operators are observables and the self-adjoint Hamiltonian generates the unitary time evolution \cite{simon1, simon2}.

Our proof is based on the following corollary  (Corollary 9.5 in \cite{Pazy_Semigroups}), and it is essentially first used in \cite{existence} for proving the existence of the self-adjoint Hamiltonian of the nonrelativistic Dirac delta potentials in two- and three-dimensional manifolds and  of the relativistic (Klein-Gordon) Dirac delta potentials on two dimensional manifolds \cite{Caglar2} (also includes the Lee model). For the sake of completeness, let us restate this corollary here. 

Let $\Delta$ be a subset of the complex plane and $E \in \Delta$. A family $J(E)$  of bounded linear operators on the
Hilbert space $\mathcal{H}$ under consideration, which satisfies
the resolvent identity
\beqs J(E_1)-J(E_2)= (E_1-E_2)J(E_1)J(E_2)\; \label{resolvent
identity}\eeqs
for $E_1, E_2 \in \Delta$ is called a pseudo resolvent on
$\Delta$ \cite{Pazy_Semigroups}. Let
$\Delta$ be an unbounded subset of $\mathbb{C}$ that does not coincide with the spectrum of $A$ and $J(E)$ be a
pseudo resolvent on $\Delta$. If there is a sequence $E_k \in
\Delta$ such that $|E_k| \rightarrow \infty$ as $k\rightarrow
\infty$ and
\beqs \lim_{k \rightarrow \infty} - E_k J(E_k)x =x \;,
\label{resolvent limit} \eeqs
for all $x \in \mathcal{H}$, then $J(E)$ is the resolvent of a
unique densely defined closed operator $A$. 

We are not going to give the first part of the proof here again since it is exactly given in \cite{Caglar2} and the reader can easily go through it by reading the relevant section given there.

If we choose the sequence $\triangle = \{E_k | E_k= -k |E_0| , k=1,2, \ldots \}$, where $E_0$ is below the lower bound on the ground state
energy that has been found in Sec. \ref{Lower Bound on the Ground State Energy}, the resolvent (\ref{renormalized resolvent}) is a pseudo resolvent on the above set. 

As for the second part of the proof, it is more involved and technical. Since the proof is not essential to be able to follow the rest of the paper, we give it in  Appendix B.

%%%%%%%%%%%%%%%%%%%%%%%%%%%%%%%%%%%%%%%%%%%%%%%%%%%%%%%%%%%%%%%%%%%%%

\section{The Bound State Wave Function For $N$ Centers}
\label{Bound State Wave Function For $N$ Centers}

The projection operator onto the subspace spanned by the eigenfunctions 
corresponding to the $k$th isolated eigenvalue (bound state energy $E_{\mathtt{bound}}^k$) is given by the following contour integral \cite{simon}:
\beqs \langle x| \mathbb{P}_{k}|y \rangle =
\psi_{B}^k(x)(\psi_{B}^{k}(y))^{*}=- {1\over 2\pi i} \oint_{\Gamma_k}
d E \;  R(x,y|E), \eeqs
where $R(x,y|E)=\langle x |R(E) |y\rangle$ is the resolvent kernel and $\Gamma_k$ is a sufficiently small contour enclosing only $E_{\mathtt{bound}}^{k}$. We note that the free resolvent kernel or Green's function
$R_0(x,y|E)$ does not contain any pole on the real axis below $m$ [spectrum of the free part is $\sigma(H_0)=[m, \infty)$].
Therefore, all the poles on the real axis smaller than $m$ must come only from the
poles of the inverse principal matrix. 
Since it has been shown that the principal matrix is a symmetric [$\Phi^{\dagger}_{ij}(E)=\Phi_{ij}(E^{*})$] holomorphic (analytic) family in Appendix A, its eigenvalues and its eigenprojections are also holomorphic on the real axis \cite{Kato}.

As a result of Hermiticity of the principal matrix on the real $E$ axis and its analytical continuation to the complex $E$ plane, we
can apply the spectral theorem to the principal matrix
\beqs \Phi_{ij}(E)=\sum_{\sigma=1}^{N} \omega^{\sigma}(E) [\mathbb{P}_{\sigma}(E)]_{ij} \;. \eeqs
Here $\mathbb{P}_{\sigma}(E)_{ij}=(A_i^{\sigma}(E))^* \; A_j^{\sigma}(E)$ and $A^{\sigma}_i(E)$ are the projection operator and the
normalized eigenvector corresponding to the eigenvalue
$\omega^{\sigma}(E)$, respectively. Similarly, we can write the spectral resolution of
the inverse principal matrix,
\beqs
 [\Phi^{-1}(E)]_{ij}=\sum_{\sigma} {1 \over \omega^{\sigma}(E)} [\mathbb{P}_{\sigma}(E)]_{ij}
\;. \eeqs
The residue of the resolvent at the simple pole $E=E_{\mathtt{bound}}^k$ (assuming that only the $k$ th eigenvalue $\omega^k$ flows to its zero at $E=E_{\mathtt{bound}}^k$) is given by 
\beqs \mathrm{Res}(R(x,y|E); E_{\mathtt{bound}}^k) = R_0(x,a_i|E_{\mathtt{bound}}^k) \; \left(
\left.{\partial \omega^k(E)\over \partial E}\right|_{E=E_{\mathtt{bound}}^k}\right)^{-1} \;
[\mathbb{P}_k(E_{\mathtt{bound}}^k)]_{ij} \; R_0(a_j,y|E_{\mathtt{bound}}^k)\;, \eeqs
where $\left.{\partial \omega^k(E) \over \partial E}\right|_{E=E_{\mathtt{bound}}^k}$ can be found from Eq. (\ref{derivative of lambda evaluated nu}). Combining all these
results yields
\beqs & & \psi_{B}^k(x) (\psi_{B}^{k}(y))^{*} =  {1\over 2 \pi i} (2 \pi i) \; 
R_0(x,a_i|E_{\mathtt{bound}}^k) \Bigg(- \left. {\partial \omega^k(E) \over \partial E}
\right|_{E=E_{\mathtt{bound}}^k}\Bigg)^{-1} \nonumber \cr \\[2ex]  & & \hspace{8cm} \times 
(A_{i}^{k}(E_{\mathtt{bound}}^k))^* A_{j}^{k}(E_{\mathtt{bound}}^k) \; R_0(a_j,y|E_{\mathtt{bound}}^k) \;.\eeqs
Then, it is straightforward to read off the bound state wave function from
the equation above,
\beqs
\begin{split}
\psi_{B}^k(x) = & \left(  - \left. {\partial \omega^k(E) \over \partial E}
\right|_{E=E_{\mathtt{bound}}^k} \right)^{-\frac{1}{2}} \; \sum_{i=1}^N A_{i}^{k}(E_{\mathtt{bound}}^k) \;
\int_{0}^{\infty} d t  \; e^{t E_{\mathtt{bound}}^k}  \; K_{t}(a_i, x) \;.
\label{wavefunction heat delta} \end{split} \eeqs
This explicit result of the bound state wave function for $N$ Dirac delta potentials is the linear combination of the bound state wave functions for each single Dirac delta center located $a_i$. In the single center case, we have only one bound state energy, namely $E_B$.  Since the principal matrix is just a single function in this case, $A_i=1$ so that we obtain
\beqs
\psi_{B}(x) = \mathcal{N} \int_{0}^{\infty} d t \; K_t(x,0) \; e^{t E_B} \;, \label{singledeltawavefunction}
\eeqs
where $\mathcal{N}$ is the normalization constant given by 
\beqs
\mathcal{N} = \bigg[ \int_{-\infty}^{\infty} d x \; \left( \int_{0}^{\infty} d t \; K_t(x,0) \; e^{t E_B} \right)^2 \bigg]^{-1/2} \;.
\eeqs
The wave function (\ref{singledeltawavefunction}) is nothing but the same formula obtained recently in \cite{Al-Hashimi}. This can be seen by first expressing the heat kernel as $K_t(x,0)=\langle 0 |e^{-t \sqrt{P^2+m^2}} | 0 \rangle$ and inserting the completeness relation $\int_{-\infty}^{\infty} {d p \over 2 \pi} |p \rangle \langle p| =1$ in front of the exponential
\beqs
\psi_{B}(x) &=& \mathcal{N} \int_{0}^{\infty} d t \; \int_{-\infty}^{\infty} {d p \over 2 \pi} \; e^{i p x} \; e^{-t \sqrt{p^2 +m^2}} \; e^{t E_B} =\mathcal{N} \int_{-\infty}^{\infty} {d p \over 2 \pi} \; {e^{i p x} \over \sqrt{p^2 +m^2} -E_B} \;. \label{singlecenterwavefunctionmomentum}
\eeqs
There is an overall minus sign difference between our result (\ref{singlecenterwavefunctionmomentum}) and the one given in \cite{Al-Hashimi}, which is physically irrelevant. The above improper integral is discussed in great detail in \cite{Al-Hashimi} by using the contour integration for three different regimes of bound states, namely weak, strong, and ultrastrong bound states. We are not going to discuss the details of these various cases since they have already been studied in \cite{Al-Hashimi}. We will consider the general behavior of the bound state wave functions in the next sections.

For consistency, let us consider the nonrelativistic limit of the bound state wave function (\ref{wavefunction heat delta}) associated with $N$ delta centers. To find the wave function in this limit, we first rewrite the wave function formula (\ref{wavefunction heat delta}) in the same way as in Eq. (\ref{singlecenterwavefunctionmomentum}),
\beqs \label{Rwavefunction2}
\psi_{B}^{k}(x) & = & \mathcal{N}^{k} \;
\sum_{i=1}^N A_{i}^{k}(E_{\mathtt{bound}}^k)  \int_{-\infty}^{\infty} {d p \over 2 \pi}  \; {e^{i p (x-a_i)} \over \sqrt{p^2 +m^2}-E_{\mathtt{bound}}^k} \;,
\eeqs 
where $\mathcal{N}^{k}$ is the normalization constant. 
Note that the integral appearing in the wave function (\ref{Rwavefunction2}) is exactly the same integral as in the principal matrix. Using (\ref{Phioffdiagonalexplicit}), the nonrelativistic limit $|E_{\mathtt{bound}}^k-m|/m =|\Delta E_{\mathtt{bound}}^k|/m \ll 1$ of the above integral becomes
\beqs \label{NRlimitintegral}
 {m \over \left(-2 m \Delta E_{\mathtt{bound}}^{k}\right)^{1/2}}  \; \exp \left[ {-\left(-2 m \Delta E_{\mathtt{bound}}^{k}\right)^{1/2} \; |x-a_i|} \right]  \;,
\eeqs
where we ignored the higher order terms in $|\Delta E_B^k|/m $. Similarly, we can find the nonrelativistic limit of the principal matrix ($|E-m|/m \ll 1$ and $|E_B^i -m|/m \ll 1$) and obtain

\beqs \label{NRlimitofPhimatrix} \Phi_{ij} (E) \sim
\begin{cases}
\begin{split}
{m \over (-2 m \Delta E_B^i)^{1/2}}  - {m \over (-2 m \Delta E)^{1/2}}
\end{split}
& \textrm{if $i = j$}  \\[2ex]
\begin{split}
-  {m \over \left(-2 m \Delta E \right)^{1/2}}  \; \exp \left[ {-\left(-2 m \Delta E \right)^{1/2} \; |a_i-a_j|} \right]
\end{split}
& \textrm{if $i \neq j$} \;.
\end{cases}
\eeqs
Let us now go back to the nonrelativistic problem. We do not need renormalization in this case, and it is straightforward to calculate the resolvent formula for $N$ dirac delta centers 
\beqs
R_\epsilon(E) = (H_0-E)^{-1} + \left( H_0 - E \right)^{-1} \left( \sum_{i,j=1}^N | a_i \rangle
\left[\Phi^{-1}(E) \right]_{ij} \langle a_j| \right) \left( H_0 - E
\right)^{-1}\;, \label{NRresolvent}
\eeqs
where $H_0={P^2 \over 2m}$ and 
\beqs \label{NRPhimatrix} \Phi_{ij} (E) =
\begin{cases}
\begin{split}
{1 \over \lambda_i}  - {m  \over (-2 m E)^{1/2}} 
\end{split}
& \textrm{if $i = j$}  \\[2ex]
\begin{split}
- \; {m \over (-2 m E)^{1/2}} \; \exp \left[ {-(-2 m E)^{1/2} \; |a_i-a_j|} \right] 
\end{split}
& \textrm{if $i \neq j$} \;.
\end{cases}
\eeqs
Since the bound state energy to the $i$ th center in the nonrelativistic case is given by $\Delta E_B^i = - m \lambda_{i}^2/2$ such that  $1/\lambda_i = - m / (- 2 m \Delta E_B^i)^{1/2}$, we show that the nonrelativistic limit of the principal matrix (\ref{NRlimitofPhimatrix}) is equal to the nonrelativistic principal matrix (\ref{NRPhimatrix}). Because of this result, the nonrelativistic limit of the eigenvectors $A_{i}^k$ of the principal matrix  is equal to the eigenvector of the nonrelativistic principal matrix (\ref{NRPhimatrix}). This guarantees that the nonrelativistic limit of the bound state wave function is  
\beqs \label{NRwavefunction}
\psi_{B}^k(x)  \sim \mathcal{N}_{nr}^k \; \sum_{i=1}^{N}  {m \; A_{i (nr)}^{k}( \Delta E_{\mathtt{bound}}^{k}) \over (-2 m \Delta E_{\mathtt{bound}}^{k} )^{1/2} }   \; \exp \left[ {- (-2 m \Delta E_{\mathtt{bound}}^{k})^{1/2} \; |x-a_i|} \right]    \;,
\eeqs
where $\mathcal{N}^{k}_{nr}$ is the normalization constant and $A_{i(nr)}^{k}$ is the $k$th eigenvector of the nonrelativistic principal matrix (\ref{NRPhimatrix}) associated with the $k$th eigenvalue $\omega_{nr}^k$. Here $\Delta E_{\mathtt{bound}}^{k}$ must be the solution of $\omega_{nr}^k(\Delta E_{\mathtt{bound}}^{k})=0$.
Hence, we show that  the nonrelativistic limit of the bound state wave function for $N$ centers (\ref{Rwavefunction2}) is actually the linear combination of the bound state wave function for single nonrelativistic Dirac delta centers.

%%%%%%%%%%%%%%%%%%%%%%%%%%%%%%%%%%%%%%%%%%%%%%%%%%%%%%%%%%%%%%%%%%%%%

\section{Pointwise Bound on the Bound State Wave Function and Expectation Value of the Free Hamiltonian}
\label{Pointwise Bound on the Bound State Wave Function and Expectation Value of the Free Hamiltonian}

The exponential decay
of the bound state wave functions of the Schr\"{o}dinger operators are known as the consequence of regularity theorems. Basically, square-integrable solutions of $(- \nabla^2 + V)\psi=E \psi $ obey
pointwise bounds of the form
\beqs |\psi(r)| \leq C e^{-a r} \;, \eeqs
if the potential energy $V$ is continuous and bounded below and
$E$ is in the discrete spectrum of $- \nabla^2 + V$ (see
\cite{simon} for the review of the subject). We
shall prove that it is still possible to get exponential pointwise
bounds for the bound state wave function of our semirelativistic problem.

It is easy to find an upper bound for the wave function
(\ref{wavefunction heat delta}) by applying Cauchy-Schwarz
inequality
\beqs |\psi_{B}^k(x)| & \leq &  |\mathcal{N}^k|  \left| \sum_{i=1}^N
A_{i}^{k}(E_{\mathtt{bound}}^k) \int_{0}^{\infty} d t \;
e^{t  E_{\mathtt{bound}}^k} K_{t}(a_i, x) \right| \cr & \leq
& |\mathcal{N}^k|  \left[ \sum_{i=1}^N \left|\int_{0}^{\infty} d
t \; e^{t  E_{\mathtt{bound}}^k} \; K_{t}(a_i,
x)\right|^2 \right]^{1/2} \cr & \leq & |\mathcal{N}^k| \sum_{i=1}^N
\int_{0}^{\infty} d t \; e^{t  E_{\mathtt{bound}}^k} K_{t}(a_i, x) \;, \eeqs
where $\sum_{i=1}^{N}|A_{i}^{k}(E_{\mathtt{bound}}^k)|^2 = 1$. Thanks to the upper bound
of the Bessel function $K_1(x)$ given in Eq. (\ref{besselupperbound}), the wave function is
pointwise bounded on the real line
\beqs   |\psi_{B}^k(x)|  & \leq &   |\mathcal{N}^k|  \int_{0}^{\infty} d t \; {m \; t \over \pi \sqrt{(x-a_i)^2 + t^2}} \left({1 \over m \sqrt{(x-a_i)^2 + t^2}} + {1 \over 2} \right)   \exp\left( t \, E_{\mathtt{bound}}^k -m \sqrt{(x-a_i)^2 + t^2} \right)  \nonumber \cr  \\[2ex]   & \leq &  |\mathcal{N}^k|  {m \over \pi |x-a_i| ({m \over \sqrt{2}}-E_{\mathtt{bound}}^k)^2}  \left( {1 \over m |x-a_i|} + {1 \over 2} \right) \; \exp \left( -{m \over \sqrt{2}} |x-a_i|\right)
\label{psiboundcompactD2} \eeqs
where we have used $(x-a_i)^2 + t^2 \geq (x-a_i)^2$ for the expressions in front of the exponential and the inequality ${a+b \over 2} \leq \sqrt{{a^2 + b^2 \over 2}}$ in the exponent (for all $a,b$). This shows that the bound state wave functions for Salpeter Hamiltonians with point interactions are also  pointwise exponentially bounded. 
Note that this upper bound blows up at the locations of Dirac delta centers $a_i$. This singular behavior of the bound state wave function is expected due to the small $t$ asymptotic expansion of the Bessel function (\ref{asymtoticBessel}). 
Nevertheless, the bound state wave function can be shown to be square integrable from its explicit expression using the semigroup property of the heat kernel (\ref{semigroupheat})
 \beqs
 \int_{-\infty}^{\infty}  d x \; |\psi_{B}^k(x)|^2 & = & |\mathcal{N}^k|^2 \int_{-\infty}^{\infty}  d x \; \sum_{i,j=1}^{N} A_{i}^{k} \; (A_{j}^{k})^* \; \int_{0}^{\infty} \int_{0}^{\infty} d t_1 \; d t_2 \; K_{t_1}(a_i,x) \; K_{t_2}(x,a_j) \; e^{(t_1+t_2)E_{\mathtt{bound}}^k} \cr & =& |\mathcal{N}^k|^2 \; \sum_{i,j=1}^{N} A_{i}^{k} \; (A_{j}^{k})^* \; \int_{0}^{\infty} d t \; t\;  K_{t_1}(a_i,a_j) \; e^{t  E_{\mathtt{bound}}^k} 
\;. \eeqs
In the second line we have made the change of variables $t=t_1+t_2$ and $u=t_1-t_2$ and then integrated with respect to the variable $u$. From the explicit expression of the heat kernel (\ref{heatkernel}) and the upper bound of the Bessel function (\ref{besselupperbound}), the above expression is finite so that the bound state wave function is square integrable, 
\beqs
\psi_B^k \in L^2(\mathbb{R})\;.
\eeqs
To understand heuristically why our problem can be
considered as a self-adjoint extension of the free Hamiltonian, which is also suggested by the 
Krein formula, let us calculate the
expectation value of the kinetic energy for the bound state,
\beqs & & \langle \psi_{B}^k | H_0 | \psi_{B}^k \rangle = |\mathcal{N}^k|^2 \int_{-\infty}^{\infty} d x \; \Bigg(
\int_{0}^{\infty} d t_1 \; e^{t_1  E_{\mathtt{bound}}^k} \sum_{i=1}^{N} (A_{i}^{k})^*
K_{t_1}(a_i,x) \Bigg) \cr & & \hspace{5cm} \times \Bigg( \int_{0}^{\infty} d t_2
 \; e^{t_2  E_{\mathtt{bound}}^k} \sum_{j=1}^{N}
A_j^k  \left(-\sqrt{P^2 +m^2} \; 
K_{t_2}(a_j,x) \right)\Bigg) \;,\eeqs
where we have suppressed the energy dependence of $A_i^k$ for simplicity. Using the heat equation (\ref{relativistic heat equation}) with its initial condition, and integration by
parts for the $t_2$ integral, we see that the above expression includes the following term:
\beqs |A_i^k|^2 \; \int_{0}^{\infty} d t_1 \; e^{ t_1 E_{\mathtt{bound}}^k} \; K_{t_1}(a_i,a_i) \;. \eeqs
This integral is clearly divergent due to the small $t$ asymptotic expansion of the Bessel function (\ref{asymtoticBessel}).
Hence we
show that the expectation value of the free Hamiltonian is
divergent,
\beqs \langle \psi_{B}^k | H_0 | \psi_{B}^k \rangle  \rightarrow  \infty \;.
\eeqs
The self-adjoint extension of the semirelativistic kinetic energy operator in the context of a single point interaction was rigorously studied in \cite{AlbeverioKurasov}. We may here heuristically
deduce that the extension of the problem to the finitely many point interactions can also be considered as a  self-adjoint extension of the free part since we have proved that the
bound state wave function $\psi_B^k(x)$ that we have found does not belong to
the domain of the free Hamiltonian $\sqrt{P^2 +m^2}$ so the self-adjoint extension of
the free Hamiltonian extends the domain of it such that the states
corresponding to the eigenfunctions $\psi_B^k(x)$ are included.

%%%%%%%%%%%%%%%%%%%%%%%%%%%%%%%%%%%%%%%%%%%%%%%%%%%%%%%%%%%%%%%%%%%%%

\section{Nondegeneracy of the Ground State}
\label{Non-degeneracy of the Ground State}

The rigorous proof of nondegeneracy and positivity of the ground
state in standard quantum mechanics is given in
\cite{simon}, which includes neither the singular
potentials nor the relativistic cases. Therefore, it is necessary to check whether a similar conclusion can be drawn
for our problem. The proof here is essentially the same as the one for the nonrelativistic case given in the previous work \cite{point Dirac on manifolds2} based on utilizing the Perron-Frobenius theorem \cite{horn}. It states that if $A$ is an $N \times N$ matrix and $A>0$ (i.e., $A_{ij} >0$), then the following statements are true:

\begin{itemize}

\item[(a)] The spectral radius $\rho(A)$ is strictly positive. (Recall that
$\rho(A)=\max\{|\omega|: \omega \; \mathrm{is}
\; \mathrm{an} \; \mathrm{eigenvalue} \; \mathrm{of} \;
A\}$);

\item[(b)] The spectral radius $\rho(A)$ is an eigenvalue of the matrix $A$;

\item[(c)] There is an $x\in \mathbb{C}^{N}$ with $x>0$ and $A x =
\rho(A)x$;

\item[(d)] The spectral radius $\rho(A)$ is an algebraically (and hence geometrically) simple
eigenvalue of $A$;

\item[(e)] $|\omega|< \rho(A)$ for every eigenvalue $\omega \neq
\rho(A)$, that is, $\rho(A)$ is the unique eigenvalue of maximum
modulus.

\end{itemize}

The first step is to find a positive ``equivalent" matrix to the principal matrix (\ref{Phimatrixheatkernel}). Let us subtract the maximum of the diagonal part, and reversing the
overall sign,
\beqs \Phi'(E)= -\left(\Phi(E)- (1+ \varepsilon)  \mathbb{I}
\max_E \; \Phi_{ii}(E)\right) > 0 
\;, \eeqs
where $\varepsilon >0$ and  $E \in [E_{gr}, \infty)$. Since $\Phi_{ii}$ is a decreasing function of $E$, $\max_E \; \Phi_{ii}(E) = \Phi_{ii}(E_{gr})$. 
Note that the results obtained by both $\Phi$ and $\Phi'$ are physically equivalent. First of all, adding a diagonal term to the principal matrix $\Phi$ does not change its eigenvectors, whereas the eigenvalues are shifted by a constant amount. Nevertheless, this shift is equivalent to a constant shift in the bound state spectrum, which is physically unobservable (we can shift the spectrum without altering its physics). Hence, this transformed matrix $\Phi'$ and $\Phi$ have the same common eigenvectors so it guarantees that there exist a strictly positive
eigenvector $A_i$ for the principal matrix $\Phi$
and $\rho(\Phi')= -\omega^{\min}(E)+(1+\varepsilon)
\Phi_{ii}(E_{gr})$.

For a given $E$, there is a
unique  $\omega^{\min}(E)$, and since we are looking
for the zeros of the eigenvalues $\omega(E)=0$, the minimum goes to zero at $E=E_{gr}$. This means that the positive
eigenvector $A_{i}$ corresponds to the ground state energy. Hence, we 
prove that the ground state energy is unique and the associated
eigenvector $A_i$ is strictly positive. Because of the positivity
property of the heat kernel, it is easy to see that the ground state
wave function is strictly positive from Eq.
(\ref{wavefunction heat delta}),
\beqs \psi_{gr}(x) & = & \mathcal{N} \int_{0}^{\infty} d t \;
e^{t  E_{gr}} \sum_{i=1}^N
A_i(E_{gr}) K_{t}(a_i, x)
> 0 \;, \eeqs
where $\mathcal{N}>0$. Despite the singular character of the
interaction, we prove that the ground state is still nondegenerate. This may seem to be inconsistent with the result discussed in Sec. \ref{On Bound States} for the case where there are twin symmetrically located delta centers. We have shown that as the distance between the centers goes to infinity, we have degeneracy in the bound states. However, this is not contradicting with our proof above since this degeneracy occurs due to the vanishing of the  off-diagonal terms in the principal matrix so that the positivity hypothesis of the Perron-Frobenius theorem breaks down. As long as the distance between the centers is finite, the ground state is always nondegenerate.

%%%%%%%%%%%%%%%%%%%%%%%%%%%%%%%%%%%%%%%%%%%%%%%%%%%%%%%%%%%%%%%%%%%%%

\section{The Scattering Problem for $N$ Centers}
\label{Scattering Problem for $N$ Centers}

The reflection and transmission coefficients of the problem for a single center case has recently been investigated in \cite{Al-Hashimi} by constructing even and odd parity scattering solutions. Here we calculate the reflection and transmission coefficients for finitely many centers using the semirelativistic version of the Lippmann-Schwinger equation \cite{Taylor} 
\beqs \label{LS}
|k^\pm\rangle=|k\rangle-R_0(E_k\pm i 0) \; V \; |k^\pm\rangle\,,
\eeqs
where $R_0(E)$ is the free resolvent or Green's operator, and $V$ represents the interaction, and $E_k=\sqrt{k^2 +m^2}$, the energy of the incoming particles. The notation $E_k + i 0$ denotes the limit of Green's function as $\varepsilon \downarrow 0$. Following the similar arguments developed in Sec. \ref{Renormalization of Relativistic Finitely Many Dirac delta Potentials through Heat Kernel}, we can write the regularized semirelativistic Lippmann-Schwinger equation by the heat kernel 
\beqs
|k^\pm (\epsilon)\rangle=  |k\rangle + \sum_{j=1}^N  \lambda_i(\epsilon) \; R_0(E_k\pm i0)|a_j^\epsilon\rangle\langle a_j^{\epsilon}|k^\pm\rangle\,.
\eeqs
Let us consider the outgoing boundary conditions and rescale the ket vectors $|f_i^\epsilon \rangle = \sqrt{\lambda_i(\epsilon)} |a_i^\epsilon \rangle$ so we have 
\beqs \label{LS2}
|k^+ (\epsilon) \rangle=  |k\rangle+ R_0(E_k + i0)|f_i^\epsilon \rangle\langle f_i^\epsilon |k^+ (\epsilon) \rangle
+ \sum_{j \neq i}^N  \; R_0(E_k + i0)|f_j^\epsilon \rangle\langle f_j^\epsilon |k^+ (\epsilon)\rangle \;,
\eeqs
where we have isolated the $j=i$th term. By acting on $\langle f_i^\epsilon|$ from the left, we can write the resulting expression in the following form:
\beqs & &
\left( 1 -  \langle f_i^\epsilon | R_0(E_k + i0) | f_i^\epsilon \rangle \right) \; \langle f_i^\epsilon |k^+ (\epsilon) \rangle \cr & & \hspace{4cm}
- \sum_{j \neq i}^N  \; \langle f_i^\epsilon | R_0(E_k + i0)|f_j^\epsilon \rangle \; \langle f_j^\epsilon |k^+ (\epsilon)\rangle  =  \langle f_i^\epsilon |k\rangle \;,
\eeqs
or it can be written as a matrix equation
\beqs \label{24}
\sum_{j=1}^N T_{ij} (\epsilon, E_k+i0) \; \langle f_j^\epsilon |k^+ (\epsilon)\rangle = \langle f_i^\epsilon |k\rangle  \qquad i=1,2,\dots,N\,,
\eeqs
where
\beqs \label{25} T_{ij}(\epsilon, E_k+i0) = \left\{   \begin{array}{ccc}  1-   
\langle f_i^\epsilon | R_0(E_k + i0) | f_i^\epsilon \rangle & {\rm if}  & i=j\,,               \\[2ex] 
- \langle f_i^\epsilon | R_0(E_k + i0)|f_j^\epsilon \rangle & {\rm if}  & i\ne j\,.        \end{array}                           \right. 
\eeqs
Hence, the solution to Eq. (\ref{24}) is given by
\beqs \label{26}
\langle f_i^\epsilon |k^+ (\epsilon)\rangle = \sum_{j=1}^N \left[T^{-1}(\epsilon, E_k+i0) \right]_{ij}\;\langle f_j^\epsilon |k\rangle \;. 
\eeqs
Substituting this result into the formula (\ref{LS2}) that we have obtained for the scattering solution, and acting on the position bra vector $\langle x |$ from the left yields
\beqs 
\langle x | k^+ (\epsilon)\rangle   =  \psi_k^+(\epsilon, x)  & = & e^{ikx}+\sum_{i,j=1}^N \langle x | R_0(E_k +i 0) | f_i^\epsilon \rangle   \; \left[T^{-1}(\epsilon, E_k+i0)\right]_{ij} \;  \langle f_j^\epsilon |k\rangle \cr & = & e^{ikx}+\sum_{i,j=1}^N \langle x | R_0(E_k +i 0) | a_i^\epsilon \rangle   \; \left[\Phi^{-1}(\epsilon, E_k+i0)\right]_{ij} \;  \langle a_j^\epsilon |k\rangle \;,
\eeqs
where
\beqs  \Phi_{ij}(\epsilon, E_k+i0) = \left\{   \begin{array}{ccc}  {1 \over \lambda_i(\epsilon)} -   
\langle a_i^\epsilon | R_0(E_k + i0) | a_i^\epsilon \rangle & {\rm if}  & i=j\,,               \\[2ex] 
- \langle a_i^\epsilon | R_0(E_k + i0)|a_j^\epsilon \rangle & {\rm if}  & i\ne j\,.        \end{array}                           \right. 
\eeqs
If we insert the choice (\ref{couplingconstantrenormalization}) and take the limit as $\epsilon \rightarrow 0$, we obtain
\beqs \label{LSscatteringsolution}
\psi_k^+(x) = e^{ikx}+\sum_{i,j=1}^N   R_0(x,a_i|E_k +i 0) \; \left[\Phi^{-1}(E_k+i0)\right]_{ij} \;  e^{i k a_j} \;,
\eeqs
where the principal matrix $\Phi(E_k +i0) \equiv \lim_{\varepsilon \rightarrow 0^+} \Phi(E_k +i \varepsilon)$ is 
\beqs \label{Phiscattering}
\Phi_{ij} (E_k+i0) = 
\begin{cases}
\begin{split} 
-{1 \over \lambda(E_k,E_B^{i})} -{i E_k \over \sqrt{E_k^2 -m^2}}
\end{split}
& \textrm{if $i=j$} \\ \\
\begin{split}
- {i E_k \over \sqrt{E_k^2 -m^2}} \; e^{i \sqrt{E_k^2 -m^2} |a_i-a_j|}  - {1 \over \pi} \int_{m}^{\infty} d \mu \; e^{- \mu |a_i-a_j|} \; {\sqrt{\mu^2 -m^2} \over \mu^2 + E_k^2 -m^2}
\end{split}
& \textrm{if $i \neq j$} \;.
\end{cases}
\eeqs
The function $\lambda(E_k,E_B^i)$ is defined as 
\beqs
{1 \over \lambda(E_k, E_{B}^i)} & = & -\bigg[
{E_k \over \pi \sqrt{E_k^2 -m^2}} \; \arctanh \left({\sqrt{E_k^2 -m^2} \over E_k}\right)  + 
{E_{B}^{i} \over \pi \sqrt{m^2-(E_{B}^{i})^2}} \; \left( {\pi \over 2} + \arcsin {E_B^i \over m} \right) \bigg] \;,
\eeqs
and called the energy dependent running coupling constant originally introduced in \cite{Al-Hashimi} for a single center.

The diagonal term of the principal matrix (\ref{Phiscattering}) is actually nothing but the analytic continuation of the formula (\ref{Phi1a}). For the scattering problem, we need to determine the asymptotic behavior of the scattering solution for large values of $x$, namely $x \gg a_i$. 
For this reason, let us first express the resolvent kernel $R_0(x,a_i|E_k +i 0)$ in the following way:
\beqs
  R_0 (x, a_i |E_k +i 0) = \langle x | R_0(E_k +i 0) | a_i \rangle &=& \int_{-\infty}^{\infty} {d p \over 2 \pi} \; {e^{i p (x-a_j)} \over \sqrt{p^2 +m^2} - (E_k+i0)}  \nonumber \cr \\[2ex]  & = & {i \sqrt{k^2+m^2} \over k} \; e^{i k |x-a_i|} + {1 \over \pi} \int_{m}^{\infty} d \mu \; e^{- \mu |x-a_i|} \; {\sqrt{\mu^2 -m^2} \over \mu^2 +k^2} \;. \eeqs
A simple asymptotic analysis applied to the above integral shows that it is exponentially damped for large values of $x$ ($x \gg a_i$) so that we may ignore it compared to the first oscillating term for the outgoing scattering problem. Putting this into Eq. (\ref{LSscatteringsolution}), we get
\beqs
\psi_{k}^{+}(x)  \sim e^{ikx}+\sum_{i,j=1}^N   {i \sqrt{k^2+m^2} \over k} \; e^{i k |x-a_i|} \; \left[\Phi^{-1} (E_k+i0)\right]_{ij} \; e^{i k a_j} \;. \label{scatteting solution}
\eeqs
This is an explicit and exact solution to the semirelativistic Lippmann-Schwinger equation, and it includes the  information about the reflection and transmission coefficients so that
we can immediately find them by simply reading the factors in front of $e^{i kx}$ for $x<a_i$ and the factors in front of $e^{i k x}$ for $x>a_i$, respectively,
\beqs
R(k) =|r(k)|^2 & = &\Bigg| \sum_{i,j=1}^N   {i \sqrt{k^2+m^2} \over k} \; (\Phi^{-1} (E_k+i0))_{ij} \; e^{i k (a_i+a_j)} \Bigg|^2 \;, \label{reflectionNdelta} \nonumber \cr \\[2ex] 
T(k) = |t(k)|^2 & = & \Bigg| 1+ \sum_{i,j=1}^N   {i \sqrt{k^2+m^2} \over k} \; (\Phi^{-1} (E_k+i0))_{ij} \; e^{i k (-a_i+a_j)} \Bigg|^2 \label{reflectionNdelta} \;.
\eeqs
Here $R(k)$ represents the reflection coefficient and $T(k)$ the transmission coefficient. It is important to notice the notational difference that the same letters have been used for the scattering amplitudes in \cite{Al-Hashimi}. Here, we prefer to stick to a more traditional notation.   
Although the above solution is exact, it is difficult to calculate the inverse of the principal matrix for any number of Dirac delta centers located arbitrarily on the line. Moreover, the off-diagonal part of the principal matrix (\ref{Phiscattering}) even includes an integral term that cannot be evaluated analytically. For this purpose, we shall first consider the simplest possible cases.

$N=1$ case:

First, we consider the case where we have a single center ($N=1$). We can assume that the Dirac delta potential is located at the origin without loss of generality. In this case, the principal matrix (\ref{Phiscattering}) is simply a function.
Hence, the reflection and transmission coefficients become
\beqs
R(k) & = & {(k^2 + m^2) \; \lambda^2(E_k, E_B) \over k^2 +\lambda^2(E_k, E_B) \; (k^2 +m^2) } \;,\cr \nonumber \\[2ex]  T(k) & = & {k^2 \over k^2 + \lambda^2(E_k, E_B) \; (k^2 +m^2)} \;.
\eeqs
This is exactly the same result that was derived in \cite{Al-Hashimi} by constructing the even-parity and odd-parity scattering solutions. In our method, the derivation for the reflection and transmission coefficients is much simpler and more general. The scattering phase shift $\delta(k)$ can simply be computed from the $S$-matrix $S(k)=r(k)+t(k)=\exp(2 i \delta)$. Further physical questions have been discussed in \cite{Al-Hashimi}.

\begin{figure}[h!]
%\centering
\begin{minipage}{5cm}
\includegraphics[scale=0.5]{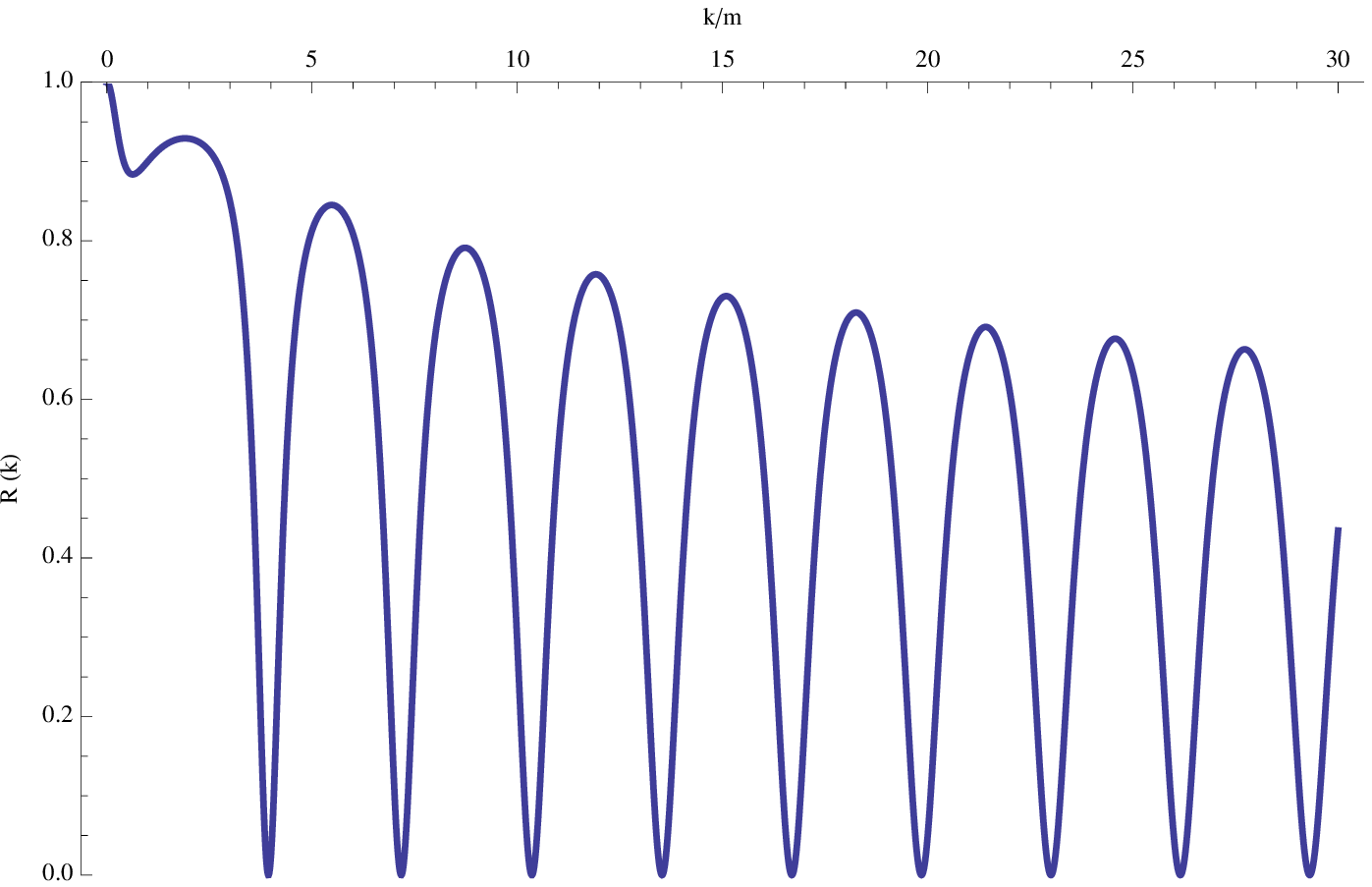}
\end{minipage}
\qquad \qquad \qquad \qquad \qquad
\begin{minipage}{5cm}
\includegraphics[scale=0.5]{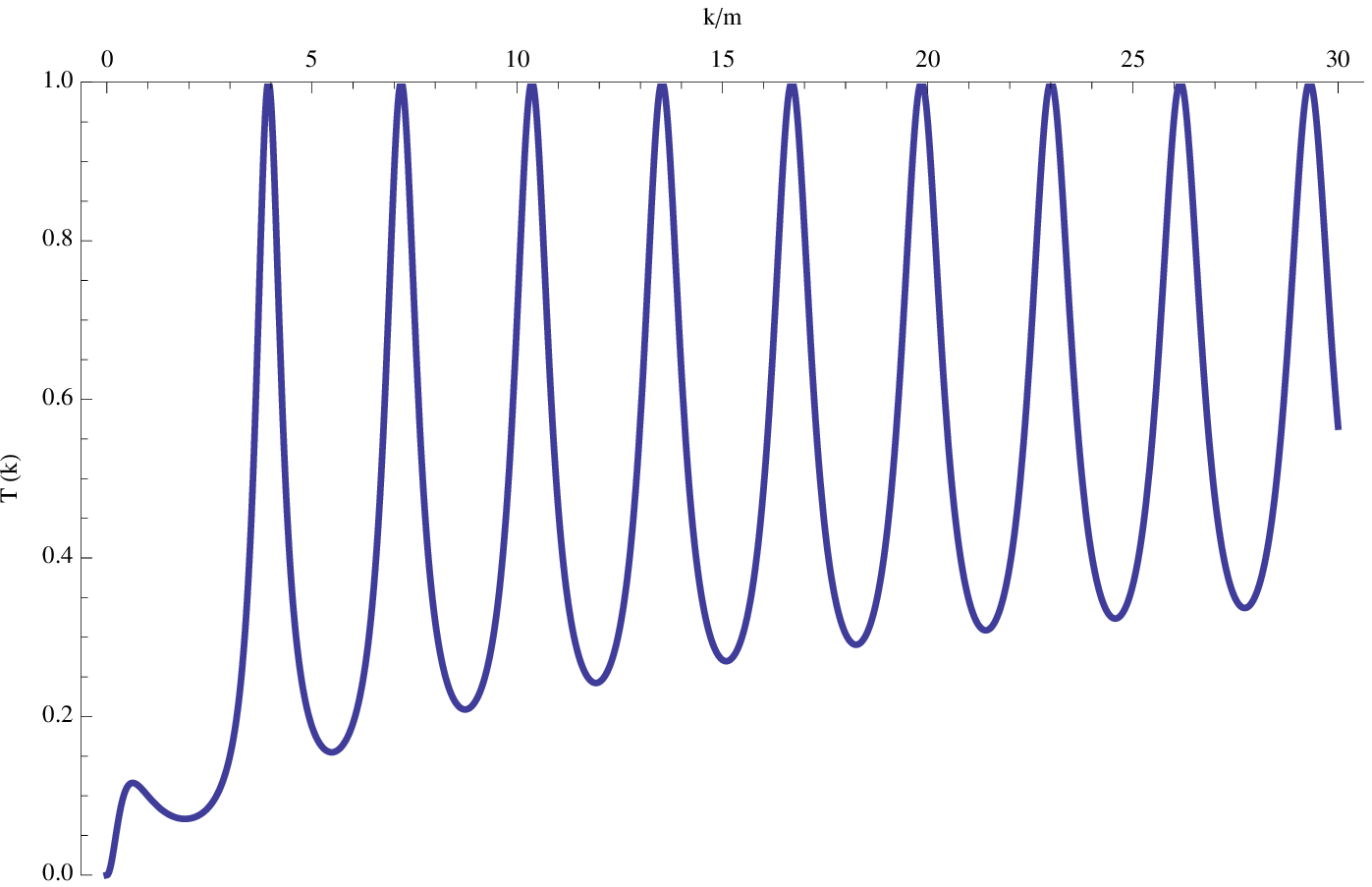}
\end{minipage}
\caption{The reflection and transmission coefficients of two symmetric twin Dirac delta centers as a function of $k/m$ for the values $E_B/m=1/2$, $2 m a=1$.} \label{RTnumerical} \end{figure}

$N=2$ case ($E_B^1=E_B^2=E_B$):

We can always choose our coordinate system such that two Dirac delta centers are located symmetrically with respect to the origin, so that $a_1=-a$ and $a_2=a$. 
Since we cannot analytically evaluate the integrals in the off-diagonal part of the principal matrix (\ref{Phiscattering}), we compute the reflection and transmission coefficients numerically with the help of \textit{Mathematica} and their graphical representations are depicted in Fig. \ref{RTnumerical}.

\begin{figure}[h!]
\begin{center}
\includegraphics[scale=0.5]{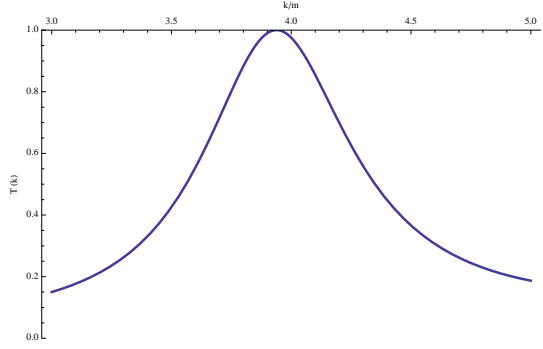}
\end{center}
\caption{The transmission coefficient as a function of $k/m$ plotted near its first peak $k/m=4$ for $E_B/m=1/2$ and $2ma=1$.} \label{Lorentzian}
\end{figure}

Let us address some issues about the behavior of the reflection and transmission coefficients. The general pattern of these coefficients as functions of $k/m$ is very similar to the one in  the nonrelativistic version of the same problem \cite{LS, L}.  
All maxima of the transmission coefficient in Fig. \ref{RTnumerical} indicate perfect transmissions. If we plot the transmission coefficient near one of those peaks, say at $ k/m \sim 4$, in a higher resolution, we can see that the peak has the form, as shown in Fig. \ref{Lorentzian}.
This is why these peaks are sometimes interpreted as resonances in \cite{L}. However, one must be careful about this terminology since these do not have to correspond to decaying states \cite{Arno}. For this reason, we prefer to call them perfect transmission energies.

\begin{figure}[h!]
%\centering
\begin{minipage}{5cm}
\includegraphics[scale=0.5]{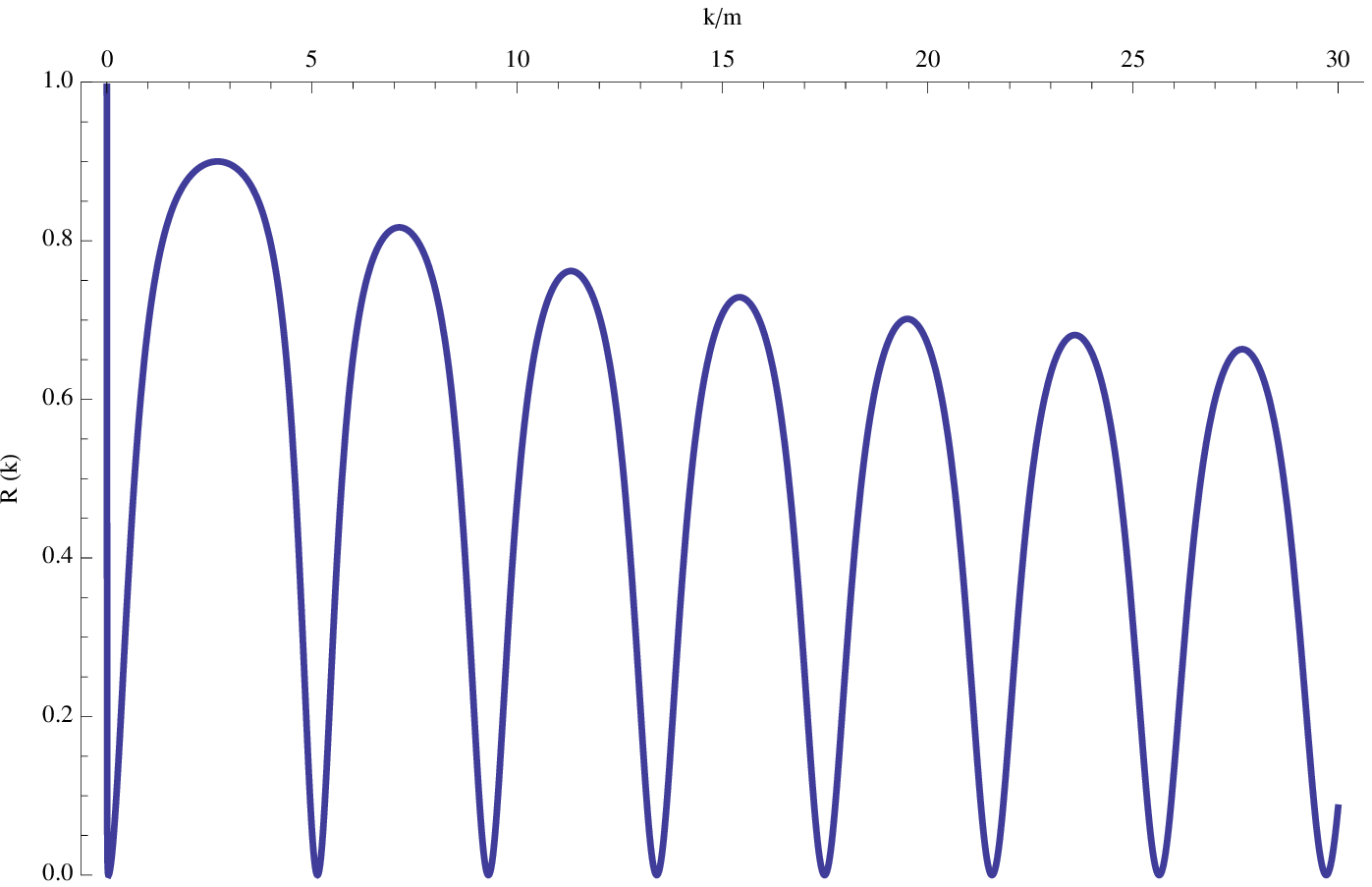}
\end{minipage}
\qquad \qquad \qquad \qquad 
\begin{minipage}{5cm}
\includegraphics[scale=0.5]{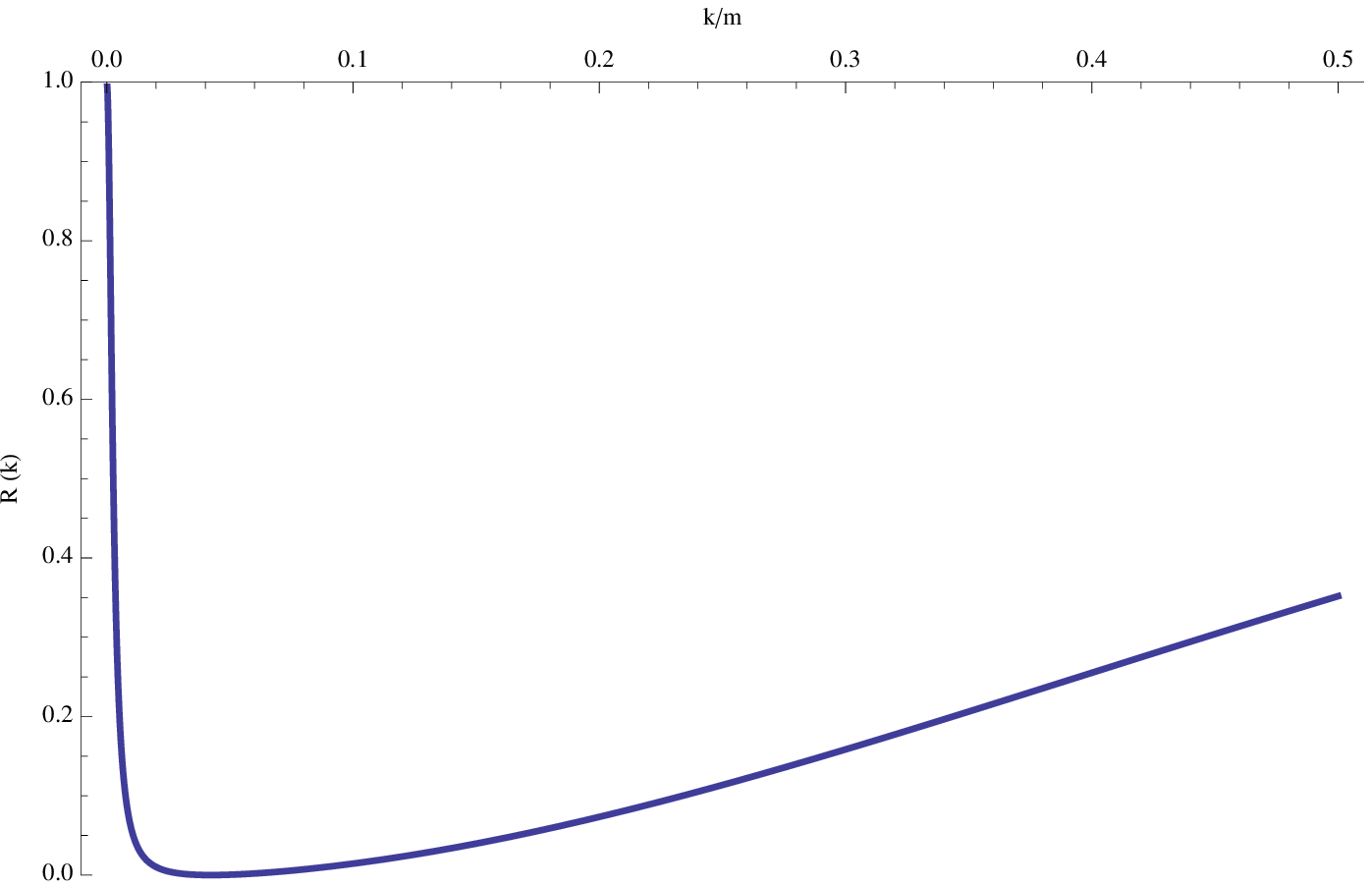}
\end{minipage}
\caption{The reflection coefficient as a function of $k/m$ in different scales for $2 m a =0.775$ and  $E_B/m=1/2$.} \label{AnomalyR1}\end{figure}

\begin{figure}[h!]
%\centering
\begin{minipage}{5cm}
\includegraphics[scale=0.5]{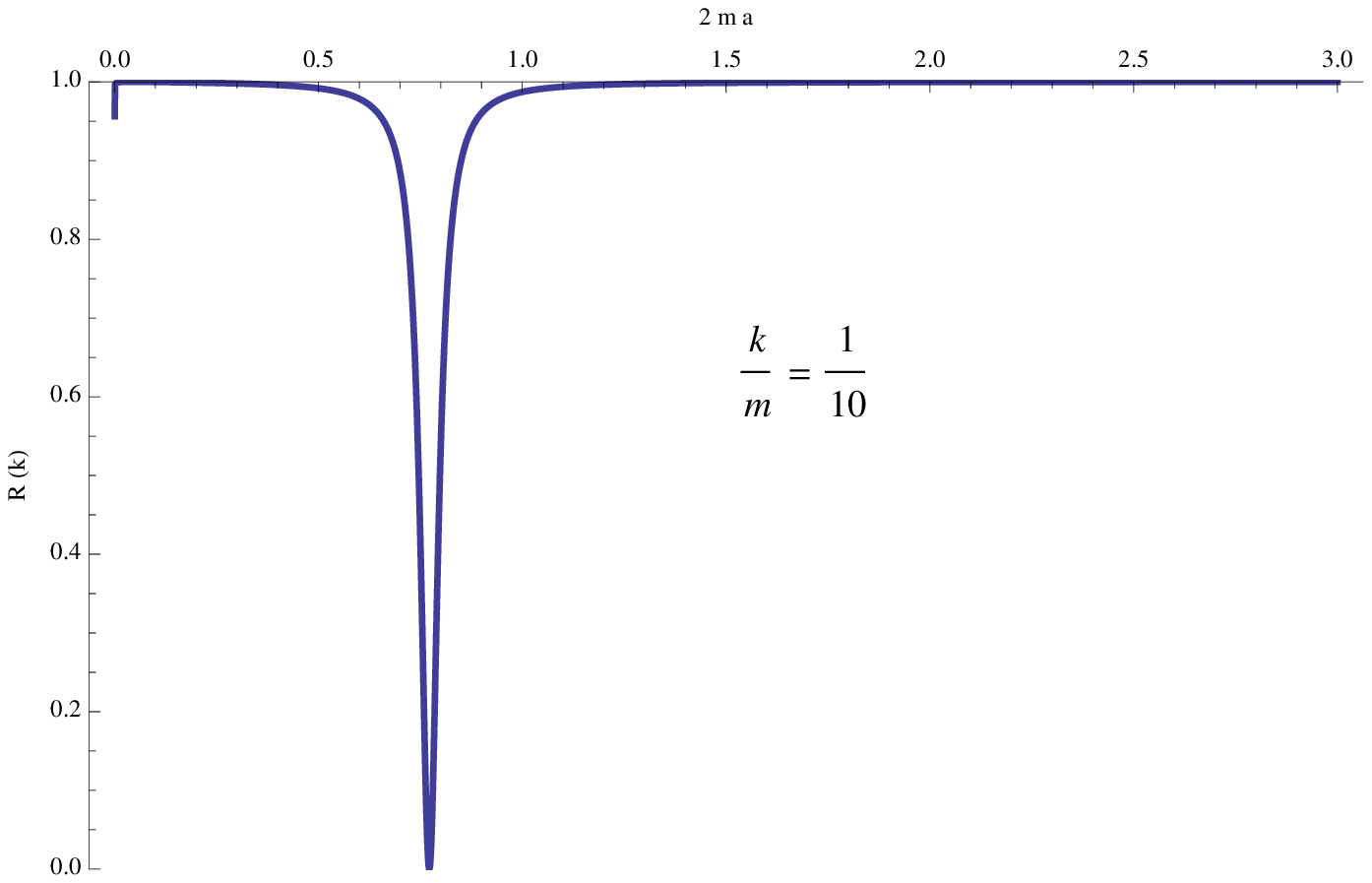}
\end{minipage}
\qquad \qquad \qquad \qquad
\begin{minipage}{5cm}
\includegraphics[scale=0.5]{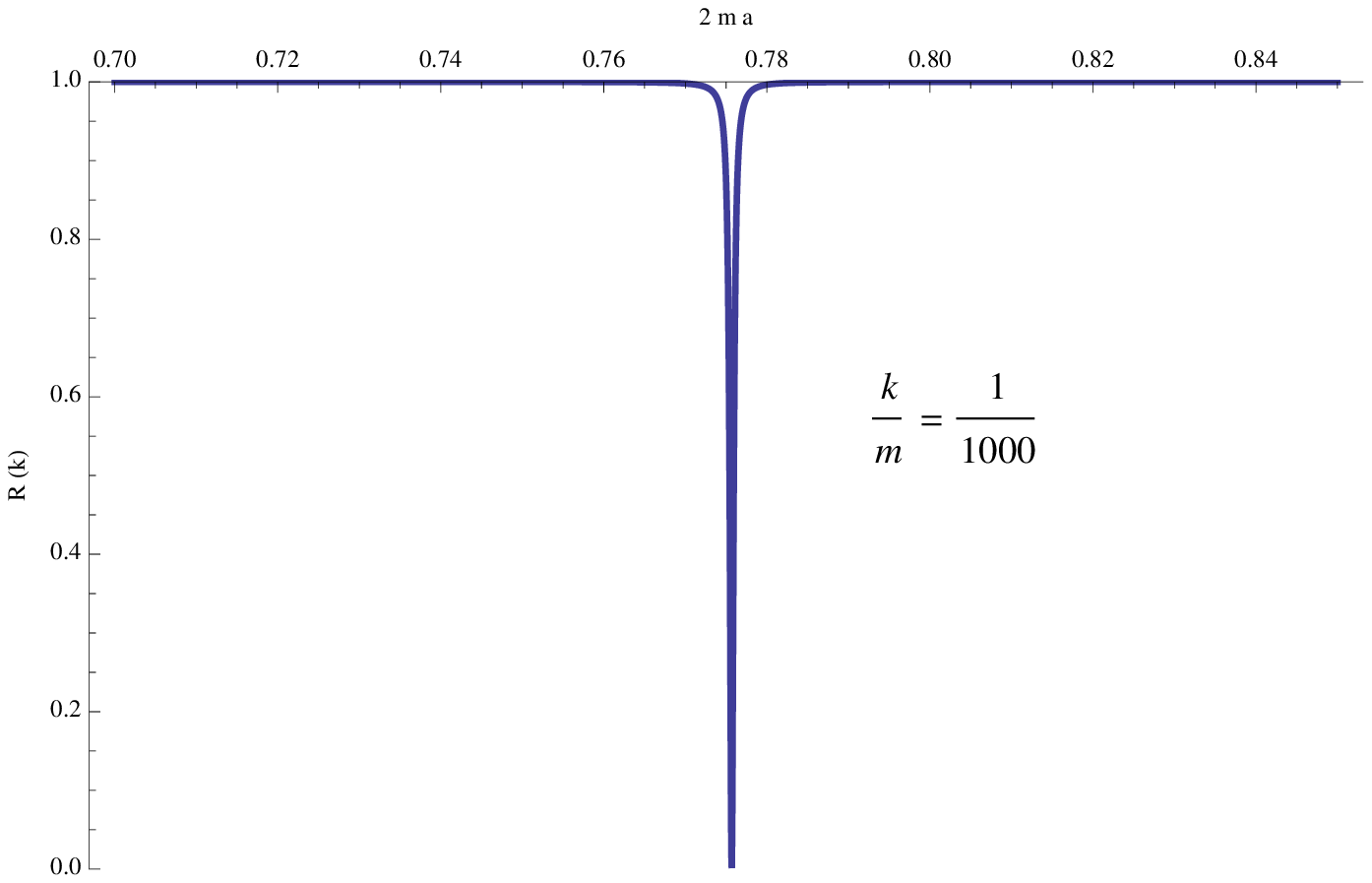}
\end{minipage}
\caption{The reflection coefficient as a function of $2 m a$ in different scales for different values of $k/m$. Here we choose $E_B/m=1/2$.} \label{AnomalyR2}
\end{figure}

There is actually a small bump around the very small value of $k/m$, and it can be more clearly observed by changing the distance between the centers $2 m a $ and $E_B/m$. To see this behavior, we plot the reflection coefficient as a function of $k/m$ for a particular value of $2m a$ and $E_B/m$ in Fig. \ref{AnomalyR1}. 
This shows that the reflection coefficient suddenly vanishes near the zero energy of incoming particles for a certain value of distance between centers ($2 m a=0.775$ in Fig. \ref{AnomalyR1}) for a given $E_B/m$. The critical value for the distance between the centers is more transparently seen if we plot the reflection coefficient as a function of $2ma$ for different small values of $k/m$, as shown in Fig. \ref{AnomalyR2}.  It is important to notice that the peak around the critical value $2ma=0.775$ becomes sharper and sharper as $k/m$ decreases.

This behavior has also been observed in the nonrelativistic case and known as a threshold anomaly \cite{Senn}. It is defined as the vanishing reflection coefficient near the threshold energy (at the border of the continuum energy spectrum), namely 
\beqs R(k) \rightarrow 0 \;, \eeqs
as $k \rightarrow 0$ for certain values of the parameters in the model. The underlying reason for threshold anomaly is basically the appearance of a bound state very close to the threshold energy for some particular choice of the parameters in the model \cite{Senn}. 
This anomaly in the nonrelativistic quantum mechanics even exists for the much more general class of potentials, and the proof is given in \cite{Senn}. Here we observe that this anomaly even exists for the semirelativistic case that includes some singular potentials requiring renormalization.

\begin{figure}[h!]
\begin{center}
\includegraphics[scale=0.5]{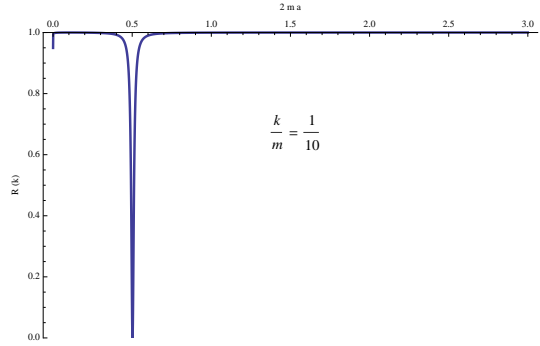}
\end{center}
\caption{The reflection coefficient as a function of $2ma$ for $E_B/m=1/10$.} \label{Rversus2maEB}
\end{figure}

We recall that the excited state of the system discussed in Sec.\ref{On Bound States} appears near $E=m$ ($k=0$) when $2ma =0.775$ and $E_B/m=1/2$, as shown in Fig. \ref{det2}. Hence, we show that the critical value of $2ma$ observed in Fig. \ref{AnomalyR2} exactly corresponds to the critical case for which the excited state appears. 
We also realize that the critical value of $2ma$ decreases as we decrease $E_B/m$ (see Fig. \ref{Rversus2maEB}). This is not surprising since we physically expect that, as we increase $E_B$, the bound state energies of the system must also increase so that the critical value for $2ma$ must be lowered.

Although the reflection and transmission coefficients can be obtained numerically for a pair of symmetrical Dirac delta centers (in principle for any finite $N$), we may ask whether there is any good approximation, where we have an explicit analytical expression for them and the above analysis can be examined analytically. The answer relies on the asymptotic expansion of the integral 
\beqs \label{integral}
 {1 \over \pi} \int_{m}^{\infty} d \mu \; e^{- \mu |a_i-a_j|} \; {\sqrt{\mu^2 -m^2} \over \mu^2 + k^2} \;,
\eeqs
in the off-diagonal part of the principal matrix (\ref{Phiscattering}). Let us first make a change of variable $\mu= s k$ so the above integral becomes $ {1 \over \pi k} \int_{m/k}^{\infty} d s \; e^{- s k |a_i-a_j|} \; {\sqrt{s^2 k^2-m^2} \over s^2 +1}$. Now we want to find the large $k|a_i-a_j|$ behavior of this integral. Note that $-s$ in the exponent has its maximum at $s=m/k$ on the interval $(m/k, \infty)$. Then, only the vicinity of $s=m/k$ contributes to the full asymptotic expansion of the integral for large $k|a_i-a_j|$. Thus, we may approximate the above integral by $ {1 \over \pi k} \int_{m/ k}^{\epsilon} d s \; e^{- s k |a_i-a_j|} \; {\sqrt{s^2 k^2-m^2} \over s^2 +1}$, where $\epsilon>m/k$ and replace the function ${\sqrt{s^2 k^2-m^2} \over s^2 +1}$ in the integrand by its Taylor or asymptotic expansion \cite{bender}. It is important to emphasize that the full asymptotic expansion of this integral as $k|a_i-a_j| \rightarrow \infty$ does not depend on $\epsilon$ since all other integrations are subdominant compared to the original integral (\ref{integral}). Hence, we find
\beqs
 {1 \over \pi k} \int_{m /k}^{\epsilon} d s \; e^{- s k |a_i-a_j|} \; {\sqrt{s^2 k^2-m^2} \over s^2 +1} & \sim & {1 \over \pi} \int_{m/k}^{\epsilon} d s \; e^{- s k |a_i-a_j|} \; {\sqrt{s-m/k} \; \sqrt{2 k m} \; k \over k^2 +m^2} \cr \nonumber \\ [2ex] & \sim & {1 \over \pi} \int_{m/k}^{\infty} d s \; e^{- s k |a_i-a_j|} \; {\sqrt{s-m/k} \; \sqrt{2 k m} \; k \over k^2 +m^2} \nonumber \\ [2ex]\cr & = & {{m^2/k^2} \over \sqrt{2\pi} (1+m^2/k^2)} \; {\exp\left( -m |a_i-a_j| \right) \over (m |a_i-a_j|^{3/2})} \;,
\eeqs
where we have used the fact that the contribution to the integral outside of the interval $(m/k,\epsilon)$ is exponentially small for any $\epsilon>m/k$. Substituting this result into Eq. (\ref{Phiscattering}) and computing the inverse of the principal matrix, we can find an explicit analytic expression for the reflection and transmission coefficients  (but they are too complicated to write them down explicitly here) as long as we have to keep in mind that these expressions are valid only in the region where $k|a_i-a_j|$ is large.

\begin{figure}[h!]
\centering
\includegraphics[scale=0.5]{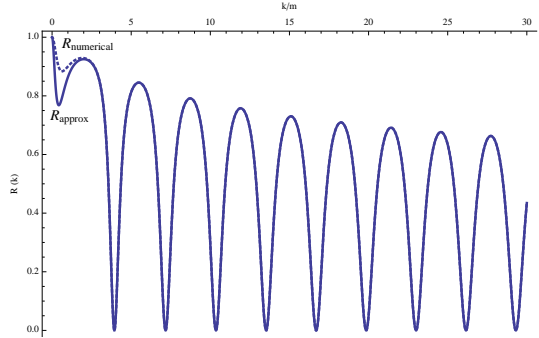}
\caption{The reflection coefficient $R_{\mathtt{approx}}$ in the asymptotic approximation and the reflection coefficient $R_{\mathtt{numerical}}$ obtained numerically for the particular values of $E_B^1/m=E_B^2/m=1/2$ and $2 m a=1$. Notice that they slightly differ only near the region when $k/m$ is zero for a fixed value of $2 m a$. This is expected since the asymptotic approximation becomes better and better as $2 k a$ takes larger values. }
\label{RappRnum}
\end{figure}

In particular, for twin ($E_B^1=E_B^2$) symmetrically oriented Dirac delta centers,  we can compare the predictions of our approximation with the numerical results. Although there is an apparent discrepancy near very small values of $k/m$ for the fixed values of $E_B/m$ and $2 m a$ given below, they are in complete agreement, as shown in Fig. \ref{RappRnum} for larger values of $k/m$. In this approximation, the appearance of a threshold anomaly occurs when $2ma=0.888$; i.e., the asymptotic approximation overestimates the critical value of $2ma$.   

Moreover, the approximation to the reflection coefficient approaches its numerically calculated values as $2 m a$ increases near the region $k/m$ are small ($2 k a = {k \over m} 2 m a$ gets bigger).

\begin{figure}[h!]
\centering
\includegraphics[scale=0.5]{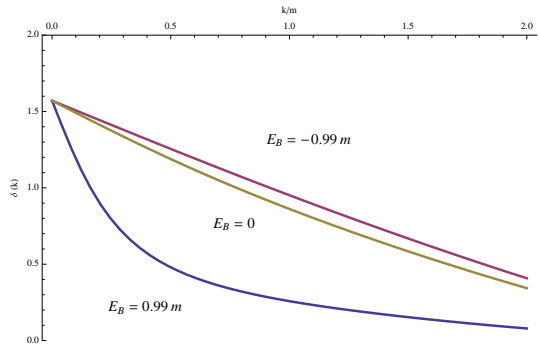}
\caption{Phase shift $\delta(k)$ as a function of $k/m$ for three different values of $E_B/m$ and for $2ma=1$.}
\label{phaseshift2}
\end{figure}

The phase shift in this particular problem can also be calculated numerically from the relation $S(k)=e^{2 i \delta(k)}$, and its graph is illustrated in Fig.  \ref{phaseshift2}. We note that $\delta(0)=\pi/2$ no matter what the values of $E_B/m$ for $2 m a =1$ are.

\begin{figure}[h!]
\begin{minipage}{5cm}
\includegraphics[scale=0.5]{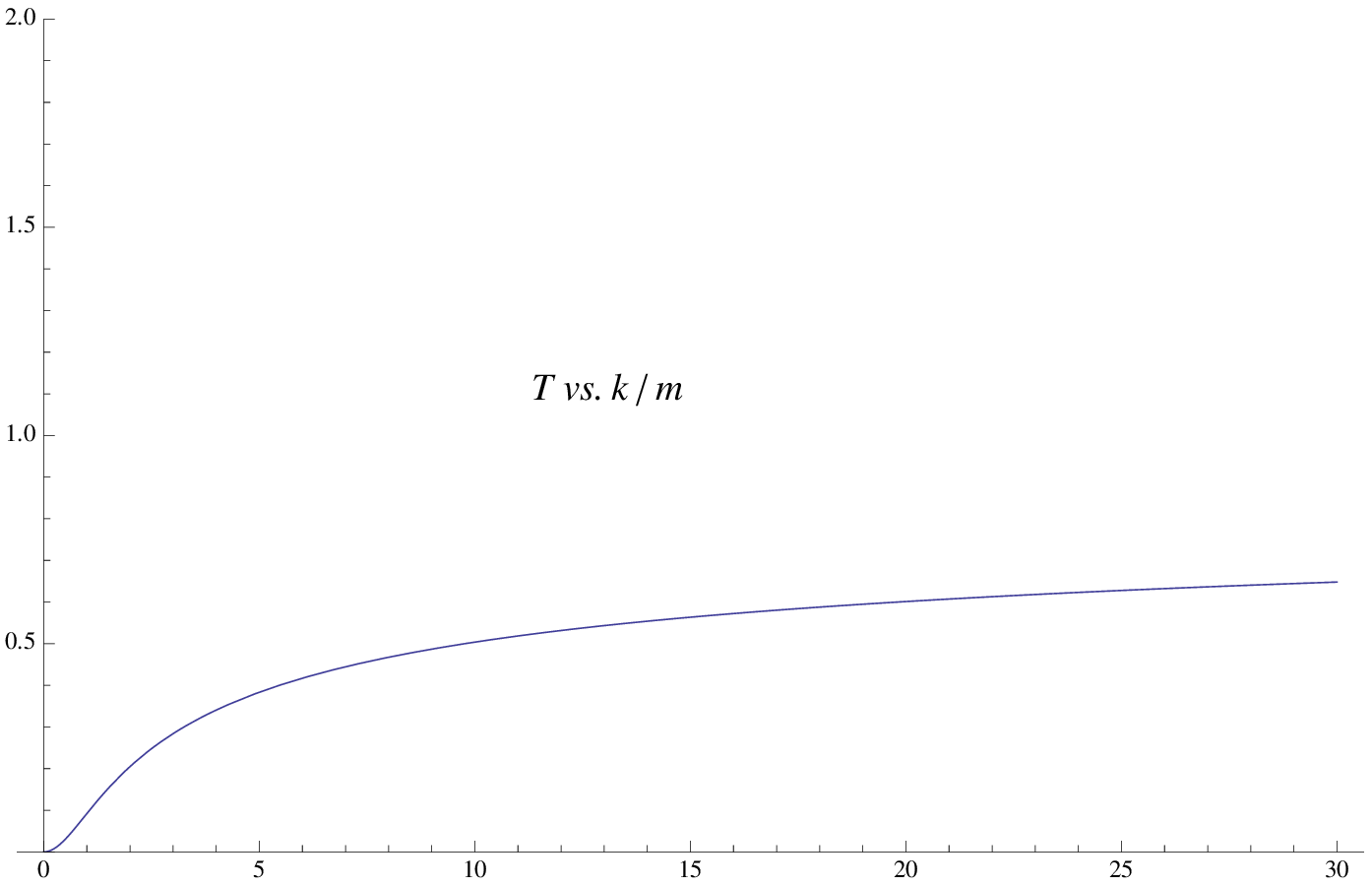}
\end{minipage}
\qquad \qquad \qquad \qquad
\begin{minipage}{5cm}
\includegraphics[scale=0.5]{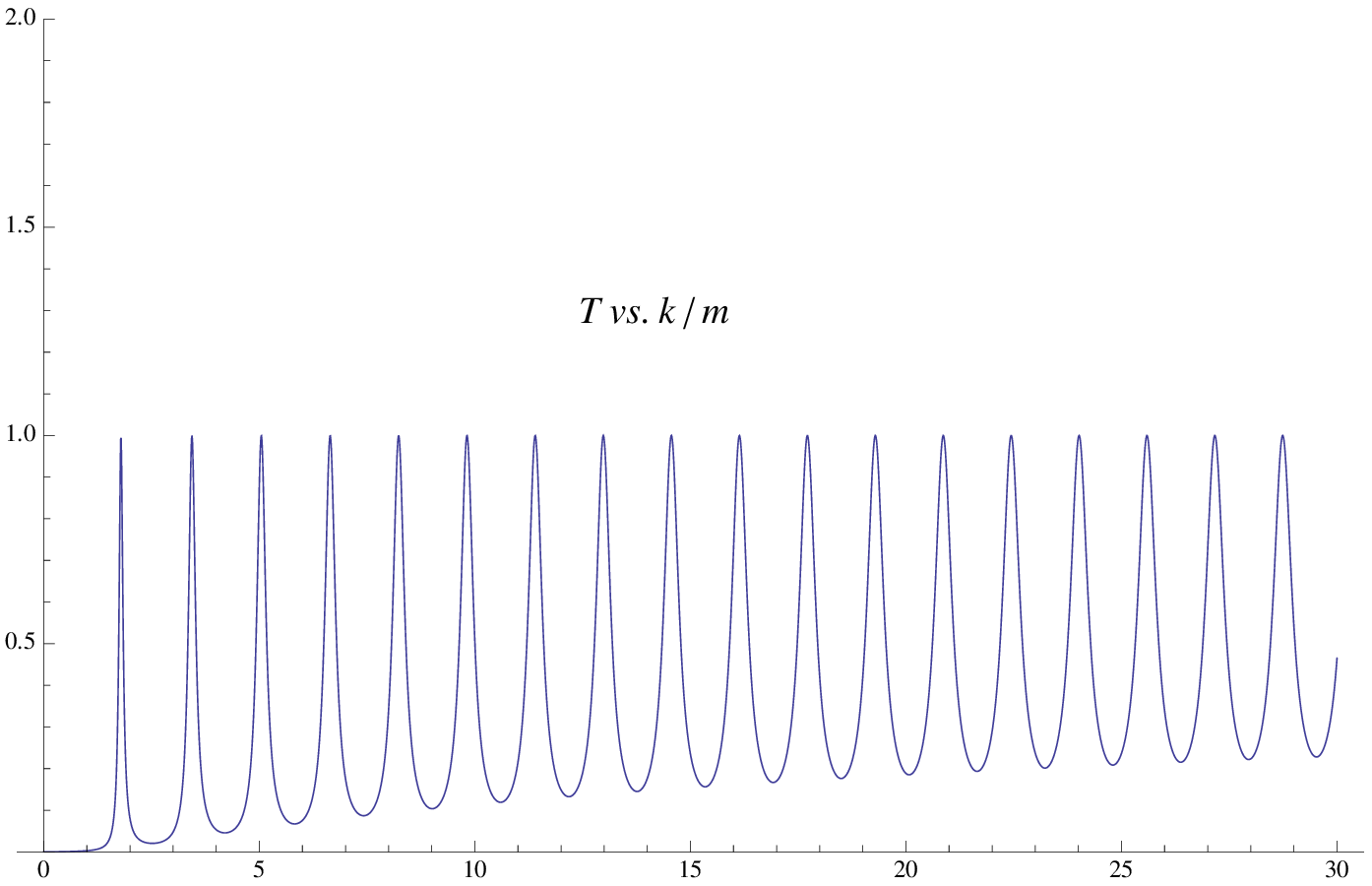}
\end{minipage} \\ 
\begin{minipage}{5cm}
\includegraphics[scale=0.5]{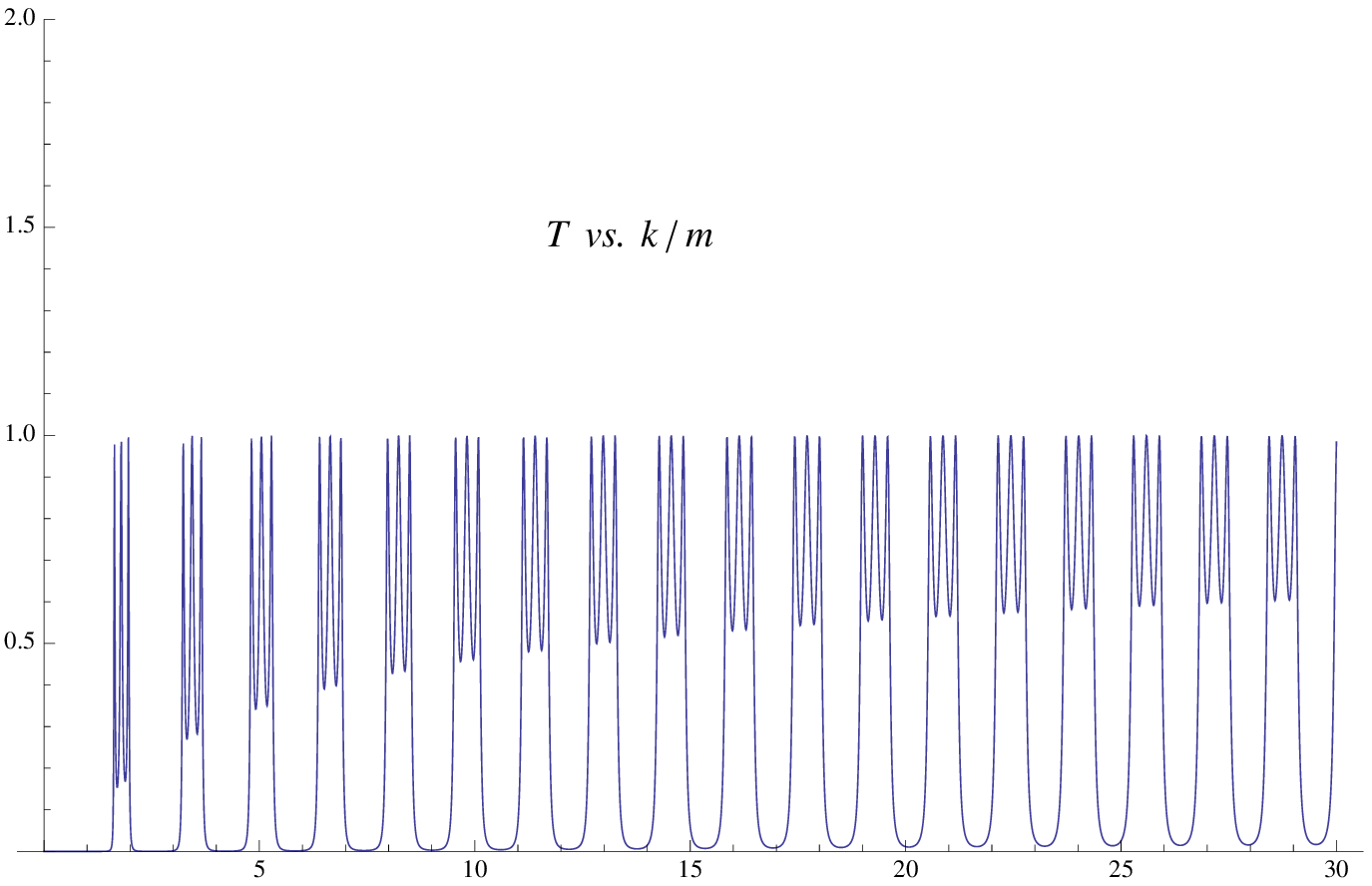}
\end{minipage}
\qquad \qquad \qquad \qquad 
\begin{minipage}{5cm}
\includegraphics[scale=0.5]{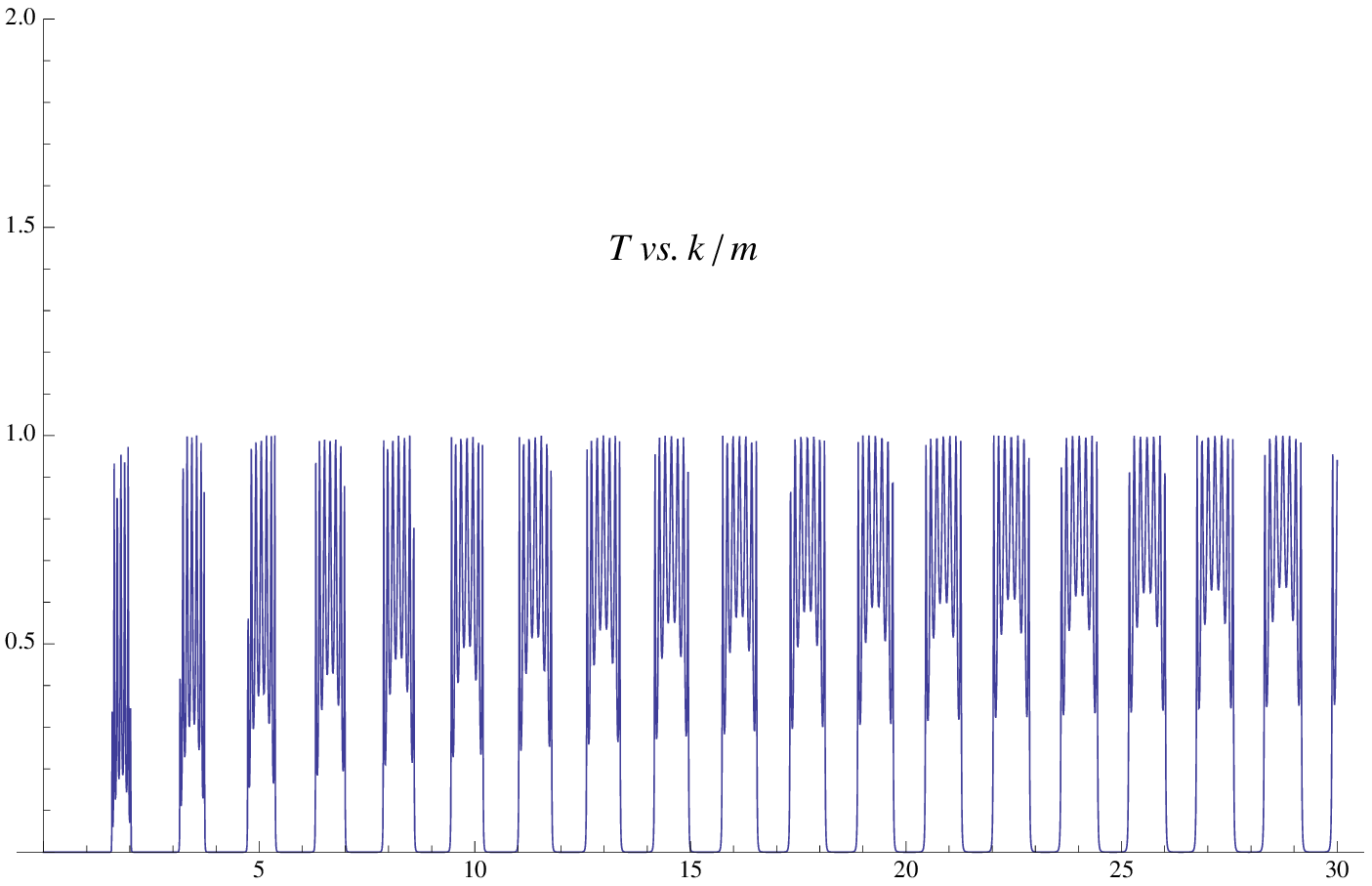}
\end{minipage}
\caption{The transmission coefficient as a function of $k/m$ for different values $N=1,2,4,8$, respectively. Here we choose that all $E_B^i$'s are the same, $E_B/m=1/10$, and $m|a_i-a_j|=2$.} \label{relativisticKPmodel}
\end{figure}

Let us consider now the array of Dirac delta potentials equally separated by some fixed distance, namely the semirelativistic Kronig-Penney model. In this case, the transmission coefficients in Fig. \ref{relativisticKPmodel} indicate the formation of the band gaps in the spectrum as we increase the number of centers. The nonrelativistic version of the problem by studying the transmission coefficient has been given in \cite{Rorres}.

To discuss the nonrelativistic limit of the reflection and transmission coefficients, we study the nonrelativistic limit (${E-m \over m} \ll 1$) of the scattering solution of the semirelativistic Lippmann-Schwinger equation (\ref{scatteting solution}).  The nonrelativistic limit of the principal matrix $\Phi_{ij}(E_k+i0)$ is
\beqs \label{NRlimitPhiscattering}
\Phi_{ij} (E_k+i0) \rightarrow
\begin{cases}
\begin{split} 
{1 \over \lambda_i} - {i m \over k}
\end{split}
& \textrm{if $i=j$} \\ \\
\begin{split}
- {i m \over k} \; e^{i k |a_i-a_j|}  
\end{split}
& \textrm{if $i \neq j$} \;,
\end{cases}
\eeqs
where we have used the fact that $-\lambda(E,E_B^i) \rightarrow \lambda_i$ in the nonrelativistic limit, which is shown for a single center in \cite{Al-Hashimi}. Here we have ignored the second integral term in the off-diagonal part of the principal matrix since 
\beqs
\Bigg| {1 \over \pi} \int_{m}^{\infty} d \mu \; e^{- \mu |a_i-a_j|} \; {\sqrt{\mu^2 -m^2} \over \mu^2 + (E_k -m)(E_k +m)} \Bigg| & = & \Bigg| {1 \over \pi} \int_{m}^{\infty} d \mu \; e^{- \mu |a_i-a_j|} \; {\sqrt{\mu^2 -m^2} \over \mu^2 + \eta(\eta+2) m^2} \Bigg| \nonumber \\ [2ex] \cr & \leq &\Bigg| {1 \over \pi} \int_{m}^{\infty} d \mu \; e^{- \mu |a_i-a_j|} \; {1 \over \sqrt{\mu^2 -m^2}} \Bigg|  = K_0(m|a_i-a_j|) \;,
\eeqs
which is of the order $O(1)$. The above limit (\ref{NRlimitPhiscattering}) is the principal matrix for the nonrelativistic version of the same problem, and it can be directly seen from 
Eq. (\ref{NRPhimatrix}). Then, we obtain the nonrelativistic limit of the scattering solution (\ref{scatteting solution})
\beqs
\psi_{k}^{+}(x)  \sim e^{ikx}+\sum_{i,j=1}^N   {i m \over k} \; e^{i k |x-a_i|} \; \left[\Phi^{-1} (E_k+i0)\right]_{ij} \; e^{i k a_j} \;,
\eeqs
where $\Phi_{ij}(E_k+i0)$ is given by (\ref{NRlimitPhiscattering}). Then, we can obtain the reflection and transmission coefficients from this solution, which is consistent with the standard results in the literature (see \cite{LS} for the two-center case).

%%%%%%%%%%%%%%%%%%%%%%%%%%%%%%%%%%%%%%%%%%%%%%%%%%%%%%%%%%%%%%%%%%%%%

\section{The Bound States and The Scattering Problem in the Massless Case}
\label{Massless Case}

\begin{figure}[h!]
%\centering
\begin{minipage}{5cm}
\includegraphics[scale=0.5]{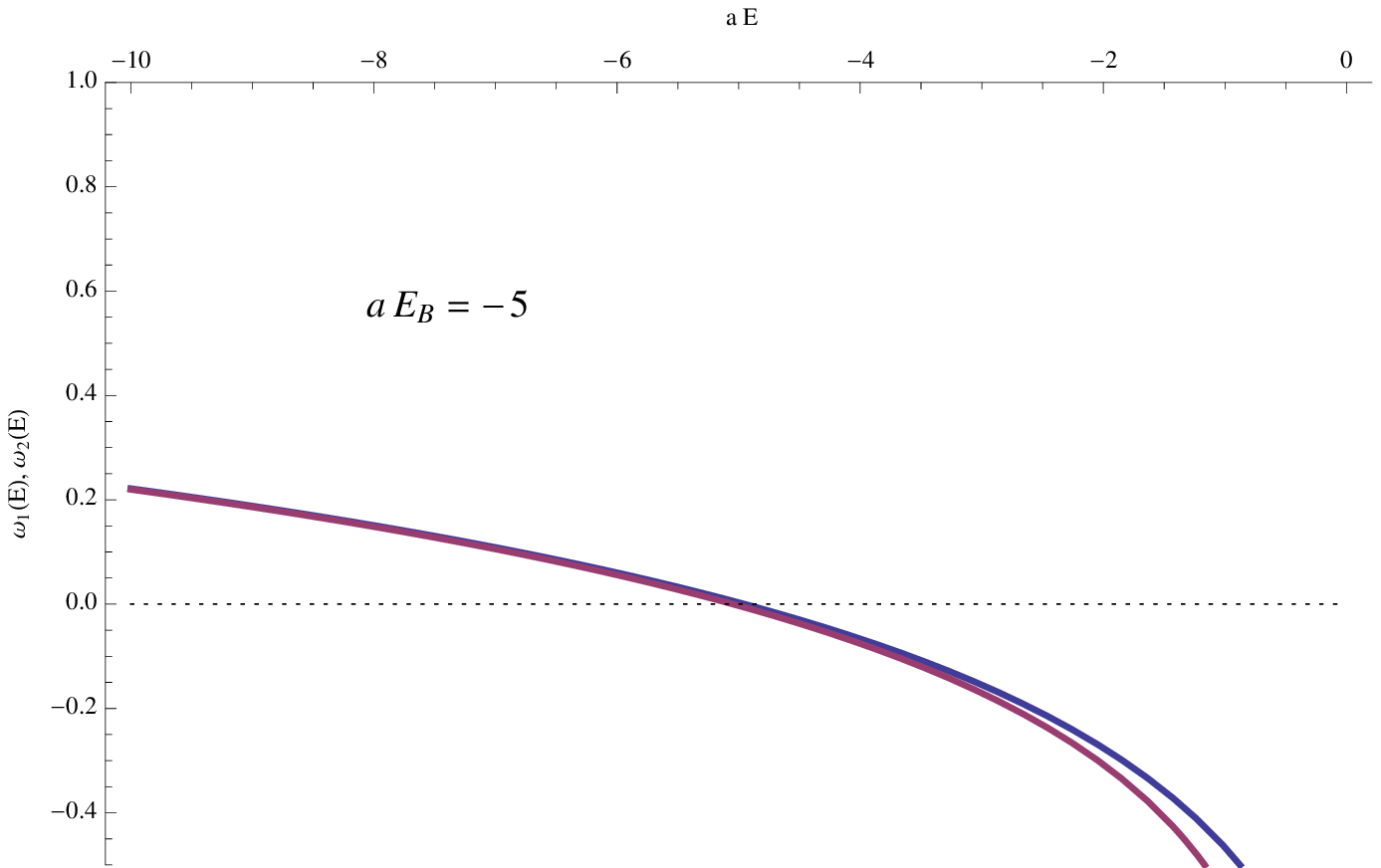}
\end{minipage}
\qquad \qquad \qquad 
\begin{minipage}{5cm}
\includegraphics[scale=0.5]{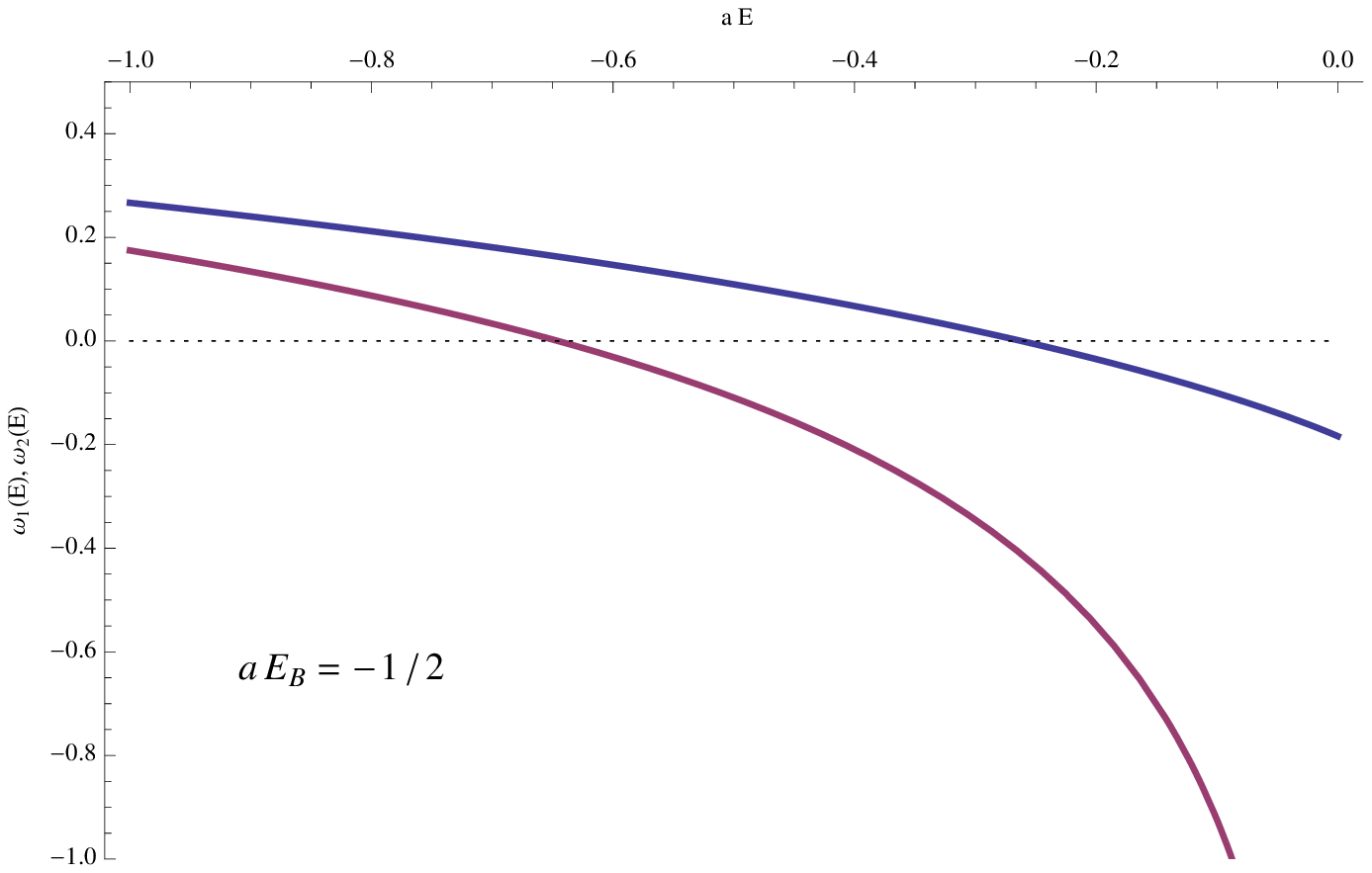}
\end{minipage} \\
\begin{minipage}{5cm}
\includegraphics[scale=0.5]{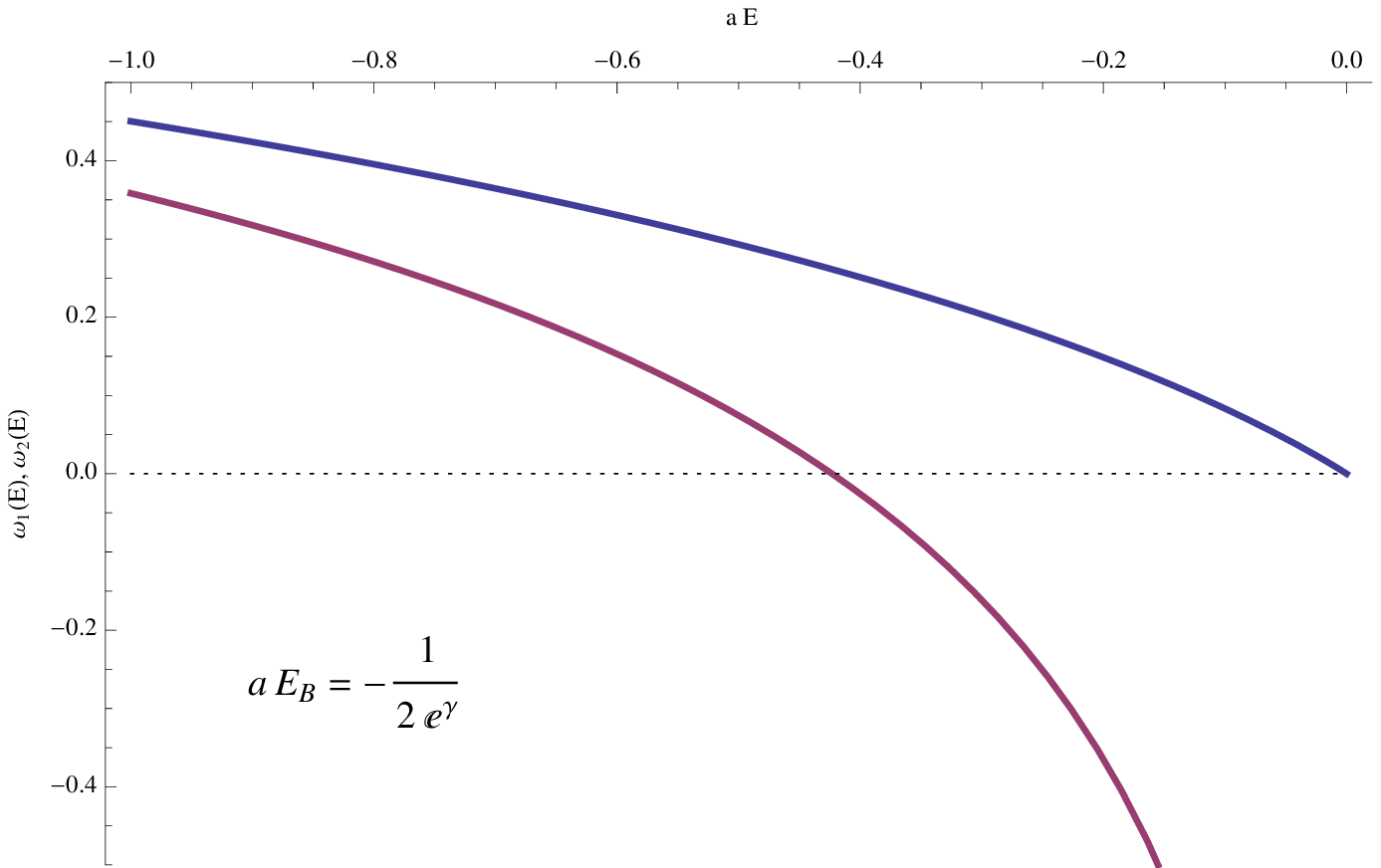}
\end{minipage}
\qquad \qquad \qquad
\begin{minipage}{5cm}
\includegraphics[scale=0.5]{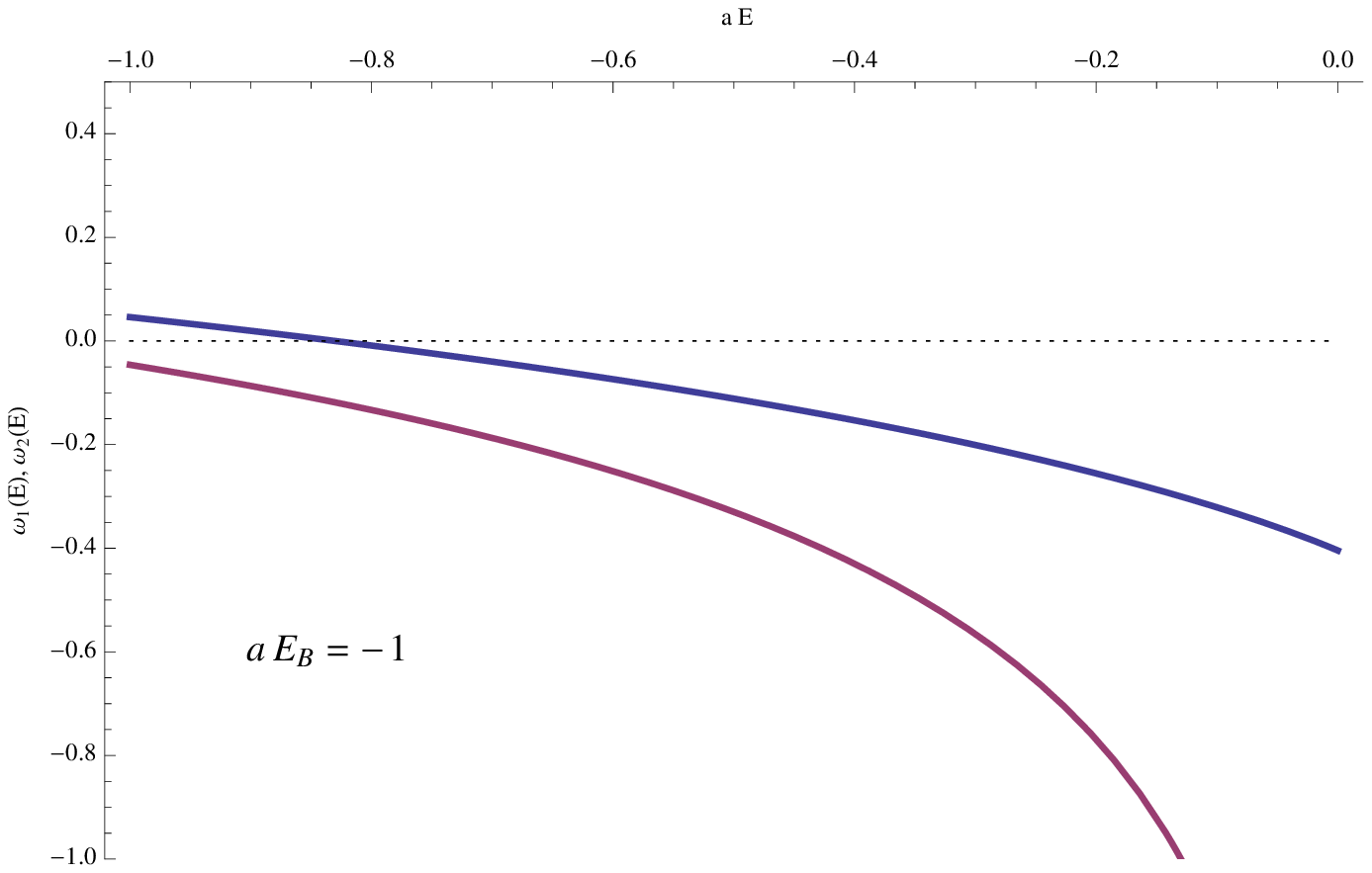}
\end{minipage}
\caption{The flow of the eigenvalues of the principal matrix as a function of $a E$ in the massless case for different values of  $a E_B$.} \label{eigenvaluesmassless}
\end{figure}

We first consider the bound state problem in the massless case $m=0$.  In this case, we have only ultrastrong bound states since they must occur in the negative $E$ axis. Using the explicit expression of the heat kernel (\ref{heatkernelmassless}), the principal matrix is
\beqs \label{Phimassless}
\Phi_{ij} (E) = 
\begin{cases}
\begin{split} 
 {1 \over \pi} \log \left( E/E_{B}^i \right)
\end{split}
& \textrm{if $i=j$} \\ \\
\begin{split}
&  {1 \over 2 \pi} \; \Bigg( 2 \cos \left(E (a_i-a_j) \right)  \; \ci \left(-E |a_i-a_j| \right) \; \\ & \hspace{4cm} + \sin \left(E |a_i-a_j| \right) \; \left(\pi + 2 \si \left(E |a_i-a_j| \right) \right)  \Bigg) \end{split}
& \textrm{if $i \neq j$} \;,
\end{cases}
\eeqs
where $E_B^i <0$ and $E$ is real and negative (for bound states). Here $\ci$ and $\si$ are the sine integral and the cosine integral functions defined by their integral representations \cite{Lebedev}
\beqs
\ci(x)  =  -\int_{x}^{\infty} d t \; {\cos t \over t} \;\;, \qquad  \si(x) & = & \int_{0}^{x} d t \; {\sin t \over t} \;.
\eeqs
For simplicity, we assume that $E_B^1= E_B^2= E_B$ and $a_1=-a_2=-a$ (twin symmetrically located centers). The bound state energies can be found from the transcendental equation $\det \Phi(E)=0$ or the zeros of the eigenvalues of the principal matrix (\ref{Phimassless}) as emphasized earlier. In contrast to the complications in the massive case, the eigenvalues can be explicitly calculated in this case and given by
\beqs
\omega^1(E) &=&\frac{ 2 \log \left(\frac{E}{E_B}\right) + 2 \cos \left(2 a E \right) \; \ci \left(-2 a E \right) +\pi  \sin \left(2 a E \right) + 2 \sin \left(2 a E \right) \; \si \left(2 a E \right)} {2 \pi } \;, \nonumber \cr \\ [2ex]  \omega^2(E) &=& \frac{2 \log \left(\frac{E}{E_B}\right) -2 \cos \left(2 a E \right) \; \ci \left(-2 a E \right) -\pi  \sin \left(2 a E \right) - 2 \sin \left(2 a E \right) \; \si \left(2 a E \right)} {2 \pi }  \;. \label{expressioneigenvaluesmassless}
\eeqs
Let us analyze the behavior of bound states for this case by plotting them as a function of $E$ for different values of $a E_B$. In Fig. \ref{eigenvaluesmassless}, one can apparently notice that the eigenvalues of the principal matrix become degenerate as we increase $|a E_B|$. 

This can be analytically justified  from the following fact: 
\beqs & & 
\lim_{a |E_B| \rightarrow \infty} 2 \cos\left(2 a |E_B| {E \over |E_B|} \right) \; \ci \left(-2 a |E_B| {E \over |E_B|} \right) +\pi  \sin \left(2 a |E_B| {E \over |E_B|} \right) \cr & & \hspace{7cm} + \, 2 \sin \left(2 a |E_B| {E \over |E_B|} \right) \; \si \left(2 a |E_B| {E \over |E_B|} \right) = 0
\eeqs
for all finite $E/|E_B|$. Hence we conclude that $\omega^1 \rightarrow \omega^2$ as $ a |E_B| \rightarrow \infty$ for all $E/|E_B|$. As a result of this, the bound states become degenerate.

It is also important to realize from Fig. \ref{eigenvaluesmassless} that the eigenvalues $\omega^1$ and $\omega^2$ are decreasing functions of $E$, as proved in Sec. \ref{On Bound States} [the proof for the massless case would be exactly the same except for the fact that the form of the heat kernel is given by (\ref{heatkernelmassless})]. 

Figure \ref{eigenvaluesmassless} also illustrates that  we may have one or two ultrastrong bound states depending on the choice of the values of $a E_B$.
It is not difficult from Eq. (\ref{expressioneigenvaluesmassless}) to show that $\lim_{E \rightarrow -\infty} \omega^1 = \infty$ and $\lim_{E \rightarrow 0^+} \omega^1 =-\infty$ for all $a$ and $E_B$. Since the eigenvalues are decreasing functions, $\omega^1$ must have exactly one zero. On the other hand, $\lim_{E \rightarrow -\infty} \omega^2 = \infty$ and $\lim_{E \rightarrow 0^+} \omega^2 =-\frac{ \gamma +\log(-2 a E_B)}{\pi }$ for all $a$ and $E_B$. Here $\gamma \approx 0.5772$ is Euler's constant. This means that $\omega^2$ could cross the $E$ axis only when 
\beqs
a |E_B| > {1 \over 2 e^{\gamma}} \;.
\eeqs
The second bound state appears only if the condition $a |E_B| > {1 \over 2 e^{\gamma}}$ is fulfilled. The point $E$ at which $\omega^1$ has a simple zero is the ground state energy.

Alternatively, this critical value can also be estimated analytically by working out the characteristic equation $\det \Phi(E)=0$, whose solutions are the bound state energies,
\beqs
\log \left(\frac{E}{E_B}\right) = \pm \Bigg[ {2 \cos \left(2 a E \right) \; \ci \left(-2 a E \right) + 
 \sin \left(2 a E \right) \; \left(\pi + 2 \si \left(2 a E\right) \right) \over 2} \Bigg] \;. \label{LHSRHSboundstates}
\eeqs

The principal matrix $\Phi(E_k+i0)$ in the scattering problem turns out to be
\beqs \label{Phiscatteringmassless}
\Phi_{ij} (E_k+i0) = 
\begin{cases}
\begin{split} 
i + {1 \over \pi} \; \log \left( {k \over |E_{B}^i|}\right)
\end{split}
& \textrm{if $i=j$} \\ \\
\begin{split}
&  {1 \over 2 \pi} \; \Bigg( 2 \cos \left(k(a_i-a_j) \right)  \ci \left(-k|a_i-a_j| \right) \; \\ & \hspace{3cm} + \sin \left(k|a_i-a_j| \right)  \left(\pi + 2 \si \left(k|a_i-a_j|\right) \right)  \Bigg)\end{split}
& \textrm{if $i \neq j$} \;.
\end{cases}
\eeqs
Then, the scattering solution to the semirelativistic Lippmann-Schwinger equation is
\beqs
\psi_{k}^{+}(x) \sim e^{ikx}+\sum_{i,j=1}^N   i \; e^{i k |x-a_i|} \; \left[\Phi^{-1} (E_k+i0)\right]_{ij} \; e^{i k a_j} \;, \label{massless scatteting solution}
\eeqs
where $\Phi(E_k+i0)$ is given by Eq. (\ref{Phiscatteringmassless}).
Hence, we can analytically find the reflection and transmission coefficient
\beqs
R(k) & = &\Bigg| \sum_{i,j=1}^N   i \; \left[\Phi^{-1} (E_k+i0)\right]_{ij} \; e^{i k (a_i+a_j)} \Bigg|^2 \label{reflectionNdeltamassless} \cr 
T(k) & = & \Bigg| 1+ \sum_{i,j=1}^N   i \; \left[\Phi^{-1} (E_k+i0)\right]_{ij} \; e^{i k (-a_i+a_j)} \Bigg|^2 \label{reflectionNdeltamassless} \;.
\eeqs
For $E_B^1=E_B^2=E_B$ and $a_1=-a_2=-a$, the behavior of the reflection coefficient as a function of $k a$ is shown below for particular values of $a |E_B|$.
\begin{figure}[h!]
\begin{center}
\includegraphics[scale=0.5]{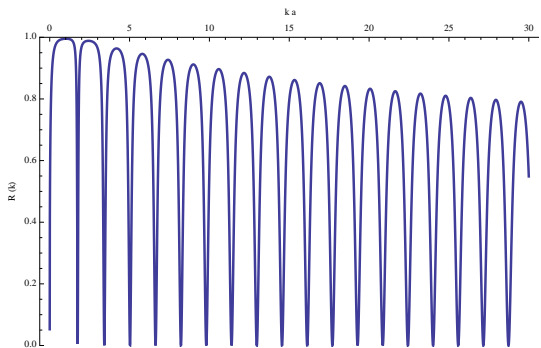}
\end{center}
\caption{The reflection coefficient as a function of $k a$ for a pair of symmetrically located centers in the massless case 
for $a|E_B|=1/2$.} \label{R versus k massless}
\end{figure}
This is also a  typical behavior of the reflection coefficient in the nonrelativistic case \cite{LS, AKSS}. The particle is fully transmitted at some certain energies that can be seen easily from Fig. \ref{R versus k massless}. Also the above graph is plotted for $a |E_B|=1/2$. In contrast to the massive and the nonrelativistic problems, the reflection coefficient is always zero for small values of $k a$ no matter what value $a|E_B|$ is. In the massive and the nonrelativistic cases, the reflection coefficient is always unity for very small values of $k/m$. Nevertheless, an anomalous behavior is also observed in this case when $a|E_B|={1 \over 2 e^\gamma}$ (this critical value corresponds to the condition for the appearance of a new bound state near $E=0$). This can easily be seen by plotting the reflection coefficient as a function  $a|E_B|$ near $k=0$ ($k a=0.01$). Around $a|E_B|$ in Fig. \ref{R versus a massless}, the reflection coefficient suddenly drops to zero at this critical value of $a|E_B|$ for small values of $k a$. Note that there is a curious sudden change near $a|E_B|=0$. However, our model is not properly defined when the centers coincide as long as $E_B$ is nonzero (recall that $a \neq 0$ as  defined in Sec. \ref{Renormalization of Relativistic Finitely Many Dirac delta Potentials through Heat Kernel}). 
\begin{figure}[h!!]
\begin{center}
\includegraphics[scale=0.5]{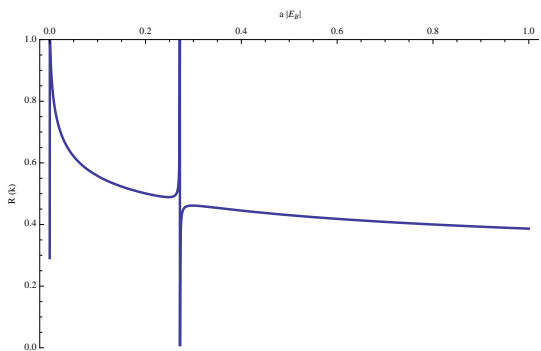}
\end{center}
\caption{The reflection coefficient as a function of $a |E_B|$ for a pair of symmetrically located centers in the massless case 
($k a=0.01$).} \label{R versus a massless}
\end{figure}
More interesting, the reflection coefficient always vanishes as $k \rightarrow 0$ in contrast to the nonrelativistic and massive case. Nevertheless, the threshold anomaly still occurs very close to the threshold energy.  

The massless problem is a simple quantum mechanical model where we have an explicit example of dimensional transmutation. Initially, the problem has no intrinsic energy scale, but we obtain an energy scale through the renormalization procedure.

%%%%%%%%%%%%%%%%%%%%%%%%%%%%%%%%%%%%%%%%%%%%%%%%%%%%%%%%%%%%%%%%%%%%%

\section{The Renormalization Group Equations and the Beta Function for $N$ centers}
\label{RGEquationssection}

One possible way for the
renormalization scheme to determine how the coupling
constant changes with the energy scale is to define the following
renormalized coupling constant $\lambda^{R}_{i}(M)$ in terms of
the bare coupling constants $\lambda_i(\epsilon)$:
\beqs {1 \over \lambda^{R}_{i}(M_i)} = {1 \over
\lambda_{i}(\epsilon)} - \int_{\epsilon}^{\infty} d t \;
{e^{- M_{i} t} \over \pi t} \;, \label{renormalized coupling constantsRG} \eeqs
where $M_i$ is the renormalization scale. Then, the renormalized principal matrix in terms of
the renormalized coupling constant is 
\beqs \Phi_{ij}^{R} (E) =
\begin{cases}
\begin{split} {1 \over \lambda_{i}^{R}(M_i)}-
\int_0 ^\infty d t  \left( K_{t}(a_i,a_i) e^{ t E} -
{e^{- M_{i} t} \over \pi t}\right)
\end{split}
& \textrm{if $i = j$} \\ \\ 
\begin{split}
- \; \int_0^\infty d t \;  K_{t}(a_i,a_j)e^{t E}
\end{split}
& \textrm{if $i \neq j$} \;,
\end{cases}
\eeqs
and the bound state energy is determined from the condition $\det
\Phi_{ij}^{R} (E) =0$ which gives  the relation between
$\lambda_{i}^{R}(M_i)$ and $M_i$. Here the integral in the diagonal part of the matrix $\Phi$ is convergent due to the short time asymptotic expansion of the Bessel function $K_1(mt) \sim {1 \over \pi t}$ as $t \rightarrow 0$. Explicit dependence on $M_i$
cancels out the implicit dependence on $M_i$ through
the renormalized coupling constant $\lambda_{i}^{R}(M_i)$. Physics is determined by the value of
the renormalized coupling constant at an arbitrary value of the
renormalization point $M_i$. However, the above choice of $\lambda_{i}^{R}(M_i)$ may not be physically  appropriate since we have to deal with more than one
renormalized coupling constant with the same type of interaction,
which essentially differ from  each other by arbitrary
constants. These constants can be determined by deciding the
excited energy levels. We instead prefer a single renormalized
coupling constant by redefining it without altering the physics of the problem. This could be performed in the following way. 

Instead of using the bound state energy to fix the flow, we may fix the
relative strengths of individual delta interactions. We know that
$E_{B}^i$ is the bound state energy for the individual $i$ th
Dirac delta center so that it corresponds to the solution
$\Phi_{ii}^{R}(E_{B}^i)=0$. Without loss of generality, let
us assume that $\Phi_{11}^{R}(E_B^1)=0$. This allows us
to choose the renormalized coupling constant as
\beqs {1 \over \lambda_{R}(M)} = {1 \over \lambda_{1}(\epsilon)} -
\int_{\epsilon}^{\infty} d t \;  {e^{- M t} \over \pi
 t } \label{renormcoupling 1} \;, \eeqs
at some scale $M$. Once the renormalized coupling constant is
fixed under this condition, we must also impose
$\Phi_{ii}^{R}(E_B^{2})=0$ for $i\neq 1$ with this choice at
the same scale $M$. This is always possible if we add a constant
term to the definition of a renormalized coupling constant. Let us
consider the $i=2$ case
\beqs \Phi_{22}^{R}(E_B^{2}) &=& {1 \over \lambda_{R}(M)} +
\int_{0}^{\infty} d t \; \left(  {e^{- M t} \over 
\pi t} - K_{t}(a_2,a_2) e^{t E_B^{2}} \right) - \Sigma_2 \nonumber \\ [2ex] &=& \int_0
^\infty d t \;  \left( K_{t}(a_1,a_1) e^{t E_B^{1}}-
 K_{t}(a_2,a_2) e^{t
E_B^{2}} \right) - \Sigma_2 =0 \;, \eeqs
where we have used Eq.(\ref{renormcoupling 1}) and
$\Phi_{11}^{R}(E_B^{1})=0$. This means that there always
exists a constant $\Sigma_i$ depending only on $E_B^i$ with
$\Sigma_1=0$ and $\Sigma_i \neq 0$ for $i\neq 1$ such that the
condition $\Phi_{ii}^{R}(E_B^i)=0$ can be fulfilled. Hence,
the renormalized coupling constant becomes
\beqs {1 \over \lambda_{R}(M)} = {1 \over \lambda_{i}(\epsilon)} -
\int_{\epsilon}^{\infty} d t \; {e^{- M t} \over
\pi
 t} + \Sigma_i \label{renormcoupling} \;, \eeqs
and the choice of $\Sigma_i$ refers to the relative strengths of
delta interactions in this new renormalization scheme.
If all $E_B^{i}$ are the same, then $\Sigma_i=0$. We can explicitly determine the renormalized constant by evaluating the integral and removing $\epsilon$,
\beqs & & 
{1 \over \lambda_R(M)} = \frac{E_B^i}{\sqrt{m^2-(E_B^i)^2}} +\frac{1}{\pi } \left( \log \left({2M \over m}\right) -1 \right) \cr & & \hspace{5cm}+ {1 \over \pi} \bigg( \,_2F_1^{(1,0,0,0)} (0, 2;3/2; {m-E_B^i \over 2m}) + \,_2F_1^{(0,1,0,0)} (0, 2;3/2; {m-E_B^i \over 2m}) \cr & & \hspace{7cm} + \,_2F_1^{(0,0,1,0)} (0, 2;3/2; {m-E_B^i \over 2m})\bigg)
 +\Sigma_i \;, \label{renormcouplingexplicit}
\eeqs
where $\,_2F_1$ is the hypergeometric function \cite{Lebedev}. The superscripts on the hypergeometric functions denote the derivative with respect to each variable; e.g., $\,_2F_1^{(1,0,0,0)} (0, 2;3/2; {m-E_B^i \over 2m}) $ is the derivative of $\,_2F_1 (x, 2;3/2; {m-E_B^i \over 2m})$ with respect to $x$ evaluated at $x=0$. Although this is a rather complicated function, we will see that this gives us a simple formula for the $\beta$ function. The renormalized coupling constant (\ref{renormcouplingexplicit})  logarithmically vanishes for large values of energy $M$ as can easily be seen from its expression so that the particle becomes free in this limit. This is a phenomenon which appears in QCD and is called asymptotic freedom.

Then,
the renormalized principal matrix is
\beqs \Phi_{ij}^{R} (E) =
\begin{cases}
\begin{split} {1 \over \lambda_{R}(M)}-
\int_0 ^\infty d t \; \left( K_{t}(a_i,a_i) e^{ t E} -
{e^{-M t} \over \pi t}\right)-\Sigma_i
\end{split}
& \textrm{if $i = j$} \\ \\
\begin{split}
- \; \int_0^\infty d t \; K_{t}(a_i,a_j)e^{t E}
\end{split}
& \textrm{if $i \neq j$}.
\end{cases} \label{renormalized principal}
\eeqs
To find the beta function, we need the renormalization group equation, given by
\beqs M \; {d \Phi_{ij}^{R}(M,\lambda_R(M),E, m, |a_i-a_j|) \over d M} =\left( M {\partial \over \partial M} + \beta(\lambda_R)
{\partial \over \partial \lambda_R}
\right)\Phi_{ij}^{R}(M,\lambda_R(M),E, m, |a_i-a_j|)=0 \;,\label{renorm cond} \eeqs
where the beta function is
\beqs \label{beta function def} \beta(\lambda_R)= M \; {\partial
\lambda_R \over
\partial M} \;. \eeqs
The renormalization group equation  essentially tells us that physics should be independent of the renormalization scale. It is worth pointing out that the renormalization condition (\ref{renorm cond})
corresponding to the problem in the two-dimensional nonrelativistic version of the problem has been written in
terms of the $T$ matrix in \cite{Adhikari}. Using Eq. (\ref{renormalized principal})
in Eq. (\ref{renorm cond}), we can find the beta function

\beqs \beta(\lambda_R) = - {\lambda_{R}^{2} \over  \pi} \;. \label{betafunction} \eeqs
It is important to note that the beta function here is formally different from the one derived for a single center case \cite{Al-Hashimi} and that our formula (\ref{betafunction}) is much simpler than the one given in \cite{Al-Hashimi}. This difference is due to the choice of the renormalization condition, and the beta function has been expressed in terms of the energy-dependent running coupling constant $\lambda(E,E_B)$ in there. However, the physics is the same. 
The negativity of the beta function (\ref{betafunction}) implies that our model
is asymptotically free and the zero of it is $\lambda_R = 0$ so that it is an ultraviolet fixed point since 
$\lambda_R \rightarrow 0$ as $M \rightarrow \infty$. 
This result is consistent with the case when there is only one center in \cite{Al-Hashimi}. We realize that our convention is more convenient and simpler to investigate for more than one center. 
By integrating
\beqs \beta(\lambda_R) = \bar{M} {\partial \lambda_R(\bar{M})\over
\partial \bar{M}} = -{\lambda_{R}^{2}(\bar{M}) \over \pi} \eeqs
from  $\bar{M}= M$ to $\bar{M}=\alpha M$ with $\alpha>0$, we can find the flow
equation for the coupling constant
\beqs \lambda_R(\alpha M) = {\lambda_R(M) \over 1+ {1 \over \pi}
\lambda_R(M)
 \log \alpha} \;. \label{coupling const evolve} \eeqs
From the explicit expression of the
renormalized principal matrix, we can easily see that 
\beqs \Phi_{ij}^{R}(M, \lambda_R(M), \alpha E, \alpha m, \alpha^{-1} |a_i-a_j| ) =
\Phi_{ij}^{R}(\alpha^{-1} M, \lambda_R(M), E, m,|a_i-a_j|) \;. \eeqs
If we take the scale-invariant derivative with respect to $\alpha$ of both sides, we find the renormalization
group equation for the principal operator $\Phi_{ij}^{R}(M, \lambda_R(M), \alpha E, \alpha m, \alpha^{-1} |a_i-a_j|)$,
\beqs \alpha {d \over d \alpha} \Phi_{ij}^{R}(M,
\lambda_R(M), \alpha E, \alpha m, \alpha^{-1} |a_i-a_j|) + M {\partial \over
\partial M} \Phi_{ij}^{R}(M, \lambda_R(M), \alpha
E, \alpha m, \alpha^{-1} |a_i-a_j|) =0 \;, \eeqs
or
\beqs \left(\alpha {d \over d \alpha} -
\beta(\lambda_R) {\partial \over
\partial \lambda_R}\right) \Phi_{ij}^{R}(M, \lambda_R(M), \alpha
E, \alpha m, \alpha^{-1} |a_i-a_j|) =0 \;. \label{rg equation} \eeqs
If we postulate the following functional form for the principal
matrix:
\beqs \Phi_{ij}^{R}(M, \lambda_R(M), \alpha E, \alpha m, \alpha^{-1} |a_i-a_j|) =
f(\alpha) \Phi_{ij}^{R}(M, \lambda_R(\alpha M), E, m, |a_i-a_j|)
\label{funtional ansatz} \;,\eeqs
and substitute into Eq. (\ref{rg equation}), we obtain an ordinary
differential equation for the function $f$,
\beqs  \alpha {d f(\alpha) \over d \alpha} =0 \;.
\eeqs
This gives the solution $f(\alpha)=1$ using the initial condition
at $\alpha=1$. Therefore, we get
\beqs \Phi_{ij}^{R}(M, \lambda_R(M), \alpha E, \alpha m, \alpha^{-1} |a_i-a_j|) =
\Phi_{ij}^{R}(M, \lambda_R(\alpha M), m, |a_i-a_j|) \;, \label{grm scaling} \eeqs
which means that there is no anomalous scaling. 
We can also verify that if the renormalized coupling constant evolves as in Eq. (\ref{coupling const evolve}), the scaling relation
(\ref{grm scaling}) is satisfied.

For the massless case, the beta function is formally the same but the renormalized coupling constant is 
\beqs
{1 \over \lambda_R(M)} = {1 \over \pi} \log(-M/E_B^i) + \Sigma_i
\eeqs
When relative strengths are the same, i.e., $\Sigma_i=0$, we obtain the beta function
\beqs
\beta(\lambda_R)=-{\pi \over \left(\log(-{M \over E_B})\right)^2}
\eeqs
which is exactly the same formula as the one given for one delta center in \cite{Al-Hashimi}. Similar to the single center case, the model has both ultraviolet and infrared fixed points. 

\section{A Possible Extension of the Model}
\label{A Possible Extension of the Model}

The method we have developed for the model of a single semirelativistic particle interacting with finitely many pointlike Dirac delta potentials can be applied to more general types of singular interactions, e.g., Dirac delta potentials supported by curves in two dimensions and supported by surfaces in three dimensions. The nonrelativistic version of this kind of interactions has been studied from several points of view \cite{Burak1,Burak2,Exner}. The renormalization is required only if the codimension is two for the nonrelativistic case, whereas the semirelativistic case needs to be renormalized when the codimension is one.

Here we only illustrate how our method of renormalization can be performed for the general kind of the singular Dirac delta interactions without going into details of their spectrum. Let us consider a semirelativistic particle interacting with finitely many singular interactions, each of which is supported by arc-length parametrized closed regular curve $\Gamma_i$ of finite length $L_i$ in two dimensions. We  assume that each curve is not self-intersecting and there is no intersection among the curves as well. Then, the semirelativistic Schr\"{o}dinger equation  is
\begin{eqnarray}
\langle \mathbf{r} | \sqrt{P^2 +m^2} | \psi \rangle  - \sum_{i=1}^{n} {\lambda_i \over L_i} \left( \int_{\Gamma_i} d l_i \; \delta (\mathbf{r},\Gamma_i(s)) \right) \; \left( \int_{\Gamma_i} d l_i \; \psi(\Gamma_i(s)) \right) = E \psi(\mathbf{r}) \;,
\end{eqnarray}
where $d l_i=|\mathbf{v}_i(s)| d s$ is the $i$th integration line element,  $\mathbf{v}_i(s)=\dot{\Gamma_i}(s)$ is the tangent vector to the curve $\Gamma_i$, and $s$ is the arc-length parameter. Here $ \psi(\Gamma_i(s))$ is the restriction of the wave function $\psi(\mathbf{r})$ to the curve $\Gamma_i$. Note that the potential energy term in the above Schr\"{o}dinger equation has a nonlocal character. 

Similar to the formal definition of pointlike Dirac delta function $\langle \delta_a, \phi \rangle := \phi(a)=``\int_{-\infty}^{\infty} d x \; \delta(x-a) \;\phi(x)"$ for any test function $\phi$, the Dirac delta function supported by a closed arc-length parametrized curve $\Gamma_i$ of length $L_i$ can be defined formally \cite{Appel}
\begin{eqnarray}
\langle \delta_{\Gamma}, \phi \rangle :=\int_{\Gamma_i} d l_i \; \phi =  \int_{0}^{L_i} d s \; |\mathbf{v}_i(s)| \; \phi(\Gamma_i(s)) = ``\iint_{\mathbb{R}^2} d^2 \mathbf{r} \; \phi(\mathbf{r}) \;\int_{0}^{L_i} d s \; |\mathbf{v}_i(s)| \; \delta(\mathbf{r},\Gamma_i(s)) " \;,
\end{eqnarray}
from which we have
\begin{eqnarray}
\langle \mathbf{r} | \Gamma_i \rangle = \int_{0}^{L_i} d s \; |\mathbf{v}_i(s)| \; \delta(\mathbf{r},\Gamma_i(s)) \;.
\end{eqnarray}
In analogy with the regularization of point Dirac delta potential with the heat kernel, we introduce
\begin{eqnarray}
\langle \mathbf{r} | \Gamma_i^{\epsilon} \rangle = \Gamma_i^{\epsilon} (\mathbf{r})= \int_{\Gamma_i} d l_i \; K_{\epsilon/2} (\mathbf{r},\Gamma_i(s)) \;.
\end{eqnarray}   
It is important to notice that as $\epsilon \rightarrow 0^+$, we obtain the delta function supported by the curve $\Gamma_i$. Moreover, we have
\begin{eqnarray}
\langle \Gamma_i^\epsilon |\Gamma_j^\epsilon \rangle = \iint_{\Gamma_i \times \Gamma_j} d l_i \; dl'_j \; K_{\epsilon/2} (\Gamma_i(s),\Gamma_j(s')) \;. 
\end{eqnarray}
We can then write the regularized semirelativistic Schr\"{o}dinger equation 
\begin{eqnarray}
\left(H_0 - \sum_{i=1}^{N} {\lambda_i(\epsilon) \over L_i}  |\Gamma_j^\epsilon \rangle \langle \Gamma_i^\epsilon | \right) | \psi \rangle = E |\psi \rangle \;.
\end{eqnarray}
Following the same line of arguments introduced in Sec. \ref{Renormalization of Relativistic Finitely Many Dirac delta Potentials through Heat Kernel} for pointlike Dirac delta potentials, we obtain the resolvent after the renormalization of the coupling constant
\begin{eqnarray}
\label{renormalized resolvent for curve}
R(E) = (H_0-E)^{-1} + \left(H_0 - E \right)^{-1} \left( \sum_{i,j=1}^N | \Gamma_i \rangle
\left[\Phi^{-1}(E) \right]_{ij} \langle \Gamma_j| \right) \left( H_0 - E
\right)^{-1}\;.
\end{eqnarray} 
Here, the principal matrix is defined as 
\begin{eqnarray} 
\label{Phimatrixheatkernelcurve} \Phi_{ij} (E) =
\begin{cases}
\begin{split} {1 \over L_i}
 \iint_{\Gamma_i \times \Gamma_i} dl_i \; dl'_i \; \int_{0}^{\infty} d t  \; (e^{t E_{B}^{i}} - e^{t E}) \; K_{t} (\Gamma_i(s),\Gamma_i(s'))
\end{split}
& \textrm{if $i = j$} \\
\begin{split}
- {1 \over \sqrt{L_i L_j}} \iint_{\Gamma_i \times \Gamma_j} dl_i \; dl'_j \; \int_{0}^{\infty} d t  \; K_{t} (\Gamma_i(s),\Gamma_j(s')) \; e^{t E}
\end{split}
& \textrm{if $i \neq j$} \;.
\end{cases}
\end{eqnarray} 
Similarly, we can apply our method to the Dirac delta potentials supported by a regular surface in three dimensions. This analysis can be even further extended to the curved manifolds; see the nonrelativistic discussion of it in \cite{Burak1, Burak2}.

%%%%%%%%%%%%%%%%%%%%%%%%%%%%%%%%%%%%%%%%%%%%%%%%%%%%%%%%%%%%%%%%%%%%%

\section{Conclusions}
\label{Conclusions}

In conclusion, we have considered in this paper the one-dimensional spinless Salpeter Hamiltonian with finitely many Dirac delta potentials. Similar to the one-center case, the problem requires renormalization. We have constructed the resolvent formula by using heat kernel regularization and renormalizing the model. 
We have discussed the bound state spectrum and proved that the ground state energy is bounded from below. Then, we have shown that there exists a unique self-adjoint operator associated with the resolvent formula. We have obtained an explicit wave function formula for $N$ centers and illustrated the fact that our problem is actually consistent with the self-adjoint extension theory in mathematics literature. We have also proved that the ground state is nondegenerate and discussed some new results on the number of bound states.
Moreover, we have solved exactly the semirelativistic Lippmann-Schwinger equation and found an explicit expression for the reflection and transmission coefficients. We have studied the behavior of the reflection and transmission coefficients for the two-center case numerically and approximately and observed the threshold anomaly that also exists in the nonrelativistic problem. We have found that this anomaly is  due to the appearance of the bound state appearing just near the threshold energy. In particular, we have analytically analyzed the bound state and scattering problem in the massless version of the problem. Finally, we have derived renormalization group equations and computed the beta function for the model.
We hope that our construction using the heat kernel techniques can be generalized to the many-body version of the problem so that all the techniques we have developed here can guide us for more complicated field theoretical problems.

%%%%%%%%%%%%%%%%%%%%%%%%%%%%%%%%%%%%%%%%%%%%%%%%%%%%%%%%%%%%%%%%%%%%%

\section*{Appendix A: A Proof of the Analyticity of the Principal Matrix}

We first recall the following theorem (theorem 1.1 in Chapter 2 of \cite{Olver}):

Assume that the function $f(z,t)$ [$z$ is a complex variable ranging over a domain $\mathcal{R}$ and $t$ is a real variable over $(0,\infty)$] satisfies: (i) $f(z,t)$ is a continuous function of both variables. (ii) For each fixed value of $t$, $f(z,t)$ is a holomorphic function of $z$. (iii) The integral $F(z)=\int_{0}^{\infty} f(z,t)\; d t$ converges uniformly at both limits in any compact set in $\mathcal{R}$. Then, $F(z)$ is holomorphic in $\mathcal{R}$ and its derivatives of all orders may be found by differentiating under the integral sign.

The above two hypotheses for the matrix elements of the principal matrix $\Phi$ are satisfied  
since the heat kernel $K_t(x,y)$ defined on $\mathbb{R} \times \mathbb{R} \times (0,\infty)$ is  $C^1$ - a continuously differentiable function with respect to the variable $t$ and exponential function $e^{t z}$ is an entire function for each fixed value of $t$.   What is left is to show that all the matrix elements converge uniformly on a compact subset of the chosen region $\mathcal{R}$. Let $\mathcal{R}$ be the complex plane with $\Re(z)<m$.
Here we choose the compact subset of the region as $\mathcal{D}=\{ z \in \mathbb{C}|  - \epsilon_2 < -{m \over 2} \leq \Re(z) \leq \epsilon_1 < {m \over 2} \; \& \; \eta_2 \leq \Im(z) \leq \eta_1 \}$.
We first prove the uniform convergence for the diagonal part of the principal matrix on $\mathcal{D}$. Using the upper bound of the Bessel function given in Eq. (\ref{besselupperbound}) we have
\beqs
|K_t(a_i,a_i) \; (e^{-t\mu_i^2} -e^{t z}) | & < & {m \over \pi} \left({1 \over m t} + {1 \over 2} \right) \left| e^{t (E_{B}^{i}-{m \over 2})} -e^{t (z-{m \over 2})} \right| \;,
\eeqs 
for all $t>0$ and $i=1, \ldots, N$. If we define the following holomorphic function  $f(z)=- {m \over \pi} \left({1 \over m t} + {1 \over 2} \right)   e^{t (z-{m \over 2})} $ for each value of $t>0$, then it is easy to show that $|f(z)-f(E_{B}^{i})| =|\int_{\gamma} f'(\zeta) d\zeta |\leq \max_{\zeta \in \mathcal{D}} |f'(\zeta)| L(\gamma)$ for any curve $\gamma$ connecting $E_{B}^{i}$ to any $z$ in the above compact region $\mathcal{D}$. Then, we can always choose $\gamma$ as a straight line on $\mathcal{D}$ connecting these points, i.e., $L(\gamma)=|z-E_{B}^{i}|$. Hence we obtain 
\beqs
|K_t(a_i,a_i) \; (e^{t E_{B}^{i}} -e^{t z}) | & < & |z - E_{B}^{i}| \;  {m \over \pi} \left({1 \over m t} + {1 \over 2} \right) \; t \; e^{-t m/2} \; \max_{\zeta \in D} e^{t \Re(\zeta) } \nonumber \\ [2ex] & < &  \sqrt{m^2 +(\eta_{2}-\eta_1)^2}\;  {m \over \pi} \left({1 \over m } + {t \over 2} \right) \; \;  e^{-t ({m \over 2} - \epsilon_1)}  \;,
\eeqs
and the right hand side of the inequality is integrable on the interval $(0,\infty)$. As for the off-diagonal matrix elements of the principal matrix, it is also integrable in the region $\mathcal{D}$ thanks to the upper bound (\ref{besselupperbound}). Hence, we show that all the matrix elements of the principal matrix are uniformly convergent on the compact subset $\mathcal{D}$ of $\mathcal{R}$ as a consequence of Weierstrass's $M$ test. Since all its matrix elements of $\Phi$ are holomorphic, the principal matrix $\Phi$ is a matrix-valued holomorphic function on $\mathcal{R}$, and 
the derivatives of all orders of $\Phi$ with respect to $z$ can be found by differentiating under the sign of integration. Then, its eigenvalues and eigenfunctions are also infinitely differentiable due to the corollary of Theorem II.6.1 in \cite{Kato}.

%%%%%%%%%%%%%%%%%%%%%%%%%%%%%%%%%%%%%%%%%%%%%%%%%%%%%%%%%%%%%%%%%%%%%

\section*{Appendix B: A Proof of the Existence of the self-adjoint Hamiltonian}

Equation (\ref{resolvent limit}) requires the following
condition to complete the second part of the proof:
\beqs || |E_k| R(E_k) |f\rangle -|f \rangle|| \rightarrow 0 \;, \eeqs
as $k\rightarrow \infty$, where $|f\rangle$ belongs to some appropriate Hilbert space and its usual $L^2$ norm is equal to one. Using the explicit expression of the full resolvent 
(\ref{renormalized resolvent}) and separating the free part, we can find an upper bound to the norm above that we are interested in, 
\beqs  || |E_k| R(E_k) |f \rangle - |f \rangle || & \leq & || \; |E_k|\;  R_0(E_k) |f \rangle - |f\rangle ||
\cr & +& |E_k| \; || \sum_{i,j=1}^{N} R_0(E_k) |a_i \rangle \left[\Phi^{-1}(E_k)\right]_{ij} \langle a_j | R_0(E_k) ||
\label{existence ham 2 3} \;, \eeqs
where we have used the triangle inequality and $||A|f\rangle|| \leq ||A||$ for bounded operator $A$. 
Let us first consider the first term in momentum representation by using the integral representation of the free resolvent $(H_0-E)^{-1}=\int_{0}^{\infty} d t \; e^{-t(H_0-E)} $.  It is easy to see that
\beqs
|| \; |E_k|\;  R_0(E_k) |f \rangle - |f\rangle || & = & |E_k|^2 \; \int_{\infty}^{\infty} {d p \over 2 \pi} \;|f(p)|^2 \; \int_{0}^{\infty} d t \; t \; e^{-t(\sqrt{p^2 +m^2}+|E_k|)}  \nonumber \\ [2ex]  & & \hspace{3cm} + \int_{\infty}^{\infty} {d p \over 2 \pi} \;|f(p)|^2 -2 |E_k| \int_{\infty}^{\infty} {d p \over 2 \pi} \; {1 \over \sqrt{p^2 +m^2} + |E_k|}\;|f(p)|^2 \nonumber \\ [2ex] & = & \int_{\infty}^{\infty} {d p \over 2 \pi} \; {(p^2 +m^2) \over (\sqrt{p^2 +m^2} + |E_k|)^2}\;|f(p)|^2 \nonumber \\ [2ex] & < & {1 \over 2 |E_k|} \int_{\infty}^{\infty} {d p \over 2 \pi} \sqrt{p^2 +m^2} |f(p)|^2 \;,
\eeqs
so that $|| \; |E_k|\;  R_0(E_k) |f \rangle - |f\rangle || \rightarrow 0$ as $k \rightarrow \infty$.

For the second term, let $A= \sum_{i,j=1}^{N} R_0(E_k) |a_i \rangle \left[\Phi^{-1}(E_k)\right]_{ij} \langle a_j | R_0(E_k)$ be a finite rank operator so that its norm
 is smaller than its Hilbert-Schmidt norm: $||A||
\leq Tr^{1/2}(A^{\dagger} A)$, where $Tr A^{\dagger} A = \int d x \; \langle x |A^{\dagger} A |x\rangle$.  Hence, we have
\beqs   |E_k|\;  ||A|| &
\leq & |E_k| \Bigg( \sum_{i,j,r,l=1}^{N} \int_{\mathbb{R}}
d x \; R_0(a_i,x|E_k) R_0(x,a_l|E_k) \nonumber \\ & &
\hspace{3cm} \times \int_{\mathbb{R}} d y \;
R_0(a_j,y|E_k) R_0(y,a_r|E_k) |\Phi_{ij}^{-1}(E_k)| \; |
\Phi_{rl}^{-1}(E_k)|\Bigg)^{1/2} \;. \label{resolvent inequality}
\eeqs
Let us first consider the diagonal case $l=i$ and $r=j$ for the
terms inside the bracket above.
\beqs  & & |E_k|\Bigg( \sum_{i,j=1}^{N} \int_{\mathbb{R}}
d x \; R_0(a_i,x|E_k) R_0(x,a_i|E_k) \cr & &
\hspace{3cm} \times \int_{\mathbb{R}} d y \;
R_0(a_j,y|E_k) R_0(y,a_j|E_k)
 |\Phi_{ij}^{-1}(E_k)| \; |\Phi_{ji}^{-1}(E_k)|\Bigg)^{1/2} \cr & & \hspace{1cm} \leq |E_k|\Bigg( N^2
\underset{1\leq i \leq N} \max \; \alpha_i(E_k) \;\;
\underset{1\leq j \leq N} \max \; \alpha_j(E_k) \;\;
\underset{1\leq i,j \leq N} \max \; |\Phi_{ij}^{-1}(E_k)|^2
\Bigg)^{1/2} \;, \label{norm bound} \eeqs
where we have defined $\alpha_i(E_k)=\int_{\mathbb{R}} d y \; R_0(a_i,y|E_k) R_0(y,a_i|E_k)$ for
simplicity. It is easy to see that $\alpha_i(E_k)$ is
\beqs \int_{\mathbb{R}} d x \;
R_0(a_i,x|E_k) R_0(x,a_l|E_k)  = 
\int_{0}^{\infty}\int_{0}^{\infty} \mathrm{d} t_1 \, \mathrm{d}
t_2 \; K_{t_1 + t_2} (a_i,a_l) e^{- (t_1 + t_2)|E_k|}  =   \int_{0}^{\infty} \mathrm{d} t \; t \; K_{t}
(a_i,a_l) \; e^{-t |E_k|} \;, \label{rzero2 heat kernel} \eeqs
by using the fact that the free resolvent kernel is just the
Laplace transform of the heat kernel. Using the explicit expression of the heat kernel (\ref{heatkernel}) and the upper bound of the Bessel function (\ref{besselupperbound}), we get
\beqs  \underset{1\leq i \leq N} \max \;
\alpha_i(E_k) < {1 \over \pi ({m \over 2}+|E_k|)} +{m \over 2 \pi ({m \over 2}+|E_k|)^2} \;. \label{upper bound of alpha} \eeqs
We have also
\beqs  \underset{1\leq i,j \leq N} \max \; |\Phi^{-1}_{ij}|^2 &
\leq & \underset{1\leq i \leq N} \max \; \sum_{j=1}^{N}
|\Phi^{-1}_{ij}|^2  = \underset{1\leq i \leq N} \max \;
(\Phi^{-1}(E_k)\Phi^{-1}(E_k))_{ii} \leq \rho(\Phi^{-2}(E_k)) \nonumber \\ [2ex]
& \leq & || \Phi^{-2}(E_k)|| \leq ||\Phi^{-1}(E_k)||^2 \eeqs
where we have used $\Phi^{\dag}(E_k)=\Phi(E_k)$ for $E_k \in
\mathbb{R}$ and $\rho$ is the spectral radius.

To find the upper bound for the norm of the inverse
principal matrix, we first decompose the principal matrix into two
positive matrices
\beqs \Phi = D - K \eeqs
where $D$ and $K$ stand for the on-diagonal and the off-diagonal
parts of the principal matrix, respectively. Then, it is easy to
see $\Phi = D(1-D^{-1}K)$. The principal matrix is invertible if
and only if $(1-D^{-1}K)$ is invertible. The matrix $(1-D^{-1}K)$ has an inverse if the
matrix norm satisfies $||D^{-1}K ||<1$. Then, we can write the
inverse of $\Phi$ as a geometric series,
\beqs \Phi^{-1}&=&(1-D^{-1}K)^{-1} D^{-1} = \left(1 +
(D^{-1}K) + (D^{-1}K)^2+ \cdots \right) D^{-1} \;,\eeqs
and the norm has the following upper bound:
\beqs || \Phi^{-1}|| & = & ||(1-D^{-1}K)^{-1} D^{-1}|| \leq
||(1-D^{-1}K)^{-1}|| \; || D^{-1}|| \leq  {1 \over 1-
||D^{-1}K||} \; ||D^{-1}|| \;. \eeqs
Since we are not concerned with the sharp bounds on the norm of
$\Phi^{-1}$ here, we can choose $|E_k|$ sufficiently
large such that $||D^{-1}K||< 1/2$ without loss of generality and
get
\beqs ||\Phi^{-1}(E_k)|| \leq 2 ||D^{-1}(E_k)|| \;, \eeqs
where $D^{-1}=
\mathrm{diag}(\Phi^{-1}_{11},\Phi^{-1}_{22},\ldots,
\Phi^{-1}_{NN})$ and 
\beqs ||D^{-1}|| = \underset{1\leq i \leq N} \max \;
|\Phi^{-1}_{ii}| \;. \eeqs
Since $D^{-1}$ and $K$ are decreasing functions of $|E_k|$, we can always make $||D^{-1} K|| <1/2$ by sufficiently large values of $|E_k|$. By using the lower bound of the Bessel function (\ref{bessellowerbound}), we find
\beqs
||D^{-1}(E_k)|| < {\pi \over \log\left( {m +|E_k| \over m-E_{B}^{i}}\right)} \;,
\eeqs
so that 
\beqs
|E_k| \; ||A|| < |E_k| \left[ 4 \pi^2 N^2 \left( {1 \over \pi ({m \over 2}+|E_k|)} +{m \over 2 \pi ({m \over 2}+|E_k|)^2} \right)^2 \sum_i {1 \over \log^2 \left( {m +|E_k| \over m-E_{B}^{i}}\right)} \right]^{1/2} \;.
\eeqs 
If we take the limit $k \rightarrow \infty$, the right hand side goes
to zero, the same analysis can be found similarly for the off-diagonal terms, and this completes the proof. Let us denote this densely defined closed operator as $H$.

Self-adjointness of $H$ is the consequence of the fact that
\beqs
H^{\dagger} -E = (R^{-1}(E^*))^{\dagger} = (R^{\dagger}(E^*))^{-1} = (R(E))^{-1}=H-E \;. \label{selfadjointH}
\eeqs
The self-adjointness also requires that the domains of $H$ and $H^*$ must be the same. This is actually the result of the above result (\ref{selfadjointH}) since the range $Ran(H-E)$ is the entire Hilbert space.

%%%%%%%%%%%%%%%%%%%%%%%%%%%%%%%%%%%%%%%%%%%%%%%%%%%%%%%%%%%%%%%%%%%%%

\section*{Acknowledgments}

The present work has been fully financed by TUBITAK from Turkey under the "2221 - Visiting  Scientist Fellowship Programme". We are very grateful to TUBITAK for this support. We also acknowledge Osman Teoman Turgut for clarifying discussions and his interest in the present research. Finally, we would like to mention that the present work follows the lines of Projects No. MTM2014-57129-C2-1-P and VA057U16 from Spain.

\end{document}